\documentclass{aa}  

\usepackage{graphicx}
\usepackage{orcidlink}
\usepackage{txfonts}
\usepackage{array}
\usepackage{booktabs}
\usepackage[section]{placeins}
\newcommand{\mdIV}{\texttt{D+IV}}
\newcommand{\mdI}{\texttt{D+I}}
\newcommand{\mdInu}{\texttt{D+I$\nu$}}
\newcommand{\mdIkappa}{\texttt{D+I$\kappa$}}
\newcommand{\mdV}{\texttt{D+V}}
\newcommand{\temprt}{T_{\rm MC}}
\newcommand{\temphd}{T_{\rm RHD}}
\newcommand{\quotes}[1]{``#1''}

\begin{document}

   \title{Monte Carlo post-processing for radiation hydro simulations of accreting planets in protoplanetary disks}

   \author{Anton~Krieger \orcidlink{0000-0002-3639-2435}
          \inst{1}
        \and
          Hubert~Klahr \orcidlink{0000-0002-8227-5467}
          \inst{2}
          \and
          Julio~David~Melon~Fuksman \orcidlink{0000-0002-1697-6433}
          \inst{2}
          \and
          Sebastian~Wolf \orcidlink{0000-0001-7841-3452}
          \inst{1}
          }

   \institute{Institut für Theoretische Physik und Astrophysik, Christian-Albrechts-Universität zu Kiel, Leibnizstra{\ss}e 15, 24118 Kiel, Germany\\
             \email{akrieger@astrophysik.uni-kiel.de} 
         \and
             Max Planck Institut f\"ur Astronomie, K\"onigstuhl 17, 69117 Heidelberg, Germany\\
             \email{klahr@mpia.de} 
             }

   \date{Received 3 August 2024 / Accepted 2 December 2024}

 
  \abstract
   {
    This paper is part of a series investigating the observational appearance of planets accreting from their nascent protoplanetary disk (PPD). We evaluate the differences between gas temperature distributions determined in our radiation hydrodynamical (RHD) simulations and those recalculated via post-processing with a Monte Carlo (MC) radiative transport (RT) scheme. Our MCRT simulations were performed for global PPD models, each composed of a local 3D high-resolution RHD model embedded in an axisymmetric global disk simulation. 
    We report the level of agreement between the two approaches and point out several caveats that prevent a perfect match between the temperature distributions with our respective methods of choice.
    Overall, the level of agreement is high, with a typical discrepancy between the RHD and MCRT temperatures of the high-resolution region of only about 10 percent. The largest differences were found close to the disk photosphere, at the transition layer between optically dense and thin regions, as well as in the far-out regions of the PPD, occasionally exceeding values of 40 percent. 
    We identify several reasons for these discrepancies, which are mostly related to general features of typical radiative transfer solvers used in hydrodynamical simulations (angle- and frequency-averaging and ignored scattering) and MCRT methods (ignored internal energy advection and compression and expansion work). This provides a clear pathway to reduce systematic temperature inaccuracies in future works. 
    Based on MCRT simulations, we finally determined the expected error in flux estimates, both for the entire PPD and for planets accreting gas from their ambient disk, independently of the amount of gas piling up in the Hill sphere and the used model resolution. 
}
  
   \keywords{Hydrodynamics --
                Radiative transfer --
                Methods: numerical --
                Protoplanetary disks --
                Planet-disk interactions --
                Accretion, accretion disks
               }

   \maketitle

\section{Introduction}
    Protoplanetary disks (PPDs) are multifaceted widely studied objects that offer the opportunity to understand the broader process of planet formation, as they bridge the early stages of PPD evolution with the later stages where planets emerge. One particularly intriguing phenomenon is that of planet--disk interactions \citep[see][for a recent review]{Paardekooper2023}, which on the basis of hydrodynamical (HD) simulations have been linked to the emergence of various substructures in the disk. Potentially planet-induced substructures have 
frequently been observed in circumstellar disks and include gaps, rings, spirals, vortices, and cavities \citep[e.g.,][]{2016A&A...588A...8G,2018A&A...620A..94G,2018ApJ...863...44A,2018ApJ...869L..41A}. To investigate their origin, a modeling procedure is often applied that involves the combination of HD simulations with Monte Carlo (MC) radiative transport (RT) simulations \citep[e.g.,][]{2005ApJ...619.1114W,2010A&A...518A..16F,2013A&A...549A..97R,Ober_2015,2016ApJ...826...75D}, which allows for simulated observations that can be compared with real observations. Using the observed disk structures as a basis to implicitly constrain the properties --- such as the mass and orbit --- of a potential protoplanet is an effective tool \citep{2018ApJ...864L..26B,2018ApJ...869L..47Z,2019MNRAS.486..453L}, but ambiguity nevertheless remains. Direct detections, such as in the cases of PDS 70\,b,c \citep{2018A&A...617A..44K,2018A&A...617L...2M,2019NatAs...3..749H,2019A&A...632A..25M} and AB AUR\,b \citep{2022NatAs...6..751C}, however, offer a more robust means of determining constraints,  yet opportunities for such detections are extremely rare. 
    
    In any case, the modeling procedure requires consistency between the HD and MCRT simulations as well as precise and reliable temperature distributions. 
    While some numerical studies of planet-disk interaction do include stellar irradiation \citep[e.g.,][]{2013A&A...560A..40M,2013A&A...549A.124B,2020A&A...637A..50Z,2020A&A...642A.219C,2024ApJ...973..153K}, many still rely on simplified temperature structures or neglect irradiation altogether. Among the recent works that incorporate irradiation via ray tracing, \citet{MelonFuksman2021} and \citet{Muley2023} have employed an M1 scheme for the reprocessed radiation. In this context, the present paper expands on the planet-disk interaction simulations by \citet{KlahrKley2006}, now incorporating the irradiation treatment described by \citet{Kuiper2010} and including accretion luminosity from the planet.

    The goal of the present study is to investigate the level of agreement between the temperature distributions predicted by our radiation hydrodynamical (RHD) and MCRT simulations of young accreting planets in their late stages of formation embedded in PPDs. Here, we consider two planetary masses ($10$ and $300\,{\rm M}_\oplus$) and vary the model resolution as well as the timescale for accretion of gas onto the planet. To this end, we performed 3D RHD simulations of these models and recalculated their temperature distributions using MCRT simulations while assuming radiative equilibrium. We quantified the relative differences between the two temperature distributions in different regions of the PPD, including the midplane, gap, photosphere, and (in particular) the Hill region of the planet. 

    In order to investigate the origin of the observed differences, we performed further simulations of axisymmetric, unperturbed protoplanetary disks (uPPDs), which consider different physical mechanisms (stellar irradiation and viscosity) or have variant numerical prescriptions (frequency-dependent irradiation and an optical depth limiter). Here, we identify various crucial aspects for future improvements, particularly with regard to a consistent treatment of opacities in RHD and MCRT simulations. 
    
    In order to assess the impact of these temperature discrepancies in the PPD models on simulated observations, we constructed ideal observations for observing wavelengths in the visual (VIS), near-infrared (NIR), and submillimeter (submm) wavelength ranges. To that end, we used MCRT flux simulations based on the RHD and MCRT temperature distributions. Instead of the frequency-averaged approach followed in the RHD simulations, MCRT explicitly incorporates the full frequency dependence of the dust. Comparing the resulting flux maps allowed us to determine and analyze the arising flux differences originating from the use of the RHD rather than the MCRT temperature distribution and to contrast these differences with the quantitative changes resulting from alteration of the model resolution. 
    
    The paper is structured as follows. In Sect. \ref{sec:methods}, we present our implemented RHD and MCRT schemes, introduce the utilized opacities, and describe the treatment of evaporation in our simulations. 
    The temperature comparison is presented in Sect. \ref{sec:temp_comparison_reference_model}. The origin of the resulting temperature discrepancies is investigated in Sect. \ref{sec:axisym}. In Sect. \ref{sec:flux_section}, we analyze the impact of these differing temperature distributions on ideal observations. A discussion regarding the origin of the main differences between the results of MCRT and RHD simulations is provided in Sect. \ref{sec:discussion_differences}. Finally, we present a summary, outline our conclusions, and  provide an outlook for future work in Sect. \ref{sec:summary}.
   
\section{Methods}
\label{sec:methods}

\subsection{Radiation hydrodynamical simulations}
A straightforward approach to initializing a hydrodynamical disk simulation is to begin with a power-law distribution for the gas density and temperature in the midplane of the disk. Under the assumption of zero radial velocity and vertical hydrostatic equilibrium, one can then derive a stable density and rotation profile. However, for a disk with alpha viscosity and temperature set by irradiation and viscous heating, no single power law adequately describes these quantities \citep{Bell1997}. Consequently, initializing a radiative hydrodynamic simulation with a simple power-law distribution for the density, for instance, requires allowing the disk to evolve viscously into a state of constant accretion rate. This process, though, is time-intensive and does not yield the desired total disk mass. Therefore, we first employed a global disk model based on a set of one-dimensional vertical disk atmospheres, as described in \citet{PfeilKlahr2019}, to initialize global axisymmetric hydrodynamic simulations. We then further relaxed this model by applying comprehensive radiative physics and viscous hydrodynamics while excluding radial mass transport. We thus obtained a global three-dimensional axisymmetric disk model with a predefined mass and viscosity and a consistent surface density profile.

\subsubsection{Global background model}
Our global axisymetric three-dimensional hydro simulation (henceforth 2.5D model)\footnote{We have two spatial dimensions, yet our velocities are three-dimensional, which means 2.5 dimensions, and this is equivalent to being three-dimensional yet axisymetric.} was initialized via the values obtained from the global 1+1D model described in \cite{PfeilKlahr2019} for a disk accretion rate of $3.18 \times 10^{-9}\,{\rm M}_\odot {\rm yr}^{-1}$ around a $0.5\,{\rm M}_\odot$ star with a radius of $R_* = 2.5\,{\rm R}_\odot$ and an effective temperature of $T_* = 4000\,$K, resulting in a luminosity of $L=1.43\,{\rm L}_\odot$. In combination with our chosen value of $\alpha = 3 \times 10^{-3}$, an inner radius at $R_\mathrm{i} = 0.1\,$au, and an exponential cutoff radius of the accretion rate at $R_c = 100\,$au led to a total disk mass of $1\%$ solar mass. At each of the 200 radial locations logarithmically spaced between $1$ and $300\,$au, the hydrostatic equilibrium equations were vertically integrated via finite differences in a cylindrical grid. Each vertical slice has 100 cells and spans from the midplane to considerably into the isothermal region $z/R = 0.6$, where the temperature is set by stellar irradiation. The benefit of this setup is that we already start from a consistent viscosity, accretion rate, and surface density profile, which typically does not follow a simple power law.
In these simulations, the initial sharply truncated disk is amended by a physical inner edge of Gaussian shape for the density structure following $\sigma_{\rm R} = 0.1 R_\mathrm{i} = 0.01 \mathrm{au}$ using
\begin{equation}
    \rho(R,\theta) = e^{-\frac{1}{2}\left(\frac{R - R_\mathrm{i}}{\sigma_{\rm R}}\right)^2}\rho(R_\mathrm{i},\theta),
\end{equation}
for $r<R_\mathrm{i}$, to account for proper irradiation of the inner rim and successive shadow casting.

This initial hydrostatic configuration was then evolved in the 2.5D model into an equilibrium configuration by solving the RHD equations described in the following section. These equations account for the effects of radial transport of radiation, which were absent in the 1+1D model. However, at this stage, we did not allow for radial mass transport, meaning that the surface density profile remains unchanged, while the temperature and vertical density structure are adjusted. 
Finally, this new 3D axisymmetric yet spherical disk structure extends from $0.04$ au to $300$ au and vertically to $34.4^\circ$ above and below the midplane. We used 512 logarithmically spaced cells in the radial direction and 256 cells in the polar direction. We show the results from these backbone simulations in Sect. \ref{sec:axisym} when comparing the differences of radiation transport in models without a planet.

Our 3D RHD with a planet cover a much smaller radial extent, for which we used the axisymmetric result for both the initial values as well as reference values for the radial damping layers.
The RHD simulations presented here span from $4$ au to $25$ au using 100, 200, or 400 cells in radial log space. The radial damping zones are four cells\footnote{We now consider using, respectively, 8 or 16 cells for the higher resolution cases to overcome some inconsistencies with associated simulations and their structure toward the inner and outer edge of the local models.} with a gentle damping time of ten local orbits, reassigning the initial density, velocity, and temperature levels from the 2.5D runs. This has two effects. For one, we damp wave reflections at the radial boundaries, and even more important, we are able to smoothly insert our 3D simulation into the global data set for the MCRT post-processing, as described in Sect. \ref{sec:model_embedement_and_parameters}.

\subsubsection{Numerical 3D radiation method}
\label{sec:num_3d_rhd}
We used the TRAMP hydro code as described in \citet{1999ApJ...514..325K} but amended by the radiation hydro scheme explored in \citet{Kuiper2010}. TRAMP employs Van Leer advection on a staggered mesh\footnote{While this approach does not conserve total energy to machine precision, it produces results comparable to those of Godunov solvers in problems involving the formation of gaps and spiral shocks, as demonstrated in the code comparison study by \cite{deValBorro2006}, which includes TRAMP (see also the discussions in \citet{Clarke2010} and \citet{BenitezLlambay2016}). Thus, we expect internal energy losses or gains due to the numerical scheme to be subdominant compared to the effects of accretion, compression, viscous dissipation, and radiative transfer.} combined with artificial viscosity to ensure the correct entropy jump across shocks.
The basic radiation transport is treated in the
flux-limited diffusion (FLD) approximation using the flux 
limiters of \citet{1981ApJ...248..321L} (see also \citet{1989A&A...208...98K}).
We used a simple one-temperature scheme here, which means that we added the equations for internal and radiation energy and assumed that
the latter is much smaller than the former, which is a good approximation for the optically thick regions of a protoplanetary disk. Thus, the equation for the internal energy is given as
\begin{equation}\label{Eq:internal_energy}
\partial_t e + \nabla \cdot \left(e {\bf u} \right) = - p \nabla \cdot {\bf u} + \Phi + L_\mathrm{acc} - \nabla \cdot {\bf F} \,,
\end{equation}
where ${\bf u}$ is the velocity vector, $p$ is the thermal pressure, $\Phi$
is the dissipation by viscosity, $\mathbf{F}$ is the radiation flux, and $L_\mathrm{acc}$ is the accretion luminosity of the planet, which we define at the end of this section. 
Tensor viscosity and shock viscosity are both 
included, where the former uses the same parameter $\alpha = 3 \times 10^{-3}$ as in the 1+1D model.
We note that the left-hand side of Eq. \ref{Eq:internal_energy} is solved explicitly, whereas the right-hand side is solved via an implicit scheme as part of the radiation transport step \citep{Kuiper2010}, as explained later in this section.
 
The radiation flux ${\bf F}$ has two components. One is calculated in the flux-limited diffusion approximation to represent the radiation emitted from the dust grains:
\begin{equation}\label{Eq:Frad}
 {\bf F}_\mathrm{FLD} = \, - \, \frac{\lambda c}{\rho \kappa_{\rm R}} \nabla E_{\rm R}
,\end{equation}
with the Rosseland mean opacity $\kappa_{\rm R}$ and the speed of light $c$.
The flux limiter $\lambda$ is defined according to \citet{1981ApJ...248..321L}. The other component, representing absorption of stellar irradiation, is computed as described in \cite{Kuiper2010} via ray tracing:
\begin{equation}\label{Eq:Firrad}
    \mathbf{F}_* = \sigma_\mathrm{SB} T_*^4 \left(\frac{R_*}{r}\right)^2 e^{-\tau(r)}\,,
\end{equation}
where $\sigma_\mathrm{SB}$ is the Stefan-Boltzmann constant and $T_*$ and $R_*$ are the effective temperature and radius of the star. The optical depth is determined using the Planck-averaged dust opacity at the stellar temperature, $\kappa_{\rm P}(T_*)$. 
The dust grains are in local thermal equilibrium with
the local diffuse radiation energy $E_{\rm R}$ and the direct stellar irradiation $\textbf{F}_\mathrm{*}$ such that
\begin{equation}
\label{eq:lte}
a T^4 = E_{\rm R} + \frac{\kappa_\mathrm{P}(T_\mathrm{*})}{\kappa_\mathrm{P}(T)}\frac{|\textbf{F}_\mathrm{*}|}{c}
,\end{equation}
where $a$ is the radiation constant.
Numerically, the radiative diffusion part is solved separately from the hydro advection step
as part of an operator splitting technique.
The diffusion equation for $E_{\rm R}$ also contains the heating and cooling sources resulting from hydrodynamics, namely, the expansion/compression work (PdV) and viscosity, which together with accretion luminosity of the planet $L_\mathrm{acc}$, add up to a source term: 
\begin{equation}
    Q^+ = - p \nabla \cdot {\bf u} + \Phi + L_\mathrm{acc}
.\end{equation}
Thus, following \cite{Kuiper2010},
\begin{equation}\label{Eq:Erad}
  \partial_t E_{\rm R} = f_c\left( \nabla \, \frac{\lambda c}{\rho \kappa_{\rm R}} \nabla E_{\rm R} - \nabla {\bf F_* } + Q^+ \right)\,\,,
\end{equation}
with $f_c=\left(c_V\rho/(4 a T^3 + 1)\right)^{-1}$, is solved implicitly to avoid time step limitations using, for instance, the standard successive over--relaxation scheme. The specific heat is computed as $c_V=\frac{k_B}{(\Gamma-1) \mu u}$, where $\Gamma=1.43$ is the assumed heat capacity ratio of the gas, $\mu=2.353$ is the mean molecular weight, $k_B$ is the Boltzmann constant, and $u$ is the atomic mass unit. We note that we find that the total integral over PdV is positive and typically on the same order of magnitude as the integrated viscous heating. However, while the latter is positive definite, the local PdV term can be positive as well as negative, with an amplitude of one order of magnitude larger than the mean PdV work. 

Gas accretion is implemented via a sink cell approach in which, based on a triangular-shaped cloud \citep[TSC;][]{Eastwood1986} interpolation, the temperature and gas density at the location of the planet is determined from 27 cells around it.
In effect, the TSC interpolation is of second order and allows planet migration in the future. In one extreme case where the planet is at the vertex between grid cells, all eight adjacent cells receive a weighting of $(1/2)^3 = 1/8$, and in the other extreme case where the planet is in the center of a cell, this cell then has a weighting of $(3/4)^3 \approx 0.42$.
The accretion luminosity of the planet $L_\mathrm{acc}$ thus follows from the conservation of total energy $E = E_\mathrm{kin} + E_\mathrm{pot} + E_\mathrm{int}$ from before $E_t$ to after removing the gas $E_{t + \Delta t}$ in a time step $\Delta t$:
\begin{equation}
    L_\mathrm{acc} = \frac{1}{\Delta t} \frac{1}{V} \int \left(E_{t + \Delta t} - E_t \right) \mathrm{d}V,
\end{equation}
which is the integral of the volume $V$ of the sink cells, including the change of mass for the individual components of $E$.
Here, $E_\mathrm{kin}=\frac{\rho}{2}\mathbf{u}^2$, $E_\mathrm{pot}=\rho \Phi_p$, and $E_\mathrm{int}=\rho c_V T$ are the kinetic, potential, and internal energy densities, where $\Phi_p$ is the gravitational potential of the star and planet as defined in \cite{KlahrKley2006}.
We note that all components of $E$ are linear in $\rho$, which is the only quantity that is affected by accretion, and all the other involved quantities, such as temperature and velocity, are kept constant. The planet potential assumes a planet of the size of Jupiter due to a lack of better constraints in our model.

For the thus obtained density and temperature, we determined a Bondi accretion rate and an accretion rate that would result from removing a certain fraction of the mass from the sink cells per time unit. From both accretion rates, we took the lower value and found the accretion rate for a ten Earth-mass planet to be limited by Bondi, as expected, and for Jupiter-like planets the accretion rate is limited by the mass able to flow into the Hill sphere (Klahr et al., in prep). As a result, the measured accretion rates do not depend on the mass removal timescale, which is typically set to $1/32$ of an orbital planet period, nor on the resolution of the simulation, unlike what was found for removing mass from wide regions of the Hill sphere \citep{Kley1999}. For the removed mass, we calculated its total energy, which is the sum of thermal, gravitational (with respect to a fiducial planet radius of one Jupiter radius), and kinetic energy, and we distributed that energy with the same TSC scheme across up to 27 cells in the vicinity of the planet. 
The new temperature was then calculated from Eq.~\eqref{eq:lte}.

We modeled the gravitational potential of the central star and an orbiting planet as in \cite{KlahrKley2006}. For the latter, we set the smoothing length to the cell size at the location of the planet.

\subsubsection{Opacities}
\label{sec:opac}
For the radiation hydro scheme 
we do not use the often applied opacities by \citet{Bell1994} unless we are at temperatures at which the dust evaporates as they neither provide the necessary Planck opacities needed for the treatment of irradiation nor specify the fundamental frequency-dependent opacities or in general optical constants needed in the subsequent MCRT simulations. Instead, we use the \citet{Bell1994} Rosseland opacities as the floor value whenever the dust would evaporate.

\begin{figure}
    \centering
    \includegraphics[width=\hsize]{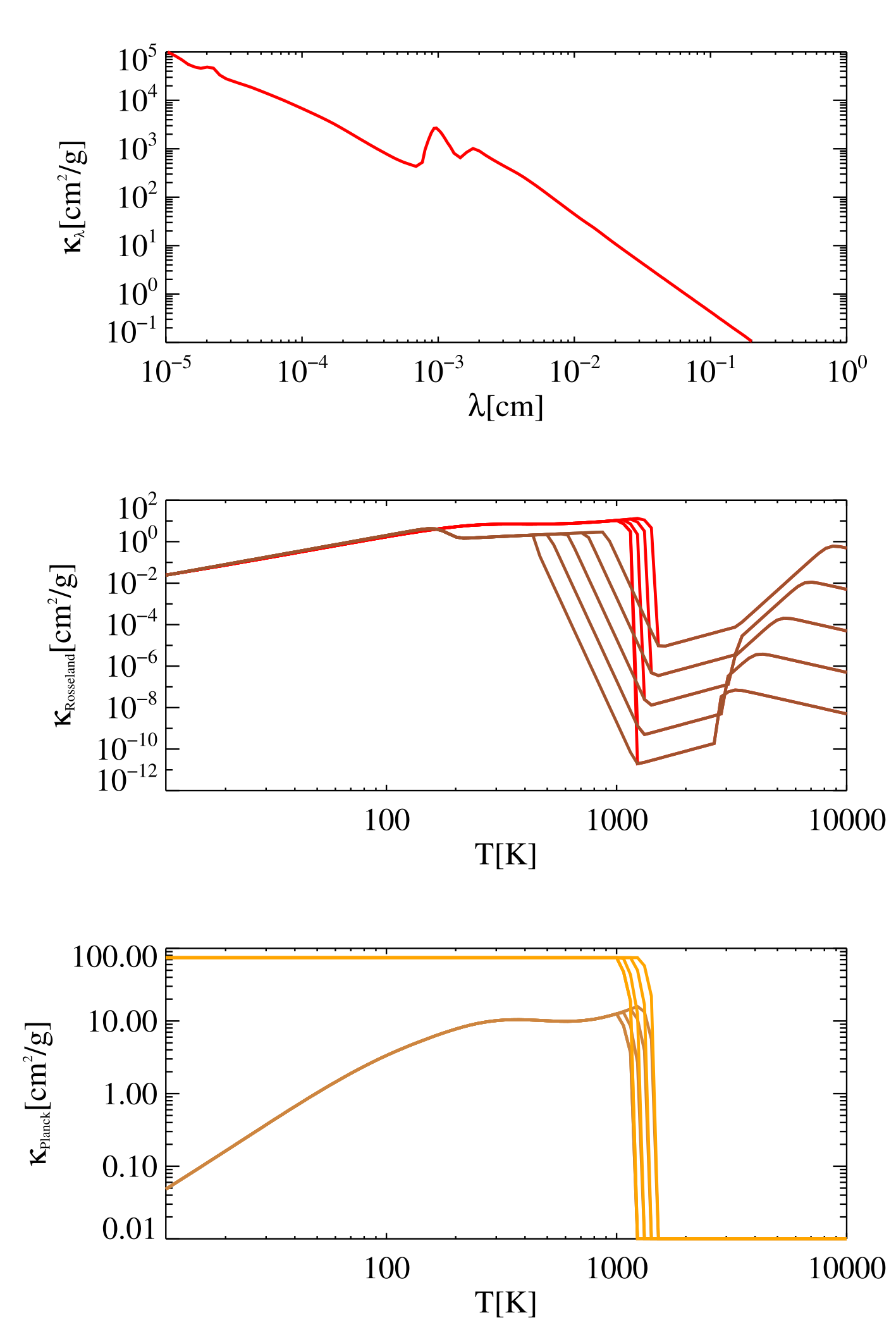}
    \caption{Wavelength and temperature dependence of the opacity. Upper plot: Dust opacity as a function of wavelength.
Middle plot: Rosseland mean opacity exemplarily for five gas densities ($10^{-18}$, $10^{-16}$, $10^{-14}$, $10^{-12}$, and $10^{-10}\,$g/cm$^3$) in ascending order in the plot \citep[red = this paper; brown = ][]{Bell1994}.
Lower plot: Planck opacities for the same gas densities, likewise in ascending order in the plot (red: $T\mathrm{rad} = T\mathrm{gas}$; orange: $T\mathrm{rad} = T\mathrm{star}$).
Mean opacities consider the evaporation of silicates following \citet{IsellaNatta2005}. But for consistency with the MCRT, ice evaporation is not considered in the opacities.
}
    \label{fig:kielopa.eps}
\end{figure}

We aimed to set our opacities to be compatible with the MCRT simulations up to the sublimation temperature of dust \citep{IsellaNatta2005}. In particular, spherical dust grains with a bulk density of $\rho_{\rm bulk}=2.5\,$g\,cm$^{-3}$ and radii $a_{\textrm g}$ ranging from $5\,$nm to $250\,$nm that follow the grain size distribution d$n_{\textrm g} \sim a_{\textrm g}^{-3.5}{\textrm d}a_{\textrm g}$ were assumed \citep{1977ApJ...217..425M}. The grains consist of $f_{\rm Si}=62.5\,\%$ silicate and $f_\text{Gr}=37.5\,\%$ graphite, which was obtained using the $1/3\,$:$\,2/3$-approximation for graphite \citep[$f_{\parallel\text{-Gr}}\,$:$\,f_{\perp\text{-Gr}}$=1\,:\,2;][]{1993ApJ...414..632D}. The corresponding wavelength-dependent refractive indices of these components \citep{1984ApJ...285...89D,1993ApJ...402..441L,2001ApJ...548..296W} were then used to calculate cross sections under the assumption of the Mie theory using the code miex \citep{2004CoPhC.162..113W}. The wavelength and grain size-dependent absorption cross sections $C_k$ for every component $k \in \{ \text{Si,}\,{\parallel}\text{-Gr,}\,{\perp}\text{-Gr}\}$ were averaged with respect to their abundance and grain size distribution to determine their mean optical properties:
\begin{equation}
\hat{C}_\lambda = \sum_k f_k \int C_k a_{\textrm g}^{-3.5} {\textrm d}a_{\textrm g}. \label{eq:averaging_c_abs}
\end{equation}
Calculating a mean grain size volume $\hat{V}_{\textrm g}$ in analogy to the calculations in Eq. \ref{eq:averaging_c_abs} then allowed us to determine the corresponding wavelength-dependent opacities $\kappa_\lambda$ as follows:
\begin{equation}
\kappa_\lambda = \frac{\hat{C}_\lambda}{\hat{V}_{\textrm g} \, \rho_{\textrm bulk}}.
\end{equation}
We plot this opacity as a function of wavelength in Fig. \ref{fig:kielopa.eps}. The values are very similar to Weingartner and Drain 2001 and steeper than the DSHARP opacities in \citep{Birnstiel2018}. We note that these opacities, $\kappa_\lambda$, are per dust mass. For the RHD simulations, we neglected scattering and assumed a dust to gas ratio of $\left. {\rm d}\,{:}\,{\rm g}\right\rvert_0 = 1\,{:}\,100$, which we used to determine the Rosseland and Planck absorption opacities for the radiation transport.

For the biggest part of our simulation space, we never reached the sublimation temperature. But inside 0.1 au from the star as well as for certain planet masses and resolutions, we reached relatively high temperatures deep in the Hill sphere around the accreting planet. Thus, when it comes to Planck opacities, we interpolated to an assumed molecular opacity of $\kappa_{\rm P} = 10^{-2}\,{\rm cm}^2/{\rm g}$, and for Rosseland, we interpolated to the values of \cite{Bell1994} (see Fig. \ref{fig:kielopa.eps}).
Coincidentally, the Rosseland mean opacities in \cite{Bell1994} for temperatures below $170$ derived for icy material is an almost perfect match to our chosen dust grain opacities. For temperatures above the ice evaporation, the grain opacities in \cite{Bell1994} are almost an order of magnitude smaller, reflecting the differences in the underlying dust model.  

We used the evaporation temperature as defined in \citet{IsellaNatta2005} \citep[see Table 3 in ][]{1994ApJ...421..615P}, which varies with the gas density roughly as a power law of the kind
\begin{equation}
    T_{\rm evp} = G \rho_g^\gamma  
    \label{eq:t_evp}
,\end{equation}
where $G = 2000\,$K and $\gamma = 1.95 \times 10^{-2}$. The transition of opacities is smoothed out over $\Delta T_{\rm evp} = 200\,$K for numerical stability, which is the same as what we applied and tested in \citet{Kuiper2010}.

These simplified opacities have the potential to be improved in the future, as more realistic opacities have already been discussed in the literature \citep[][]{Semenov2003,Malygin2017,Birnstiel2018}. But for our studies, the goal was to be consistent in opacities between the RHD and the subsequent MCRT simulations, as only this will allow for consistency of the disk structure in terms of density, temperature, and velocities as well as the derived observational appearance -- a goal that to our knowledge has not been reached in any other study so far.

As the frequency dependency and therefore the resulting Planck opacities are left unconstrained for the case of evaporating dust, we defined a floor value of $10^{-2}\,{\rm cm}^2 {\rm g}^{-1}$. This value can be orders of magnitude larger than the Rosseland opacities provided by \citet{Bell1994}, which is the first cause of the discrepancy between temperatures of RHD and MCRT simulations, as we show below with our results. The second effect of the chosen floor value in the RHD simulations is the setting of the ratio of Planck opacities for dust versus stellar radiation temperature to unity, whereas in the MCRT runs, the opacity ratio is kept constant, extending the "color" of the grains. 
Lastly, for numerical performance issues, we generally restricted even the Rosseland mean values to a level where the local optical depth of a cell does not fall below a value of $10^{-2}$. As in \citet{Kuiper2010}, the effect of multi-frequency irradiation is neglected in the present study. As later discussed, these issues lead to some inconsistencies between the temperature determination in our MCRT and HD-FLD methods. 
Likewise, the role of missing dynamics in the MCRT runs is discussed later.

\subsection{Monte Carlo radiative transport simulations}
\label{sec:rt_simulations}
   \begin{figure}
       \centering
       \includegraphics[width=0.8\hsize]{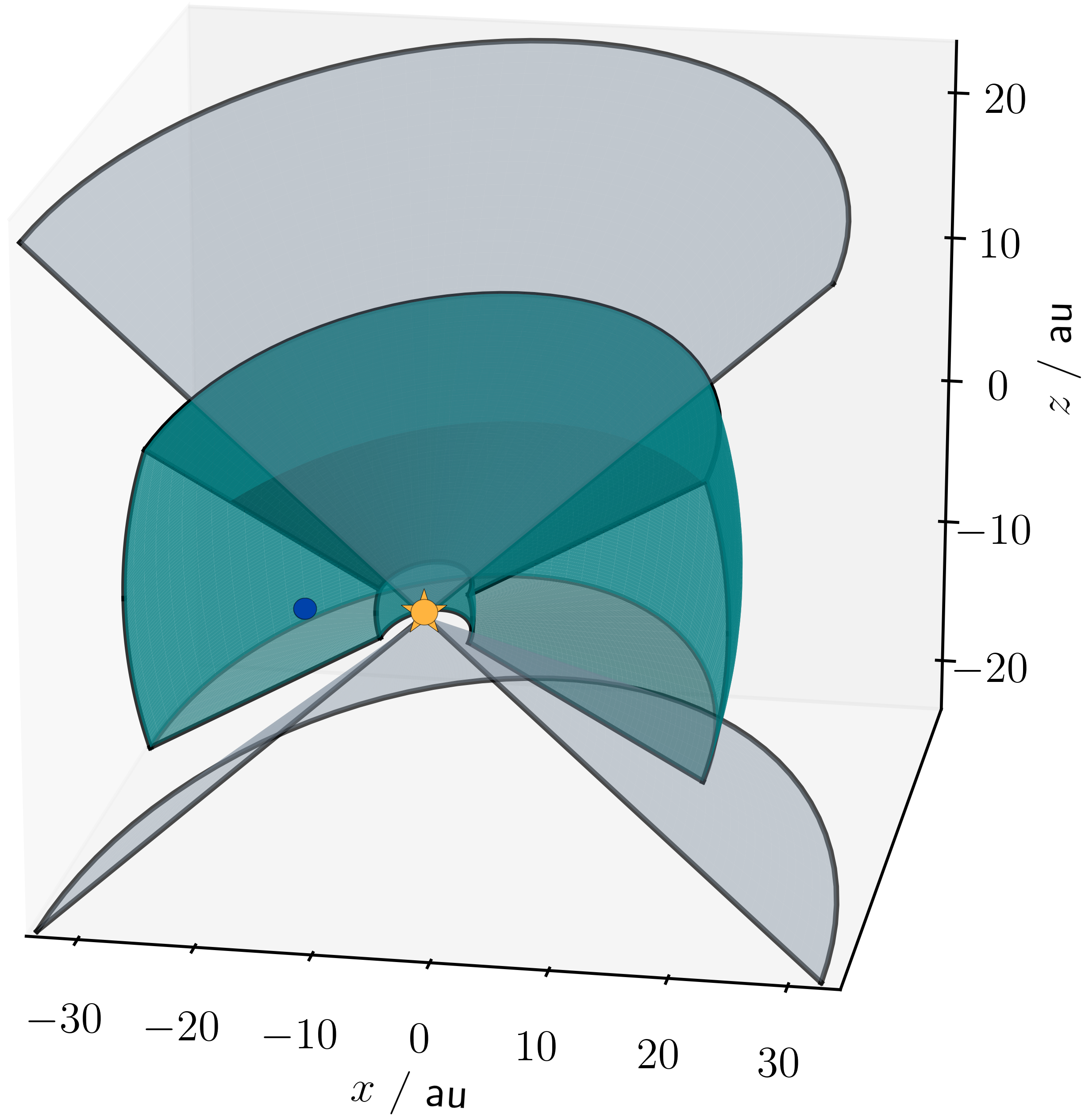}
          \caption{Vertical cut through the simulated model space depicting the embedment of a 3D high-resolution model (green region) into an axisymmetric global background model (in between gray regions) with the star (yellow) at the origin of the coordinate system and a planet (blue) located at a distance of $10\,$au from the star.}
          \label{fig:model_sketch}
   \end{figure}
    
    Our radiative transfer simulations were performed using the MCRT code Mol3D \citep{Ober_2015}, which simulates the wavelength-dependent propagation of radiation through the model space, and taking into account the optical properties of the medium and Mie theory. MCRT allows for the calculation of the underlying temperature distribution of the medium and the determination of ideal flux maps. For the temperature calculation, we considered the contribution of stellar radiation, the accretion luminosity ($L_{\rm acc}$) emitted by the forming planet, viscous heating ($\Phi$), and adiabatic compression of the gas phase ($-p\nabla \cdot \mathbf{u}$). During a simulation, radiation is represented by photon packages performing random walks that are defined by random variables, which themselves are sampled randomly using their corresponding probability distribution functions. Regions traversed by photon packages absorb a part of their carried energy, which under the assumption of a local thermal equilibrium is used to determine a corresponding location-dependent temperature of the medium. Additionally, photon packages that leave the model space in the direction of a simulated observer are used for the determination of realistic wavelength-dependent flux maps, which take into account the full complexity of the density distribution of the system as provided by the RHD simulations. Due to the generally high computational demand of such simulations, we applied a method of locally divergence-free continuous absorption of photon packages \citep{1999A&A...344..282L} coupled with an immediate reemission scheme according to a temperature-corrected spectrum \citep{2001ApJ...554..615B}. Our code is further enhanced by the utilization of a large database of precalculated photon paths in optically thick dusty media, which is particularly well-suited for simulating PPDs \citep{2020A&A...635A.148K,2022A&A...662A..99K}.
     
    During the MCRT simulations, we used the density distribution of our RHD simulations with a cell-dependent dust-to-gas ratio, which is described later in this section. Additionally, we considered different radiation sources. As before, stellar radiation is described by a black body spectrum with an effective temperature of $T_*=4000\,$K, assuming a stellar radius of $R_*=2.5\,{\rm L}_\odot$. The other three sources of radiation were assumed to be thermalized and emitted by dust, leading to the location-dependent emission of radiation corresponding to their power density $Q^+$. This quantity serves as a source term in the diffusion equation, given in units of Jansky per second per cubic meter, and was calculated during the RHD simulations (for details, see Sect. \ref{sec:num_3d_rhd}). In general, depending on the region in the model space, the dominant source of radiation may strongly vary. However, when conducting a test without considering the influence of adiabatic changes during our MCRT simulations, we found its impact on the resulting temperature distribution to be relatively small. This was especially the case in the vicinity of the planet, which is likely attributable to our models reaching a steady state. In the following analysis, it has therefore been neglected, but the impact is discussed further in Sect. \ref{sec:discussion_differences}.

\paragraph*{Consistency of optical properties:}
\label{sec:opacities_introduction}

    To assess the quality of the radiation scheme that is applied in our RHD simulations, we performed MCRT simulations based on snapshots of the simulated systems in order to determine the underlying temperature distribution, $\temprt$. These distributions were then compared with temperature distributions obtained from our RHD simulations, $\temphd$. 
    For that purpose, it was crucial to ensure consistency between both simulation types, particularly with regard to the assumed opacities, which in the RHD simulations are a function of the position-dependent temperature (see Fig. \ref{fig:kielopa.eps}). While it is generally possible to account for temperature-dependent opacities in MCRT simulations, such an endeavor necessitates an iterative solution procedure \citep[e.g.,][]{2023A&A...678A.175M}. However, in the case of high-resolution simulations of optically extremely thick systems, the computational demand for such a procedure becomes practically infeasible. During our self-consistent MCRT simulations, we instead kept the opacities, in particular the Rosseland opacities, of each cell fixed at their values $\kappa_{\rm R}(\temphd)$ as provided by the RHD simulations for the given snapshot. The validity of our approach is shown a posteriori. 
    With regard to our dust radiative transfer code Mol3D, the opacities were matched by adjusting (i.e., lowering) the dust content in each cell such that it reaches the exact value provided by our RHD simulations. This adjustment was done by changing the dust-to-gas-ratio of each cell according to
    \begin{equation}
        \left. {\rm d}\,{:}\,{\rm g}\right\rvert_{\rm cell} = \left. {\rm d}\,{:}\,{\rm g}\right\rvert_0 \, \frac{\kappa_{\rm R}(\temphd)}{\kappa_{\rm R}^{\rm dust}(\temphd)},
    \end{equation}
    where $\kappa_{\rm R}^{\rm dust}$ is the Rosseland opacity of the dust, assuming no sublimation occurs. As a result, even cells with temperatures above the sublimation limit retain non-zero opacities that are equal to their values given by the RHD simulations, which is crucial for consistency. It is important to note that after the sublimation temperature is reached inside a cell, its opacity is reduced by multiple orders of magnitude (see Fig. \ref{fig:kielopa.eps}), which often results in the cell becoming optically thin. At this point, the exact value to which its opacity has dropped hardly affects its assigned temperature. The only region in which it may affect temperature estimates is inside the sublimation radius of the protoplanetary disk, close to the inner radial boundary of the simulated model space. 
    In this partial sublimation approach, we are able to distinguish between three opacity zones in total: the dust-dominated zone ($T<T_{\rm evp}$), the gas-dominated zone ($T\geq T_{\rm evp} + \Delta T_{\rm evp}$), and the transition zone ($T_{\rm evp}\leq T < T_{\rm evp} + \Delta T_{\rm evp}$). Given the sharp decline in opacity values beyond the sublimation threshold, this approach is justified only if the temperatures derived from MCRT simulations, $\temprt$, closely align with those from the RHD simulations, $\temphd$, regarding their associated zones or, at the very least, if the classification of cells as dust-dominated or not dust-dominated matches. The fulfillment of this condition is confirmed in Sect. \ref{sec:temp_comparison_reference_model}.

\subsection{Physical and numerical model parameters} 
\label{sec:model_embedement_and_parameters}
    \begin{table*}[!htb]
        \caption{Grid properties of the MCRT models and the covered spatial domain of the corresponding RHD models.}
        \centering
        \footnotesize
        \begin{tabular}{l>{\centering\arraybackslash}m{3.2cm}>{\centering\arraybackslash}m{4.5cm}>{\centering\arraybackslash}m{2.cm}>{\centering\arraybackslash}m{2.cm}}
        \toprule\toprule
                            & 2.5D background model & 3D model & \multicolumn{2}{c}{Planet and Hill region}\\
                            &  & (embedded in 2.5D model) &  $M_{\rm p}=10\,{\rm M}_\oplus$ & $M_{\rm p}=300\,{\rm M}_\oplus$\!\!\tablefootmark{*} \\
        \midrule
        \underline{Grid properties:}      \vspace{4pt}& $N_{\rm cells}\,\left(N_r\times N_\theta \times N_\phi \right)$  & $N_{\rm cells}\,\left(N_r\times N_\theta \times N_\phi \right)$ & $N_{\rm cells}$ & $N_{\rm cells}$ \\
                    \quad\verb|N|&$0.1\,{\rm M}\,\left(512\times258\times 1\right)$&--&--&-- \\
                    \quad\verb|N1|\tablefootmark{*}&--&$22\,{\rm M}\,\left(509\times131\times 339\right)$& 15 & 446 \\
                    \quad\verb|N2|&--&$70\,{\rm M}\,\left(609\times171\times 681\right)$& 112 & 3450 \\
                    \quad\verb|N3|&--&$281\,{\rm M}\,\left(809\times257\times 1353\right)$&  446 & 28027\\
        \midrule
        \underline{Spatial domain:}      \vspace{4pt}&&&\multicolumn{2}{c}{Planet location} \\
        \quad Radial              & $<300\,$au & $4-25\,$au & \multicolumn{2}{c}{$10\,$au}\\
        \quad Polar               & -$0.6-0.6\,$rad & -$0.4-0.4\,$rad & \multicolumn{2}{c}{0\,rad} \\
        \quad Azimuthal \vspace{3pt}& $0-2\pi\,$rad & $0-2\pi\,$rad & \multicolumn{2}{c}{$\pi$\,rad} \\
        \quad Selected features   &  axisymmetric & gaps, spirals, crescents &\multicolumn{2}{c}{accretion (\texttt{A4}, \texttt{A32}\tablefootmark{*}, \texttt{A64})}\\
        \bottomrule
        \end{tabular}
        \tablefoot{
        Columns specify properties of the 2.5D background model (first column), the embedded 3D high-resolution model (second column), and the planet and Hill region inside the 3D model (third column). The latter is subdivided into two columns, depending on the planetary mass $M_{\rm p}$. Grid properties refer to the number of cells, and grid resolution refers to the specified MCRT model or region. The spatial domain denotes the parameter range (first and second column) or location (third column) of the respective RHD model. The last row highlights selected features of these models. Values of varied physical and numerical parameters that are marked by an asterisk specify the reference model.\\
        \tablefoottext{*}{Reference model.} M = Million. }
        \label{tab:grids}
    \end{table*}

    Each RHD model is defined on a grid, the cells of which are described by lists of spherical coordinates ($r$, $\theta$, and $\phi$). The number of cells in the radial ($N_r$), polar ($N_\theta$), and azimuthal ($N_\phi$) direction as well as the total number of cells ($N_{\rm cells}$) of our therewith constructed MCRT models and the spatial ranges covered by the RHD models are specified in Table \ref{tab:grids}. We considered two types of MCRT models, namely the so-called 2.5D and 3D models. The 2.5D (MCRT) background models are based on snapshots of our axisymmetric hydrodynamical models. The 3D models that were used during our MCRT simulations, however, were constructed by embedding a snapshot of a high-resolution RHD model into the axisymmetric background model, which is illustrated in Fig. \ref{fig:model_sketch}. Specifically, cells in our 3D model that fell within the spatial domain of the high-resolution RHD model were defined using the properties of the corresponding high-resolution RHD model cells, while those outside this domain were based on the background model. Consequently, the constructed 3D model grid contains all cells from the high-resolution RHD model plus additional cells to cover the full domain of the background model. We tested three different grid resolutions, which in ascending order of resolution are hereafter referred to as \verb|N1|, \verb|N2|, and \verb|N3|. The planet is located in the midplane (i.e., $z=0\,$au) at a distance of $10\,$au from the central star. We note that as a consequence, the number of cells encompassed by the Hill sphere varies among different models depending on the resolution and planetary mass $M_{\rm p}\in\left\{10\,{\rm M}_\oplus, 300\,{\rm M}_\oplus \right\}$ (see Table \ref{tab:grids}). Moreover, we considered and analyzed the impact of three different mass accretion parameters, more specifically the timescale of mass removal from the sink-cell, on the temperature distribution. Model \verb|A4| uses an accretion timescale of $1/4$ planet orbits, \verb|A32| $1/32$ planet orbits, and \verb|A64| $1/64$ planet orbits. This means that while all models reach the same steady state accretion rate (Klahr et al., in prep), they show different levels of mass accumulation around the planet inside the Hill sphere. The measured accretion rates approach a mean value of  $3\times 10^{-3}\,{\rm M}_\oplus {\rm yr}^{-1}$ for the $300\,{\rm M}_\oplus$ case and $6.7\times 10^{-4}\,{\rm M}_\oplus {\rm yr}^{-1}$ for the $10\,{\rm M}_\oplus$ planet. With the assumption of accreting this mass on a planet with a radius equivalent to Jupiter, we can estimate a  resulting luminosity with an average value of $3.4\times 10^{-3}\,{\rm L}_\odot$ ($4.8\times 10^{-5}\,{\rm L}_\odot$) for models with a planetary mass of $300\,{\rm M}_\oplus$ ($10\,{\rm M}_\oplus$). As mentioned in the introduction, this accretion rate and luminosity does not depend on the resolution nor on the accretion timescale of the models, which means the fluctuations of the accretion rate in one model can be larger than the deviation of the mean accretion rates between the different models.

\section{Results and discussion}
\label{sec:results_and_discussion}
    The MCRT simulations were employed to calculate temperature distributions corresponding to 18 high-resolution RHD models (see Table \ref{tab:grids}) of PPDs with embedded accreting planets, as described in Sect. \ref{sec:rt_simulations}. In Sect. \ref{sec:temp_comparison_reference_model}, the results of these simulations are presented in detail for a reference model with a grid resolution of \texttt{N1}, assuming a planetary mass of $M_{\rm p}=300\,{\rm M}_\oplus$, and the accretion parameter \texttt{A32}. We investigate the similarity between the results generated by the MCRT and RHD simulations in various regions inside the PPD, such as the disk midplane, the gap, the Hill region, and the photosphere. 
    Following this, we assess the impact of varying numerical and physical parameters on the resulting temperature distributions in Sect. \ref{sec:temp_comparison_parameter_impact}. Through this examination, we aim to determine the consistency between MCRT and RHD temperature estimations and understand how changes in model parameters affect the computed temperatures, particularly in the high-resolution region and in the vicinity of the planet. The resulting differences in temperature estimates, arising from the inclusion of various physical processes in our simulations, are then investigated in Sect. \ref{sec:axisym}. For this purpose, we simulated and analyzed axisymmetric uPPDs, which consider different physical mechanisms that determine the radiative state of these systems. 
    Subsequently, we showcase constructed flux maps of our PPD models for the VIS, NIR, and submm wavelength ranges in Sect. \ref{sec:flux_section}. We assess existing differences between flux maps derived from temperature distributions of our RHD simulations and those obtained from our MCRT simulations, and we investigate the impact of the grid resolution and planetary mass on the resulting differences. 
    Lastly in Sect. \ref{sec:discussion_differences},  we discuss the origin of the temperature differences, and we explore potential solutions for their mitigation.

\subsection{Temperature comparison}
\label{sec:temp_comparison_reference_model}

    During the temperature calculation, a total of $N_\gamma=10^8$ photon packages were used for the simulation of stellar emission. Thermal dust emission was simulated by sending out photon packages from each grid cell according to its dust content, temperature-dependent spectrum, and $Q^+$-value. The number of photon packages each individual cell sent out was chosen such that the energy content carried by a single photon package was less than or equal to that carried by simulated photon packages of the stellar radiation. 

\subsubsection{Reference model}
    \begin{figure*}
    \centering
    \includegraphics[width=\hsize]{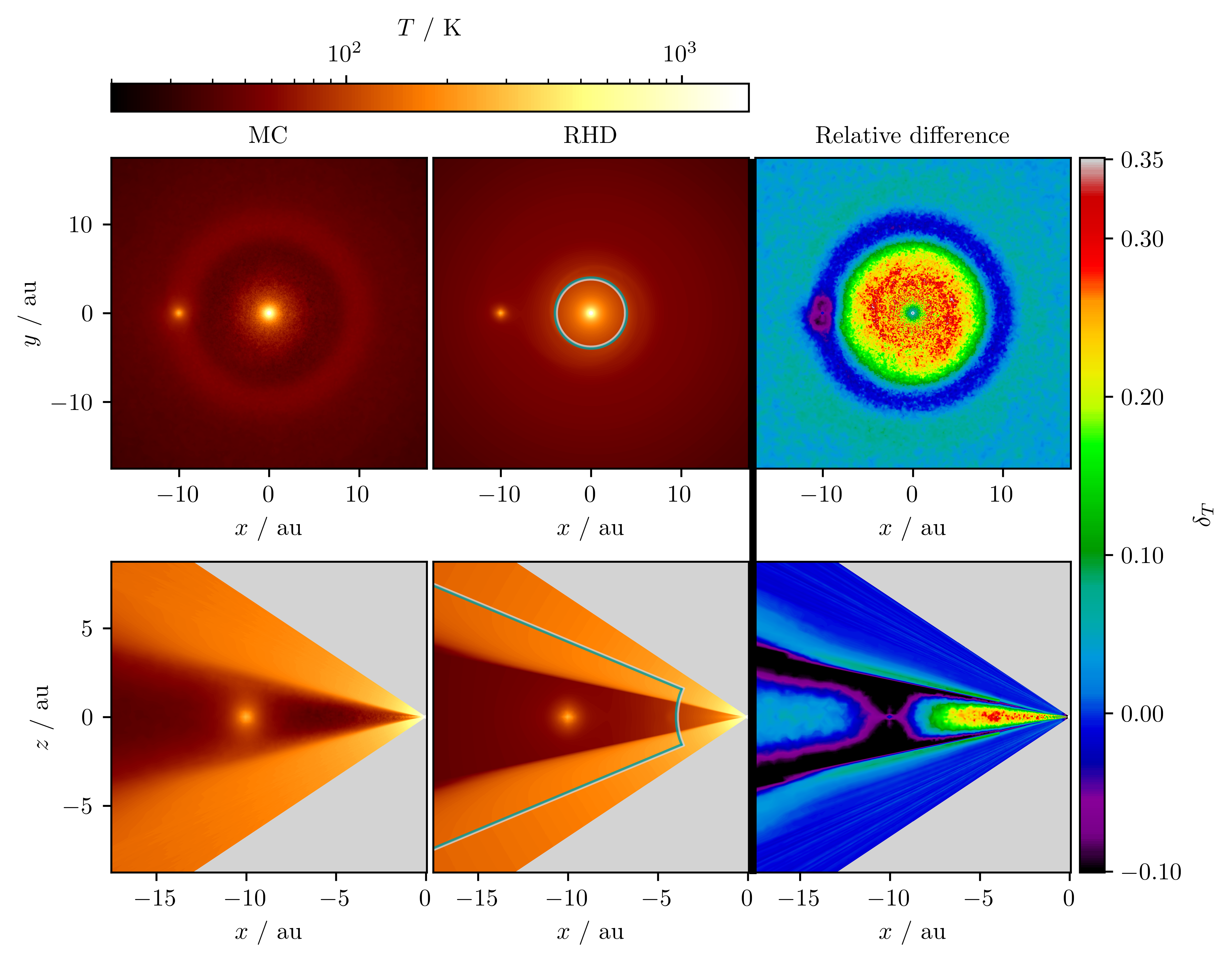}
        \caption{Cuts through the temperature distributions derived by the MCRT (left column) and RHD simulations (central column) and their relative differences $\delta_T$(right column). The upper and lower rows show results for the midplane and a vertical cut through it at the azimuthal position of the planet, respectively. The color bar for the temperature distributions (relative differences) is displayed above (to the right). Additionally, the high-resolution and the low-resolution regions are divided by a green-gray line that is shown in the plots of the RHD simulations (i.e., in the central column). Here, the high-resolution (low-resolution) region is located on the green (gray) side of the line. }
    \label{fig:temp}
    \end{figure*}

    \begin{figure*}
    \centering
    \includegraphics[width=0.8\hsize]{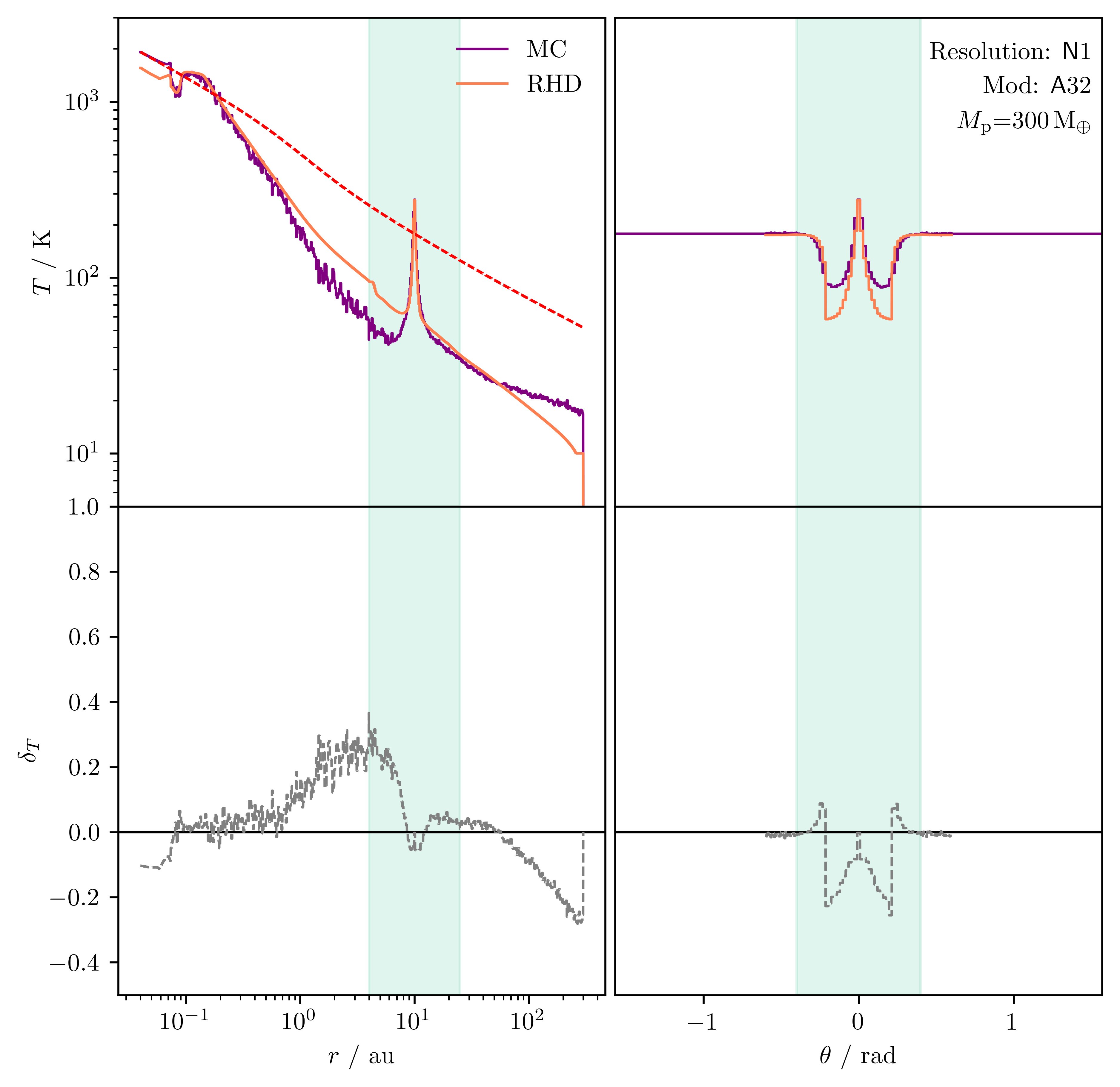}
        \caption{Comparison of temperature profiles (upper row) determined using the MCRT (purple lines) and RHD (orange lines) simulations and their relative differences (lower row, gray lines). Left: Radial temperature profiles of the midplane in the azimuthal direction of the planet. Right: Polar temperature profiles of the planetary region. Green colored regions mark the spatial domain covered by the 3D high-resolution RHD model. Model details are shown in the upper right.}
    \label{fig:temp_profiles}
    \end{figure*}
    
    Figure \ref{fig:temp} shows cuts of the temperature distribution of the reference model inside the inner ${\approx}20\,$au based on the MCRT simulations (left) and according to the corresponding RHD simulation (center) and their relative differences $\delta_T$ (right), given by
    \begin{equation}
        \delta_T = \frac{\temphd - \temprt}{\temphd + \temprt},
    \end{equation}
    where positive values correspond to the case of higher RHD temperatures, that is, where $\temphd > \temprt$.
    The upper (lower) row shows results for the midplane (vertical cut through the midplane in the direction of the planet). 
    
    As expected, the MCRT and RHD simulations result in qualitatively similar temperature distributions, with a clear impact of the star and planet on the temperatures in their respective local environments. Moreover, the transition from the high-resolution region to that of the global background model is shown to be smooth in all plots. However, in the $\temprt$ distribution, the gap feature appears significantly more prominent not only in the midplane but also in the vertical cut. Overall, the $\temphd$ distribution estimates less variation of temperature values within the whole PPD region than the MCRT simulations. The $\temprt$ distribution also appears to be affected more strongly by local density changes, which in the case of the vertical cut through the gap region leads to higher temperature gradients. Furthermore, the $\temphd$ distribution exhibits a sharp cut at higher scale heights close to the photosphere that is not present in the $\temprt$ distribution. 
    
    The relative difference plot for the midplane (upper right plot) additionally shows that the gap region in the $\temphd$ distribution and, even more so, in the circumplanetary region is colder by $\delta_T{\approx} 0-0.05$ and ${\approx} 0.05-0.1$, respectively. The regions radially just inside or outside the gap, on the other hand, are warmer than estimated by the MCRT simulations at the order of ${\approx} 0-0.05$. Furthermore, the relative differences in the midplane grow radially inward of the planetary region toward the star, where they reach values of even ${\approx} 1/3$, before dropping significantly very close to the star. Less prominent differences can also be found, first, at about $1\,$au radially outside the gap region, where a slightly greenish region indicates higher $\temphd$ temperatures with differences of ${\approx} 0.1$, and second radially inward of the gap region, where a reddish region with a spiral-like shape can be found slightly outside the inner rim of the high-resolution region. The locations of these latter spiral features correspond to those of very faint spiral features in the temperature maps of the MCRT and RHD simulations. 
    Furthermore, the lower-right plot of Fig.~\ref{fig:temp} displays a clearly stratified structure in the vertical direction. From the highest displayed scale heights toward the midplane, the discrepancy between the two simulation types is rather small (blue region). Then at about the photosphere of the PPD, where the density of the disk is still comparably low, the RHD temperatures clearly exceed the MCRT temperatures (light-blueish, slightly greenish region) before rapidly dropping (black region), followed by a purple region where $\temphd$ is lower by about $\delta_T{\approx} 0.05-0.1$, which is probably linked to an optical depth effect (compare with Fig. \ref{fig:app:opt_depth_kappa_rosseland}, which shows an optical depth map of the reference model). Overall, when comparing the $\temphd$ distribution with the $\temprt$ distribution, we found that within the depicted range, the midplane is typically warmer for the RHD case, and the gap and circumplanetary region is usually colder. Furthermore, these differences exhibit a systematic nature; they cannot be ascribed to the stochastic variability inherent to MC simulations, suggesting a systematic origin rather than limitations in numerical precision.

    To better assess the similarity of the results of both simulation types, Fig. \ref{fig:temp_profiles} shows radial (left) and polar (right) temperature profiles of the reference model based on MCRT (purple) and RHD (orange) simulations in the upper plots and their relative differences (gray, dashed) in the lower plots. The radial profiles extend from the star through the position of the planet along the midplane, and the polar profiles extend in the $\theta$-direction through the position of the planet. The red dashed line shows the temperature profile in the $z$-direction in the optically thin regions of the PPD. The green shaded regions indicate the range covered by the high-resolution model. Compared to Fig. \ref{fig:temp}, this plot displays results for the full radial (left plots, logarithmic scaling) and polar (right plots, linear scaling) ranges of the simulated model space and allows for a quantitative comparison. 
    Radial profiles of both simulation types estimate relatively flat temperature profiles close to the star and a narrow dip feature within the first $0.1\,$au. From there, up to about $1\,$au, which is well inside the optically extremely thick region of the PPD, the estimates of temperatures are in great agreement with differences typically below $0.05$. Further out $\temprt$ drops quicker than $\temphd$, leading to increasing differences. In the vicinity of the planet, however, we found these difference in the midplane drop, in absolute terms, well below values of $0.05$. Here, the temperatures rise rapidly as a result of the released accretion luminosity of the planet. Further out, the slope of the MCRT temperature profile is much greater than that of the RHD simulations, leading to (negative) relative differences of up to about ${-}1/3$. 
    The polar profiles (right plots) in Fig. \ref{fig:temp_profiles} also indicate rapidly growing temperature differences in the upper PPD layers of the region close to the planet, with $\temphd$ being $\delta_T{\approx}0.05-0.2$ lower than $\temprt$. The aforementioned abrupt increase in $\temphd$ at the photosphere appears in the lower-right plot in the form of a flip of a sign of the relative differences. At even greater scale heights, the relative differences reduce to values very close to zero. 
    
   \begin{figure}
   \centering
   \includegraphics[width=\hsize]{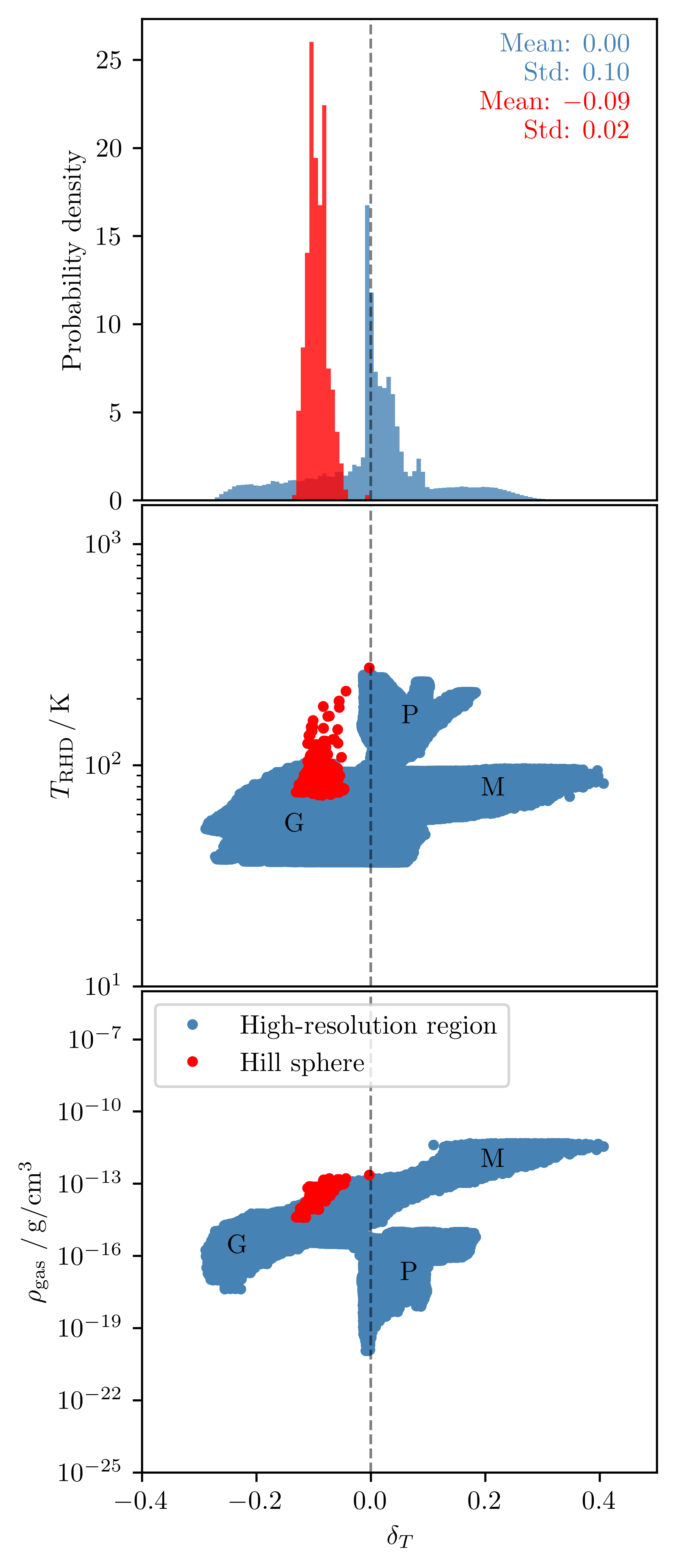}
      \caption{Relative temperature differences between RHD and MCRT simulations of the reference model. Top: Normalized histogram. Center: Scatter plot for derived RHD temperatures and corresponding relative temperature differences. Bottom: Scatter plot for derived RHD gas densities and corresponding relative temperature differences. Data for the high-resolution (Hill) region is color coded in blue (red). The mean and std of the displayed distributions are indicated in the top-right corner. Labels in the scatter plots mark areas that roughly correspond to the following regions: the midplane (indicated by the letter \quotes{M}), the gap (\quotes{G}), and the photosphere and higher layers (\quotes{P}).}
         \label{fig:histogram}
   \end{figure}
   
    To investigate the origin of the systematic differences, Fig. \ref{fig:histogram} shows a normalized histogram of relative temperature differences for the reference model (upper plot) as well as scatter plots in order to reveal potential correlations with physical quantities, such as the gas temperature ($\temphd$, center plot) and density ($\rho_{\rm gas}$, lower plot). Blue (red) color coded data correspond to the high-resolution region (Hill region). The mean and standard deviation (std) are indicated using the same color as the corresponding histogram in the top-right corner of the upper plot. 
    The histogram for the high-resolution region is non-Gaussian, exhibits a prominent peak at a value close to zero, and has a low mean-to-std-ratio of ${<}0.01$, suggesting an overall good agreement between both temperature distributions. In comparison, the distribution for the Hill sphere is much more narrow, with an std of only $0.02$, and the $\temphd$ values are on average $\delta_T{\approx}0.09$ lower than the corresponding $\temprt$ values. While the central scatter plot indicates no clear correlation between the relative differences and $\temphd$ inside the Hill sphere, it is possible to roughly identify general locations in the $\temphd$-$\delta_T$-plane that correspond to cells of three different broadly defined regions in the PPD: the midplane (indicated by the letter \quotes{M}), gap (\quotes{G}), and the photosphere and higher layers (\quotes{P}). The fact that different regions in the PPD populate different areas hints at a dependence of the quality of temperature estimates on the type of radiation source, which is investigated in Sect. \ref{sec:axisym} and further discussed in Sect. \ref{sec:discussion_differences}. This is likely linked to the fact that their environments are fundamentally very different. While a large portion of the stellar radiation only interacts with an optically thin medium, radiation due to viscous dissipation originates from the optically thickest regions, similar to the accretion luminosity of the planet. The lower plot in Fig. \ref{fig:histogram}, which shows the distribution of data points in the $\rho_{\rm gas}$-$\delta_T$-plane, provides further support in this regard. Cells of the photosphere, for instance, populate a very distinct region compared to those inside the PPD. Furthermore, denser regions in the PPD result mostly in positive temperature differences (i.e., $\temphd{>}\temprt$), while regions of low density, especially inside the gap, appear to have predominantly negative temperature differences (i.e., $\temphd{<}\temprt$). The same trend can be found inside the Hill sphere, that is, an increase in density leads to a shift of relative differences in the positive direction toward zero. However, when comparing the location of (red) data points representative for the Hill sphere with the (blue) data points corresponding to cells with similar gas density values, we found the former to be located at the far left side of the distribution, where $\delta_T<0$. This suggests a complex position dependence of the similarity of the results of our MCRT and RHD simulations, which may be dependent on the optical depth emitted radiation has to overcome in order to escape the system.  

   \begin{figure}
   \centering
   \includegraphics[width=\hsize]{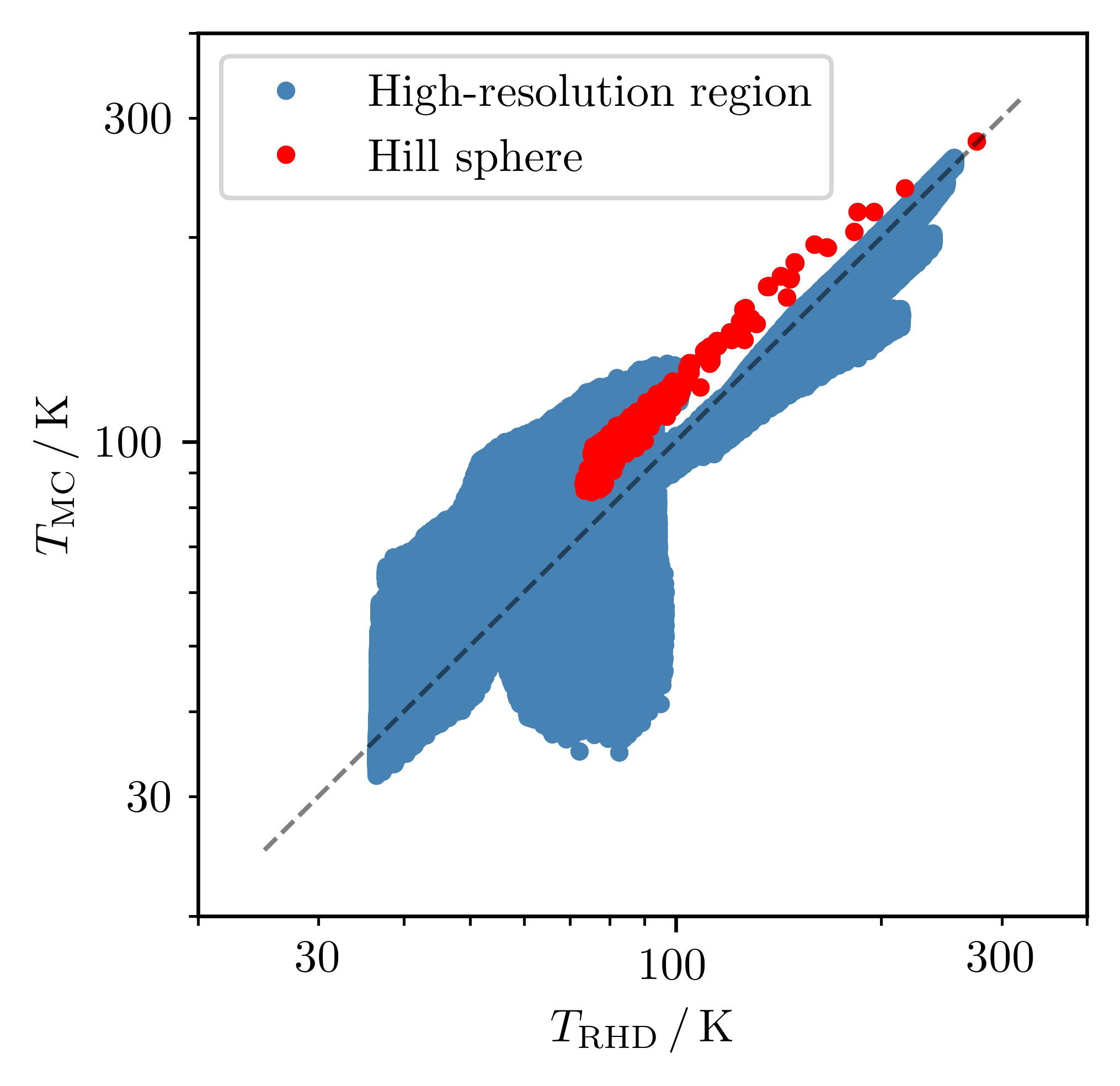}
      \caption{Scatter plot for derived RHD and MCRT temperatures. Data for the high-resolution (Hill) region is color coded in blue (red). The gray dashed line marks the region of matching temperature estimates.}
         \label{fig:TTscatterplot}
   \end{figure}

    Next, Fig. \ref{fig:TTscatterplot} shows a $\temprt$-$\temphd$ scatter plot where matching temperature estimates of the two simulation types result in a distribution of data points along the displayed gray dashed line. The plot clearly shows that higher derived RHD temperatures generally correlate with higher estimated MCRT temperatures. The spread of data points, though, is indicative of systematic differences that have also been identified in the central and lower plots of Fig. \ref{fig:histogram}. 

   As a side note, it is important to consider that during the MCRT temperature calculation step, we used a non-iterative approach (refer to Sect. \ref{sec:rt_simulations}). The validity of this approach, however, depends on the consistency of the results of the two simulation types with regard to their classification of cells undergoing sublimation or being dust dominated. In our reference model, we found that $98\,\%$ of the cells that were classified as dust dominated by the RHD simulations are also classified as such by our MCRT simulations. Likewise, nearly $100\,\%$ agreement exists for cells undergoing complete or at least partial sublimation, that is, they belong to the gas-dominated or transition zone (see Sect. \ref{sec:opacities_introduction}), respectively. Given this high level of agreement between both simulation types, opting for an iterative procedure can be expected to result in an outcome that is qualitatively very similar.

\subsubsection{Impact of model parameters}
\label{sec:temp_comparison_parameter_impact}
    In this section, we quantify the impact of the planetary mass, the resolution, and the accretion parameter on the similarity of temperature estimates of RHD and MCRT simulations. To achieve this, we conducted a comprehensive analysis, similar to that presented for the reference model in Sect. \ref{sec:temp_comparison_reference_model}, for all 18 models. Qualitatively, the results of all the models are consistent with our previous findings. Hence, in the following we provide a brief summary of the most notable differences, with a particular focus on the planetary Hill region. We then give an overview of relative temperature difference histograms of all models, effectively encapsulating their quantitative differences. 

   \begin{figure*}
   \centering
   \includegraphics[width=0.8\hsize]{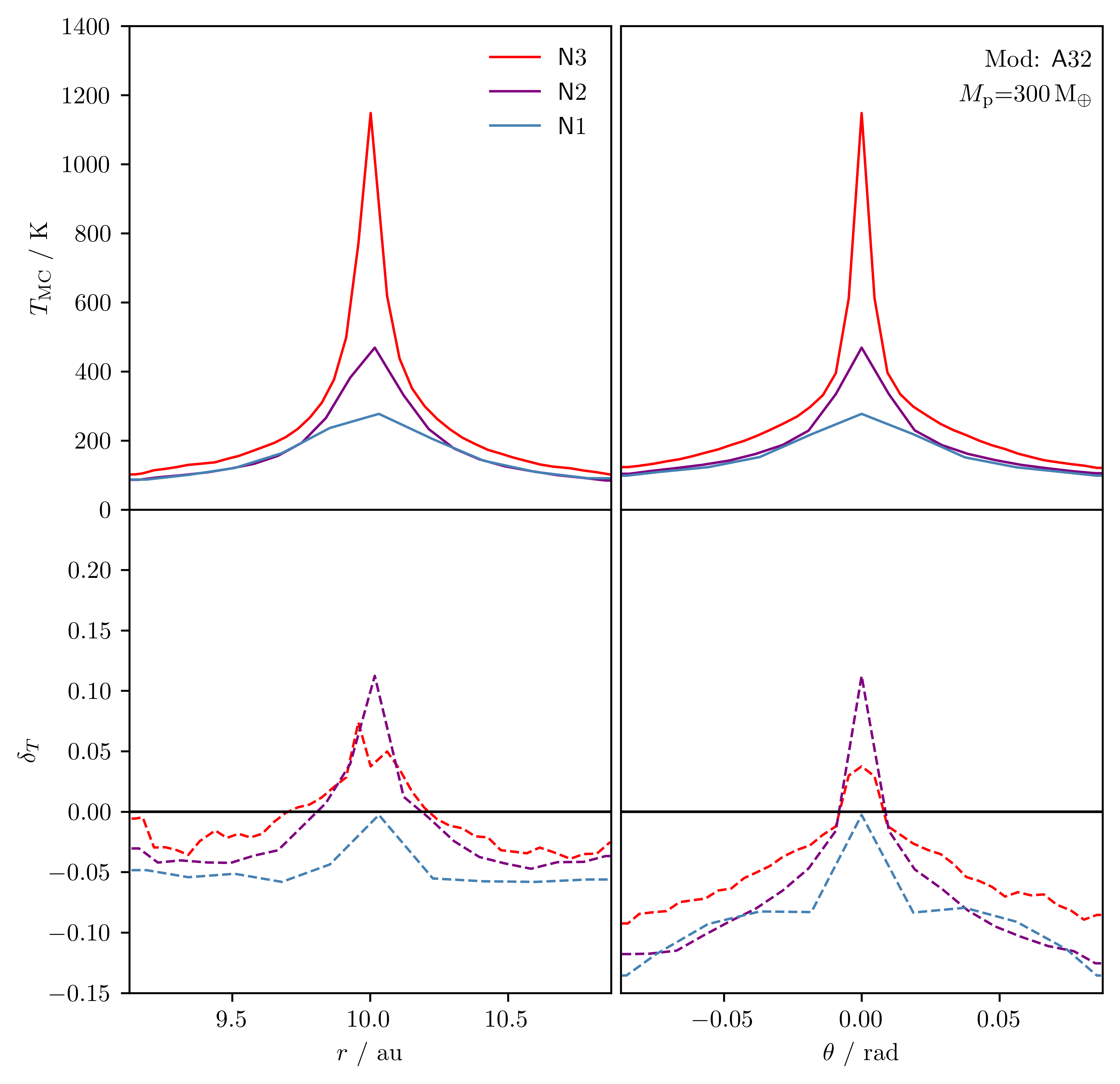}
      \caption{Comparison of temperature profiles inside the planetary Hill sphere determined using MCRT simulations (upper row) and their differences (lower row) relative to the corresponding RHD temperature profiles. Each plot contains three color coded curves that correspond to three different model resolutions (\texttt{N1}, \texttt{N2}, and \texttt{N3}). Left: Radial temperature profiles of the midplane in the azimuthal direction of the planet. Right: Polar temperature profiles of the planetary region. The planetary mass, $M_{\rm p}$, and the accretion parameter, \texttt{A}, are specified in the upper-right corner of this figure.}
         \label{fig:hill_sphere_temp_profiles}
   \end{figure*}
    
    Figure \ref{fig:hill_sphere_temp_profiles} shows a comparison of temperature profiles for different model resolutions (\texttt{N1}, \texttt{N2}, and \texttt{N3}) that were obtained with MCRT simulations (upper row) as well as their differences relative to corresponding RHD simulations (lower row) for $M_{\rm p}=300\,{\rm M}_\oplus$ models using the accretion parameter \texttt{A32}. Results for all other model parameters are shown in Figs. \ref{fig:app:hill_sphere_temp_profiles_1} to \ref{fig:app:hill_sphere_temp_profiles_5}. Here, the displayed radial and polar range is defined by the size of the corresponding planetary Hill sphere.  
    In general, the shape of the radial and polar temperature profiles inside the Hill sphere of the planet are similar across all models in that the temperature gradient is very high, and they peak at the position of the planet. Additionally, the determined temperature profiles in the radial direction appear to be very similar to the profiles in the polar direction. We find, as can be expected, that the peak temperature increases with higher planetary masses. As can be seen, it also rises with increasing resolution, which in the case of \texttt{N3} simulations with a $300\,{\rm M}_\oplus$ planet may even lead to the exceedance of the sublimation temperature (Eq. \eqref{eq:t_evp}). The former effect can be expected, as the gravitational potential of the $300\,{\rm M}_\oplus$ planet is much deeper than that of the $10\,{\rm M}_\oplus$ planet. The increase of the peak temperature with a higher resolution, on the other hand, can be explained by the fact that the temperature gradient, especially of the $300\,{\rm M}_\oplus$ models, is relatively high in this region. The temperature value that is assigned to a cell, however, corresponds to an average value of the entire spatial region that it covers. As a result, smaller cells resolve the underlying temperature distribution better, while the temperature distribution in a grid of lower resolution may appear more smeared. 
    
    Figure \ref{fig:hill_sphere_temp_profiles} also shows that inside the Hill region, the relative differences between derived MCRT temperatures and RHD temperatures tend to decrease in absolute terms with a higher resolution. The only exception is the planetary cell itself at the resolution \texttt{N1}, for which both temperature estimates almost match. Whether a change of the peak temperature and the Hill region has an impact on the derived flux maps of our models is investigated later in Sect. \ref{sec:flux_section}. 
    An assessment of the impact of the accretion parameter showed that it primarily affects the peak temperature and temperature gradient in the immediate environment of the planet (see Figs. \ref{fig:app:hill_sphere_temp_profiles_1} to \ref{fig:app:hill_sphere_temp_profiles_5}). In the case of the $10\,{\rm M}_\oplus$ models, its impact is relatively small. For the $300\,{\rm M}_\oplus$ models, its increase typically results in a decrease in peak temperature. Moreover, we find that the \texttt{A32} models are generally associated with the lowest corresponding relative temperature differences and that \texttt{A64} models typically achieve a very similar level of agreement. 
    
\begin{figure*}
   \centering
   \includegraphics[width=0.9\hsize]{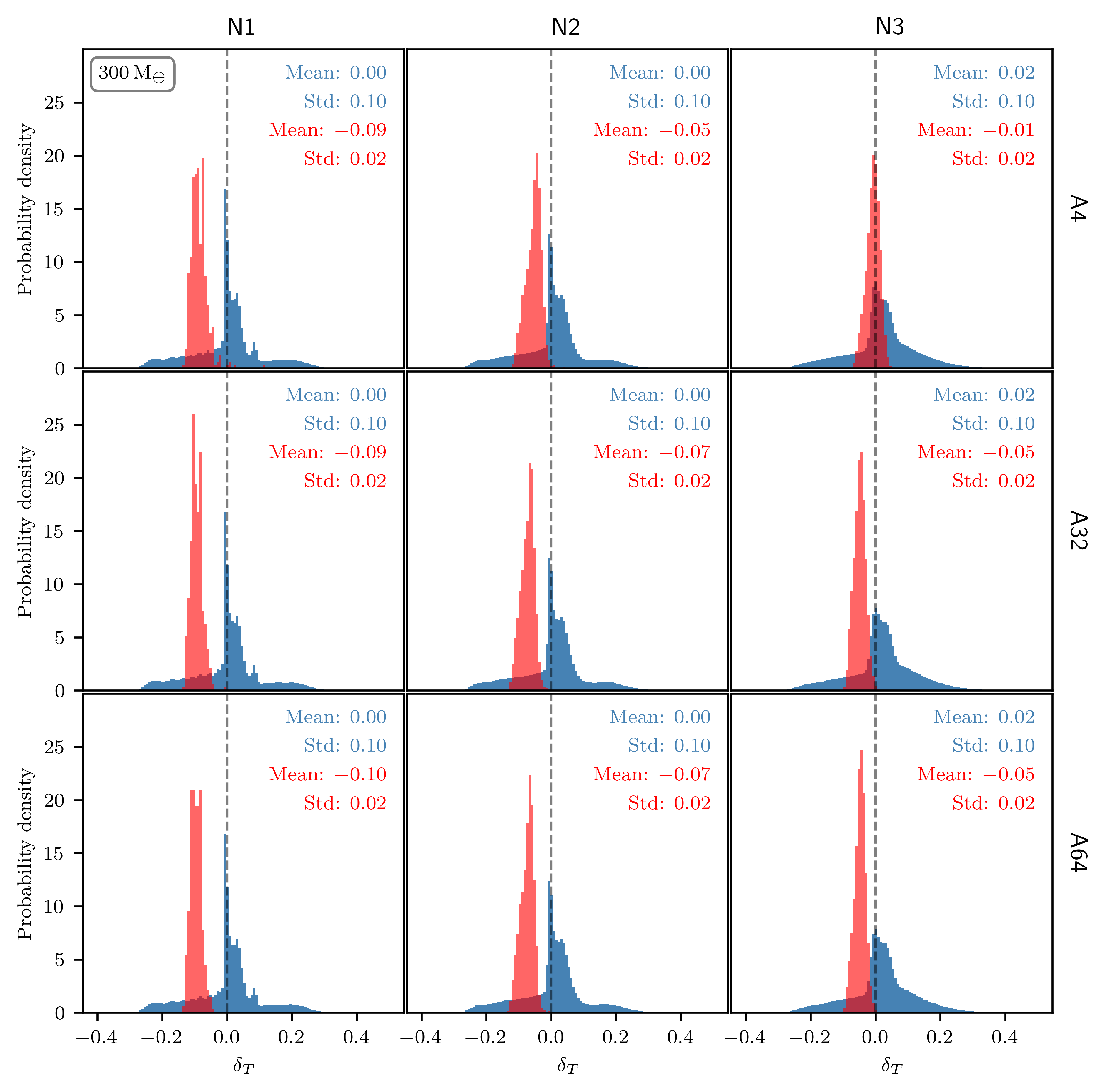}
      \caption{Overview of normalized histograms illustrating relative temperature differences for the $300\,{\rm M}_\oplus$ models, as in the upper plot of Fig. \ref{fig:histogram}. Here, each column represents data for a different model resolution (\texttt{N1}, \texttt{N2}, and \texttt{N3} from left to right), while different rows present data for different accretion parameters (\texttt{A4}, \texttt{A32}, and \texttt{A64} from top to bottom). For the purpose of better comparability, histograms have been clipped at a probability density value of 30.}
         \label{fig:histograms_P300}
   \end{figure*}

    To further study the effects of these parameters, Figs. \ref{fig:histograms_P300} and \ref{fig:histograms_P10} provide comprehensive views of normalized histograms illustrating relative temperature differences for the $300$ and $10\,{\rm M}_\oplus$ models, respectively, similar to the upper plot of Fig. \ref{fig:histogram}. Each column represents data for different resolutions (\texttt{N1}, \texttt{N2}, and \texttt{N3} from left to right), while each row shows data for different accretion parameters (\texttt{A4}, \texttt{A32}, and \texttt{A64} from top to bottom).    
    Overall, the (blue) distributions for the high-resolution regions are only marginally impacted by both numerical parameters. This is particularly the case with respect to the std, which has a value of ${\approx}0.1$ for all 18 models. The mean of the distribution, on the other hand, exhibits a slight shift toward larger relative differences for an increasing resolution, albeit relatively minor compared to the std. 
    
    The (red) distribution for the Hill region has a stronger dependence on the numerical parameters, which mostly affects its mean position. We find that the mean relative differences diminish with increasing resolution to values between ${-}0.05$ and ${-}0.01$ at resolution \texttt{N3}. Therefore, even at the highest tested resolution, $\temphd$ values are on average lower than corresponding $\temprt$ values in the Hill region, with a highest absolute value of the mean-to-std-ratio of ${\lesssim}2.5$. Furthermore, as anticipated, the effect on the mean is more pronounced for the $300\,{\rm M}_\oplus$ models, and its value is rather small for the models with a lower mass, which likely stems from the fact that models with higher resolutions do not adequately sample the regions with large temperature gradients near the $300\,{\rm M}_\oplus$ planet. The accretion parameter only barely affects the mean in the case of a $10\,{\rm M}_\oplus$ planet, with the most notable effect observed at the lowest resolution. Here, increasing values of the accretion parameter shift the distribution further away from zero toward higher negative temperature differences. In the case of the $300\,{\rm M}_\oplus$ planet, we likewise find that the (red) distribution shifts in the positive direction toward zero as the accretion parameter decreases. 
    
    Altogether, our findings suggest that while the peak temperature in the Hill region exhibits the highest similarity between RHD and MCRT simulations for larger accretion parameters, particularly for \texttt{A32}, the overall relative temperature difference distribution inside the planetary Hill region is closest to zero for the lowest accretion parameter, \texttt{A4}. Using higher resolutions leads to a general decrease in the mean difference between temperature estimates of our RHD and MCRT simulations; however, the systematic nature of the differences remains. Consequently, this issue does not appear to arise from the noise that is inherent to MC simulations. Instead, resolving it may require the development of more comprehensive numerical methods.

\subsection{Flux-limited diffusion: Axisymmetric models}
    \label{sec:axisym}
    To further investigate the origin of derived temperature differences, we performed the same analysis as presented in the previous parts of this section on the basis of axisymmetric uPPD models, which incorporate different parts of the physics. Model \mdIV{} considers the effect of stellar irradiation and viscosity, while model \mdI{} (\mdV) only considers stellar irradiation (viscosity) with the other effect excluded. All of these models use the same Planck opacity and Rosseland optical depth limiter as the 3D simulations.
    
    Additionally, we added two models in which we used reduced Planck opacities, as in the MCRT simulations, and did not apply an optical depth limiter (see Sect. \ref{sec:opac}). The iterative procedure of the FLD scheme necessitated a smoother transition around the evaporation temperature, and for this we chose a sigmoid function with a width of $100\,$K.
    In model \mdInu{}, we computed the irradiation flux $\mathbf{F}_*$ via frequency-dependent ray tracing, while in model \mdIkappa{} we computed it using Planck-averaged opacities, as in the hydrodynamical simulations.
    We post-processed the resulting HD-FLD simulation data using MCRT simulations in which the same sources of radiation have been considered. The analysis of these models with regard to arising differences in temperature estimates allowed us to further limit their potential origins. 

\subsubsection{Stellar irradiation and viscosity}
   \begin{figure*}
   \centering
   \includegraphics[width=\hsize]{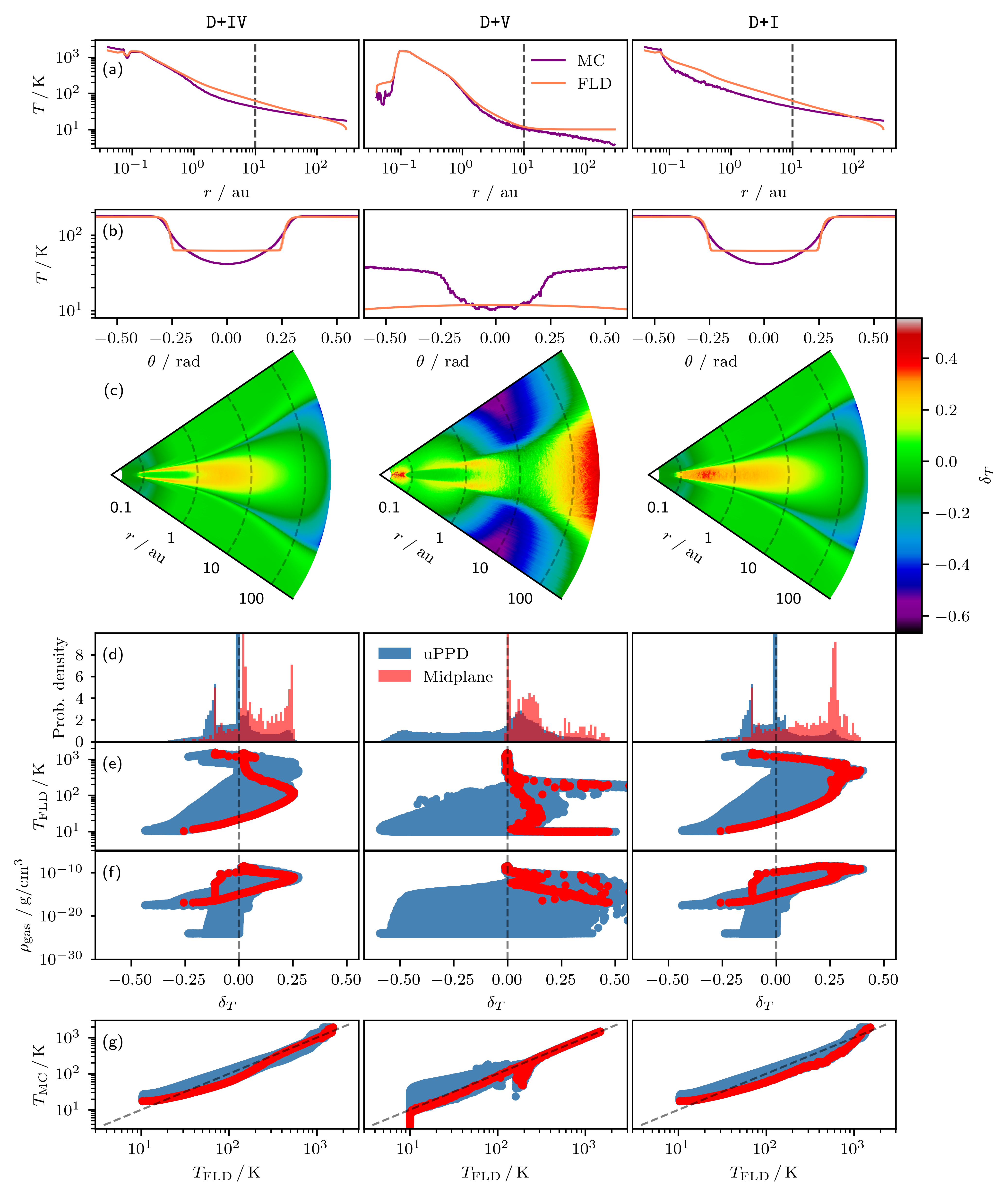}
      \caption{Temperature comparison overview for three axisymmetric uPPD models. The model is denoted at the top of each column. Different rows highlight different aspects of the comparisons between the FLD and MCRT temperatures. Row labels (a - g) at the top left of each row indicate the following: (a) Radial temperature profiles of the midplane, (b) polar temperature profiles at $r=10\,$au, (c) relative differences in a vertical cut through the midplane, (d) histogram of relative temperature differences, (e) scatter plot of FLD gas temperatures and relative temperature differences, (f) scatter plot of gas densities and relative temperature differences, and (g) scatter plot of MCRT and FLD temperatures.}
         \label{fig:temp_statistics_axisym_models}
   \end{figure*}

    Figure \ref{fig:temp_statistics_axisym_models} presents an overview of results for various performed analyses (rows a - g) of the models \mdIV{}, \mdV{}, and \mdI{} (left to right column). Here, the quantity $\delta_T$ refers to the determined (antisymmetric) relative temperature difference between the FLD and MCRT simulations.
    The radial temperature profiles in the midplane (row a) show a great level of agreement within the first $0.1\,$au for the models that include the effect of stellar irradiation (i.e., \mdIV{} and \mdI{}). The model \mdV{} does not include its effect and exhibits large differences in this region. A similar effect can be seen in the further-out regions at $r>10\,$au. In between, at $0.1\,{\rm au}<r<10\,{\rm au}$, the models show an overall good agreement between derived FLD and MCRT temperature estimates. Interestingly, the transition from a high level of agreement in the midplane to large differences concurs with the midplane gas density falling below a threshold at roughly $10^{-12}\,{\rm g}/{\rm cm}^2$. Moreover, in the lower-density regions in the midplane, the derived FLD temperatures consistently exceed the MCRT temperatures. However, it is important to note that in the FLD case, temperatures are bounded both inside the domain and at ghost cells by the value of the radiation energy in equilibrium at $10\,$K, which explains the encountered discrepancies in the outer regions of model \mdV{}. Since viscous heating dominates the radiation field in the densest regions of the midplane, the model \mdIV{} likewise shows a high level of agreement of both temperature estimates. 
    Apart from that, we find that models that consider stellar irradiation are characterized by higher derived MCRT temperatures than FLD temperatures in the far-out regions of the PPD. This is likely a consequence of using frequency-averaged opacities in the FLD, as discussed in Sect. \ref{sec:discussion_differences}. Another factor resulting in this discrepancy is the fact that MCRT simulations take the scattering of photons into account, which contributes to the radiation field in these shadowed low-density regions. The FLD simulations, in contrast, consider a single radiative flux proportional to the gradient of the energy density, and they are thus unable to reproduce angle-dependent scattering phenomena. In the FLD case, the inclusion of stellar irradiation appears to result in a greater increase in midplane temperatures than in the MCRT case. This can be seen in both models (i.e., \mdI{} and \mdIV{}).

    A similar effect can be seen in the polar temperature profiles at $r=10\,$au (row b). At this distance, stellar irradiation dominates the radiation field, as the determined temperatures are significantly lower in the case of model \mdV{} compared to model \mdI{}. We find that in the FLD case, the treatment of stellar radiation thus leads to notable plateaus in the far-out regions of the uPPD, with a rapid temperature change occurring at the photosphere. However, the model \mdV{} does not exhibit such a pronounced temperature variation. 
    
    Regarding the two-dimensional relative difference plots (row c), we find that the results for the model \mdIV{}, for the most part, resemble the results for the model \mdI{} except for the densest regions in the uPPD. This likely arises because stellar radiation dominates the radiation field in these regions while only marginally contributing to the radiation field deep inside the uPPD, where viscous heating dominates. Furthermore, the model \mdV{} exhibits interesting features in the photosphere and higher layers of the uPPD approximately between 3 and 80\,au, where MCRT temperatures exceed FLD temperatures (purple and black region). Since the only source of radiation in this model is viscous heating and considering that derived FLD temperatures do not fall below their MCRT counterparts in the midplane, we can likely attribute this drop in the relative difference to the simulation of scattering in the MCRT simulations. In particular, radiation that leaves the hot inner regions of the PPD gets scattered at the photosphere and upper layers and eventually contributes to radially further out optically thin regions of the PPD (compare with Fig. \ref{fig:app:temp_axisym_DV}). 
    
    In accordance with our previous results (Fig. \ref{fig:temp}), we find the transition from the inner part of the uPPD to its optically thin upper layers to be more abrupt in the FLD case, particularly if stellar irradiation is included, which manifests in the form of a sudden change of relative differences (from green to light blue). 
    In general, we can attribute this to a few factors. Firstly, the use of Planck-averaged opacities for stellar irradiation in the FLD simulations implies a single $\tau=1$ surface for stellar photons (see Sect. \ref{sec:discussion_differences}).
    Secondly, the transition in the MCRT case can be expected to be smoother due to the wavelength-dependence of the vertical optical depth of escaping photons. Thirdly, the interaction between the flux entering and leaving the disk in the FLD leads to a different functional form of the radiation field than in the MCRT, as discussed in Sect. \ref{sec:discussion_differences}.
    Lastly, photons emitted at inner regions of the uPPD may scatter at the photosphere and higher layers and warm up the radially far-out regions closer to the midplane. 
    We also find that the inner roughly $0.3\,$au of the models \mdIV{} and \mdI{} are generally colder in the FLD case than in the MCRT case, which is not the case for the model \mdV{}. This is a numerical feature resulting from the assumed minimum optical depth per cell of $10^{-2}$, as we verify in Sect. \ref{sec:optdepthlimiter}.

    The histograms of relative temperature differences (row d) show results for the entire uPPD (blue) and a single thin layer of cells in the midplane (red). Overall, they reveal a good agreement in temperatures for the models that include stellar irradiation. However, the midplane distributions of these models (\texttt{D+IV} and \texttt{D+I}) exhibit a local maximum at about $\delta_T=0.25$, which does not show for the model \mdV{}. This local peak is again a result of the stronger heating of the dense region that we observed for the FLD simulations in comparison to the MCRT case. 
    This effect additionally shows in the scatter plot of the FLD temperatures and relative differences (row e), as the distribution of points corresponding to the midplane is mostly found on the very right side of the uPPD distribution for the model \mdI{}. The larger relative differences that can be found in this model are notably reduced by the additional consideration of viscous heating, which is reflected by the narrower distribution of data points in the scatter plot of the model \mdIV{}.
    Likewise, a reduction of relative differences is also found in the scatter plot of gas densities and relative differences (row d). Moreover, these plots reveal that a large portion of low-density cells have a much lower FLD than MCRT temperatures. This is in line with our finding of potentially missing contributions to the radiation field due to a neglect of angle-dependent transport and scattering in the FLD, together with the use of frequency-averaged opacities and the mentioned boundary-related phenomena in the FLD. 
    Lastly, the scatter plots of the MCRT and FLD temperatures (row g) show a very clear trend in that the data points generally align very well with the desired linear distribution indicated by the gray dashed line. The most noticeable spreading of the distribution can be found for the model \mdV{}, particularly at very low temperatures, which is at the very least partly attributed to the temperature bound of $10$ K imposed both inside the domain and at the boundaries.

\subsubsection{Optical depth limiter and wavelength dependence}
\label{sec:optdepthlimiter}

    To investigate the impact of the optical depth limiter (Sect. \ref{sec:opac}) and the inclusion of a wavelength-dependent radiation scheme, Fig. \ref{fig:temp_statistics_axisym_models_IRR}  presents a similar overview of results as shown in Fig. \ref{fig:temp_statistics_axisym_models} but for the models \mdI{}, \mdInu{}, and \mdIkappa{} (left to right column). To remove the impact of the noise inherent to the MCRT simulations from this analysis, we used the same MCRT temperature distribution for the generation of all of the results. Overall, we find that the removal of the optical depth limiter and the inclusion of wavelength-dependence lead to an overall improvement of the FLD radiation scheme, especially in the midplane of the uPPDs and in the region close to the star at $r{\lesssim}0.3\,$au. In particular, the latter effect is caused by the removal of the minimum optical depth bound. In all rows, these improvements show a reduced difference in the radial profiles (row a), smoother transitions in the polar profiles (row b), increased similarity at various regions in the vertical cut plots (row c), a significantly lower probability for negative relative differences in the uPPDs (blue histograms in row d), and much narrower regions populated in the scatter plots (rows e, f, and g), especially for regions of very low or particularly high gas densities. 

   \begin{figure}[!htb]
   \centering
   \includegraphics[width=\hsize]{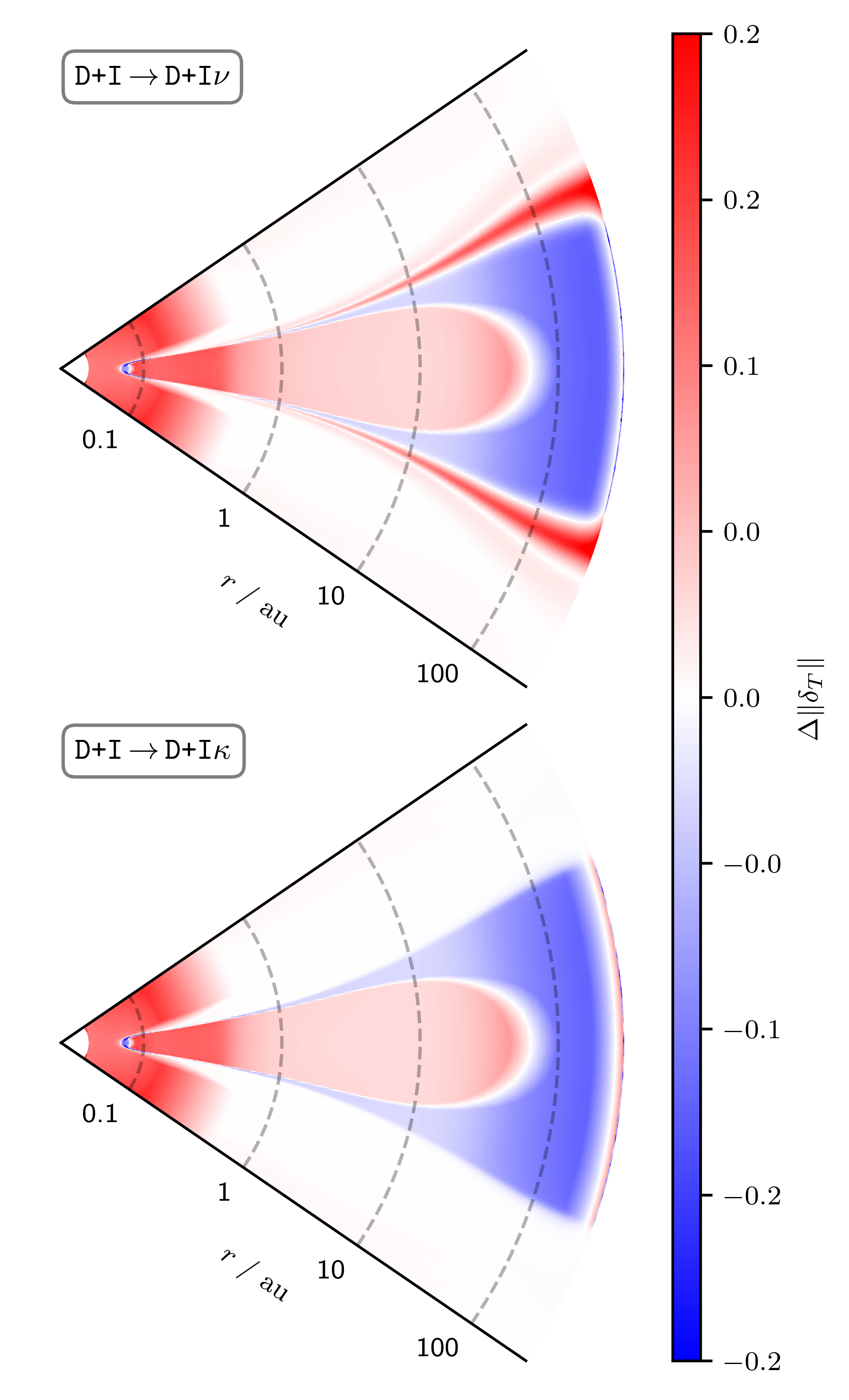}
      \caption{Change of similarity, $\Delta\|\delta_T\|$, in a vertical cut through the midplane of the uPPD. The upper (lower) plot shows the results for switching from model \mdI{} to model \mdInu{} (\mdIkappa{}). In regions where $\Delta\|\delta_T\|$ is positive (negative), the change led to an increased (decreased) agreement between FLD and MCRT temperatures. (For details, see Sect. \ref{sec:optdepthlimiter}.)}
         \label{fig:temp_statistics_axisym_models_diffIRR}
   \end{figure}

    To better assess in which regions of the uPPDs the changes to the optical depth limiter (model \mdIkappa{}) and wavelength-dependence (model \mdInu{}) mostly affect the similarity of the derived FLD and MCRT temperatures, we determined cell-wise the quantity $\Delta\|\delta_T\| = \lvert \delta_T({\rm \texttt{D+I}}\kappa)\rvert - \lvert \delta_T({\rm \texttt{D+I}}x)\rvert$, where $x$ is either $\nu$ or $\kappa$. In regions where it is positive (negative), the change led to a better (worse) agreement between the FLD and MCRT temperatures. Figure \ref{fig:temp_statistics_axisym_models_diffIRR} shows $\Delta\|\delta_T\|$ for a vertical cut through the midplane of model \mdInu{} (upper plot) and \mdIkappa{} (lower plot). In both cases, the change to the treatment of radiation significantly improved the similarity (red regions) in the inner region close to the star, which includes the optically thin upper layers of the uPPD, the optically thickest regions of the disk, and a large part of the denser regions inside the uPPD close to the midplane. Regions where the similarity decreased (blue regions) encompass a spatially small region slightly within $0.1\,$au, a thin layer enveloping the aforementioned denser region of the uPPD, and spatially large regions of lower density extending far out to the outer boundary of the simulated uPPD. Additionally, model \mdInu{} exhibits a significant improvement at the photosphere and at slightly higher layers of the PPD, which also showed in the form of a smoother transition of temperatures in row b of Fig. \ref{fig:temp_statistics_axisym_models_IRR}. 
    
    Based on the analysis of the axisymmetric models, we conclude that understanding the observed temperature discrepancies requires considering various factors. These include the treatment of optically thin transport and scattering and the wavelength-dependence of optical properties, among other numerical implementation details. A detailed discussion is presented in Sect. \ref{sec:discussion_differences}.

    Nonetheless, it is important to consider that differences in the estimated temperature distribution do not necessarily translate to observable differences in derived flux maps. Therefore, it is also important to quantify the impact of these differences on resulting flux maps at various wavelength ranges.

\subsection{Flux map comparison}
\label{sec:flux_section}
    Based on the derived temperature distributions $\temphd$ and $\temprt$, we determined the ideal flux maps through MCRT simulations using Mol3D (see Sect. \ref{sec:rt_simulations}) for the VIS ($\lambda = 0.652\,\mu$m), NIR ($\lambda = 2.78\,\mu$m), and submm ($\lambda = 345\,\mu$m) wavelength ranges, which correspond to typical observing wavelengths of instruments such as the Spectro-Polarimetric High contrast imager for Exoplanets REsearch (SPHERE) of the Very Large Telescope (VLT) \cite{2019A&A...631A.155B}, the James Webb Space Telescope (JWST; \citealt{2022A&A...661A..80J}), and the Atacama Large Millimeter/submillimeter Array (ALMA; \citealt{2002Msngr.107....7K}), respectively. Building on the consistent findings from previous temperature distribution analyses, we limited this investigation to models characterized by the accretion parameter \texttt{A32} and resolutions \texttt{N1} or \texttt{N3}. Per model and wavelength, a total of $N_\gamma=10^8$ photon packages were simulated for the stellar contribution to the flux map. We simulated $\gtrsim 10^8$ ($10^{10}$) photon packages for the self-scattered thermal dust emission in the VIS and submm range (in the NIR), and we applied a ray tracer routine to construct flux maps corresponding to the direct unscattered thermal emission of the dust. The increased photon package number used for the simulations in the NIR are necessary to decrease the noise in relative flux difference maps, which are determined in later parts of this section. Additionally, the NIR flux maps were convolved with a Gaussian kernel using a full width at half maximum value of 5 pixels. At this observing wavelength, the resulting point spread function is significantly narrower than the real point spread function, which is achieved even by the 8.2\,m Unit Telescopes of the VLT, assuming a typical distance of $140\,$pc to the simulated system. Despite the convolution procedure, the resulting flux maps therefore still have a resolution that is significantly higher than what modern instruments can achieve. In our MCRT simulations, we additionally made use of the composite-biasing technique \citep{2018ApJ...861...80C,2021A&A...645A.143K} and the minimum scattering order method \citep{2024A&A...682A..99K} to further increase the quality of our results. 

\subsubsection{Reference model}
\label{sec:flux_reference}

\begin{figure*}[!htb]
   \centering
   \includegraphics[width=0.9\hsize]{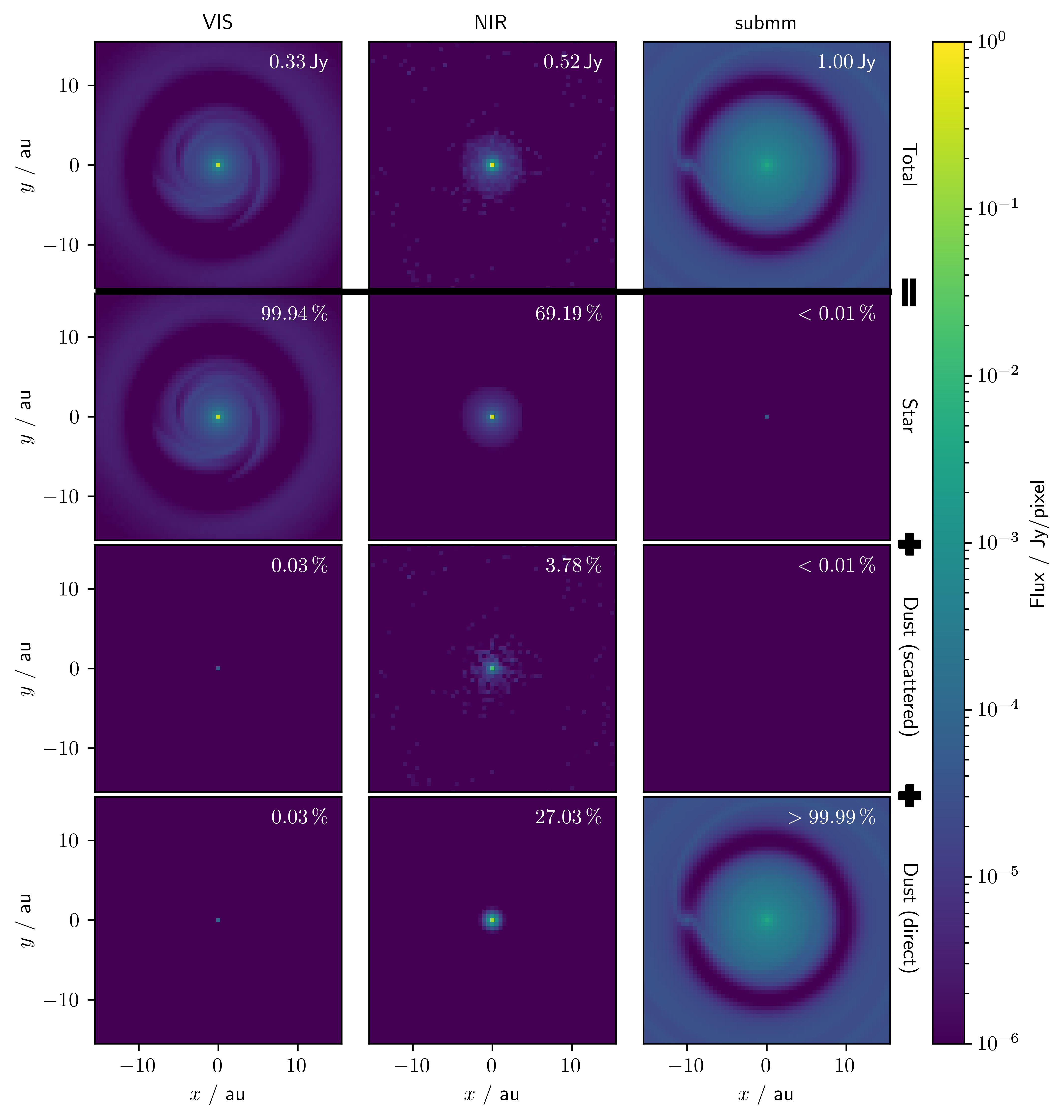}
      \caption{Total flux maps (first row from the top) at three different wavelengths, the VIS (left column), NIR (central column), and submm (right column) range, for the reference model based on the temperature distribution $\temprt$. The total flux maps were calculated as the sum of three individual flux maps corresponding to the contributions of the star (second row), self-scattered thermal dust emission (third row), and direct unscattered thermal dust emission (fourth row). Respective labels of the individual maps and their interrelationship are indicated at the right of each row. The total flux of each model, given in Jansky, is indicated in the upper-right corner of each total flux map. Percentage values in the upper-right corner of the flux maps below (rows 2, 3, and 4), denote the contribution of the respective radiation source to the total flux of the model.}
         \label{fig:flux_maps_reference_model}
   \end{figure*}

    Figure \ref{fig:flux_maps_reference_model} shows the results of our simulations in terms of total flux maps (first row from the top) at three different wavelength ranges, VIS (left column), NIR (central column), and submm (right column), for the reference model using the temperature distribution $\temprt$. Each total flux map has been calculated as the sum of three individual flux maps that correspond to the contribution of the star (second row), self-scattered thermal dust emission (third row), and direct unscattered thermal dust emission (fourth row). The corresponding labels of the individual maps and their interrelationship are signified at the right side of each row. For the purpose of comparability, the color bar applies to all displayed flux maps. The summed flux of the entire system, given in Jansky, is indicated in the upper-right corner of each corresponding total flux map. Percentage values in the upper-right corner of each flux map below (rows 2, 3, and 4) denote the contribution of the corresponding radiation source to the total flux of the system. 

    As can be expected, the appearance of the systems in the VIS wavelength range is dominated by (scattered) stellar radiation ($99.94\,\%$). Consequently, density variations at the photosphere of the PPD have a significant impact on the appearance, giving rise to the emergence of spiral features, a clear gap, and a ring feature. At such a short wavelength, the dust phase is extremely opaque, which, in combination with the high dust densities that can be found near the planet, damps down the contribution of the accretion luminosity significantly. In the NIR, about one-third of the total flux stems from thermal dust radiation (sum of rows 3 and 4); however, its contribution is for the most part confined to a small region near the central star. As a result, compared to the VIS wavelength range, the central region becomes significantly brighter than the outer regions of the PPD. In the submm range, contributions from stellar and scattered thermal radiation become negligible compared to the direct unscattered thermal dust emission, which accounts for almost the entirety (${>}99.99\,\%$) of the observed total flux. Since dust in the vicinity of the planet becomes heated due to the release of planetary accretion luminosity, the ideal flux map shows a clear feature that is indicative of the presence of circumplanetary material, which is located inside a distinct gap feature. Furthermore, faint spiral features can be spotted both inward and outward of the gap region, which roughly trace the density distribution in the disk midplane.

    \begin{table*}[]
        \caption{Results for the analyzed flux maps.}
        \centering
        \footnotesize
        \begin{tabular}{ccc|c|cccc|c}
        \toprule\toprule
        Resolution & Mass & Type & F$_{\rm VIS}\,({\rm F}_{\rm VIS}^{\rm star}/{\rm F}_{\rm VIS})$ & F$_{\rm NIR}$ & F$^{\rm star}_{\rm NIR}$ & F$^{\rm dust,\, scattered}_{\rm NIR}$ & F$^{\rm dust,\, direct}_{\rm NIR}$ & F$_{\rm submm}\,({\rm F}^{\rm dust,\, direct}/{\rm F}_{\rm submm})$\\
        \midrule
    \texttt{N1} & $300\,{\rm M}_\oplus$ & RT & $0.33\,{\rm Jy}$\,($99.94\,\%$) & $0.52\,{\rm Jy}$ & $0.36\,{\rm Jy}$ & $0.02\,{\rm Jy}$ & $0.14\,{\rm Jy}$ & $1.00\,{\rm Jy}$\,(${>}99.99\,\%$) \\
    \texttt{N1} & $300\,{\rm M}_\oplus$ & RHD & $0.34\,{\rm Jy}$\,($99.99\,\%$) & $0.79\,{\rm Jy}$ & $0.36\,{\rm Jy}$ & $0.06\,{\rm Jy}$ & $0.37\,{\rm Jy}$ & $0.87\,{\rm Jy}$\,($99.99\,\%$) \\
    \texttt{N1} & $10\,{\rm M}_\oplus$ & RT & $0.33\,{\rm Jy}$\,($99.94\,\%$) & $0.52\,{\rm Jy}$ & $0.36\,{\rm Jy}$ & $0.02\,{\rm Jy}$ & $0.14\,{\rm Jy}$ & $1.09\,{\rm Jy}$\,(${>}99.99\,\%$) \\
    \texttt{N1} & $10\,{\rm M}_\oplus$ & RHD & $0.34\,{\rm Jy}$\,($99.99\,\%$) & $0.79\,{\rm Jy}$ & $0.36\,{\rm Jy}$ & $0.06\,{\rm Jy}$ & $0.37\,{\rm Jy}$ & $1.02\,{\rm Jy}$\,(${>}99.99\,\%$) \\
    \midrule
    \texttt{N3} & $300\,{\rm M}_\oplus$ & RT & $0.34\,{\rm Jy}$\,($99.94\,\%$) & $0.55\,{\rm Jy}$ & $0.36\,{\rm Jy}$ & $0.02\,{\rm Jy}$ & $0.17\,{\rm Jy}$ & $1.03\,{\rm Jy}$\,(${>}99.99\,\%$) \\
    \texttt{N3} & $300\,{\rm M}_\oplus$ & RHD & $0.34\,{\rm Jy}$\,($99.98\,\%$) & $0.87\,{\rm Jy}$ & $0.36\,{\rm Jy}$ & $0.05\,{\rm Jy}$ & $0.45\,{\rm Jy}$ & $0.98\,{\rm Jy}$\,(${>}99.99\,\%$) \\
    \texttt{N3} & $10\,{\rm M}_\oplus$ & RT & $0.34\,{\rm Jy}$\,($99.94\,\%$) & $0.55\,{\rm Jy}$ & $0.36\,{\rm Jy}$ & $0.02\,{\rm Jy}$ & $0.17\,{\rm Jy}$ & $1.08\,{\rm Jy}$\,(${>}99.99\,\%$) \\
    \texttt{N3} & $10\,{\rm M}_\oplus$ & RHD & $0.34\,{\rm Jy}$\,($99.98\,\%$) & $0.87\,{\rm Jy}$ & $0.36\,{\rm Jy}$ & $0.05\,{\rm Jy}$ & $0.45\,{\rm Jy}$ & $1.08\,{\rm Jy}$\,(${>}99.99\,\%$) \\
        \bottomrule
        \end{tabular}
        \label{tab:fluxes}

        \tablefoot{
Model parameters (resolution, planetary mass, and simulation type) are listed in the first three columns. Total flux values for the VIS (F$_{\rm VIS}$), NIR (F$_{\rm NIR}$), and submm (F$_{\rm submm}$) wavelength ranges are shown in the fourth, fifth, and ninth column, respectively. Bracketed values denote the contributions of the dominant flux sources to the total flux in percent. The flux value F$^{s}_{w}$ corresponds to the simulated contribution of the source $s$ (e.g., the star, scattered thermal emission from the dust, or direct unscattered thermal dust emission) to the total flux observed in the wavelength range $w$. For the NIR, a complete breakdown of the contributions of three different sources to the total flux are presented in the sixth to eighth columns. 
}
    \end{table*}
    
    Table \ref{tab:fluxes} presents a summary of selected results, with model parameters listed in the first three columns. The total flux values for the VIS (F$_{\rm VIS}$), NIR (F$_{\rm NIR}$), and submm (F$_{\rm submm}$) wavelength ranges are shown in the fourth, fifth, and ninth column, respectively. Bracketed values denote the contributions of the dominant flux sources to the total flux in percent. Additionally, the flux value F$^{s}_{w}$ corresponds to the contribution of the source $s$ (e.g., the star, scattered thermal emission from the dust, or direct unscattered thermal dust emission) to the total flux in the wavelength range $w$. For the NIR, a complete breakdown of the contributions of three different sources to the total flux is presented in the sixth to eighth columns. 

    Overall, we find the flux in the VIS wavelength range to be consistent among all models. This is in agreement with our expectations, as it is dominated (${>}99.9\,\%$) by scattered stellar light, which is independent of the temperature distribution. For the NIR, we also find consistent results for the contribution of scattered stellar light to the total flux of $0.36\,$Jy; however, the total contribution of the thermal dust emission (direct plus scattered) may strongly vary depending on the simulation type. In particular, using $\temphd$ as a basis for determining flux maps leads to total flux estimates that are consistently $0.27-0.31\,$Jy higher, which corresponds to an increase of roughly $160\,\%$ in the observed total thermal dust emission and an increase of $51-58\,\%$ in total flux. This is also consistent with the finding that regions of higher density in the high-resolution region of the PPD often exhibit a positive relative temperature difference (compare with the lower plot in Fig. \ref{fig:histogram}). In contrast to the VIS and NIR, we find the flux in the submm range to be dominated by direct unscattered thermal dust emission. Here, the total flux slightly increases with higher resolution by $0.05-0.15\,$Jy, which corresponds to an increase of $5-17\,\%$. Furthermore, temperature distributions derived by the means of MCRT simulations consistently lead to increased total flux values, which are up to $0.13\,$Jy (i.e., $15\,\%$) higher than their RHD counterparts, despite the latter exhibiting often higher temperatures in the densest regions of the PPD. For observing wavelengths in the submm, however, even the coldest parts of the PPD may effectively contribute to the observed total flux of the system. Therefore, the fact that flux estimates derived from $\temphd$ are lower is consistent with our finding of colder dust in the far-out regions ($r{\gtrsim}60\,$au) of the midplane within the PPD (compare with Fig. \ref{fig:temp_profiles}). In short, we find that the temperature differences between distributions derived with the RHD and MCRT method can result in significant discrepancies for observing wavelengths in the NIR and longer. Given the systematic nature of these differences, we find that a reliable flux map determination therefore requires having the most comprehensive and precise understanding of the radiative transfer mechanisms occurring throughout the entire PPD and relying on as few approximations as possible.

\subsubsection{Impact of model parameters}
\label{sec:flux_map_comparison}
    In this section, we investigate the impact of using dust temperature distributions derived by the means of either RHD or MCRT simulations when determining flux maps in the VIS, NIR, and submm wavelength ranges using full MCRT simulations. To that end, we calculated total flux maps for all models with resolutions \texttt{N1} and \texttt{N3}, characterized by the accretion parameter \texttt{A32},  based on the temperature distribution $\temprt$, and we repeated these simulations assuming the temperature distribution $\temphd$. Then we calculated per pixel the relative flux difference of the resulting total flux maps. Here, the quantity $F_{t,\,n}$ describes the flux, with units of Jansky per pixel, assuming type $t\in\left\{{\rm RHD},\,{\rm MC}\right\}$ and resolution $n\in\left\{{\rm \texttt{N1}},\,{\rm \texttt{N3}}\right\}$.\footnote{Results for resolution \texttt{N2} can be found in Appendix \ref{sec:app:flux_compare}.}

\paragraph*{Similarity at different resolutions:}
   \begin{figure*}
   \centering
   \includegraphics[width=0.9\hsize]{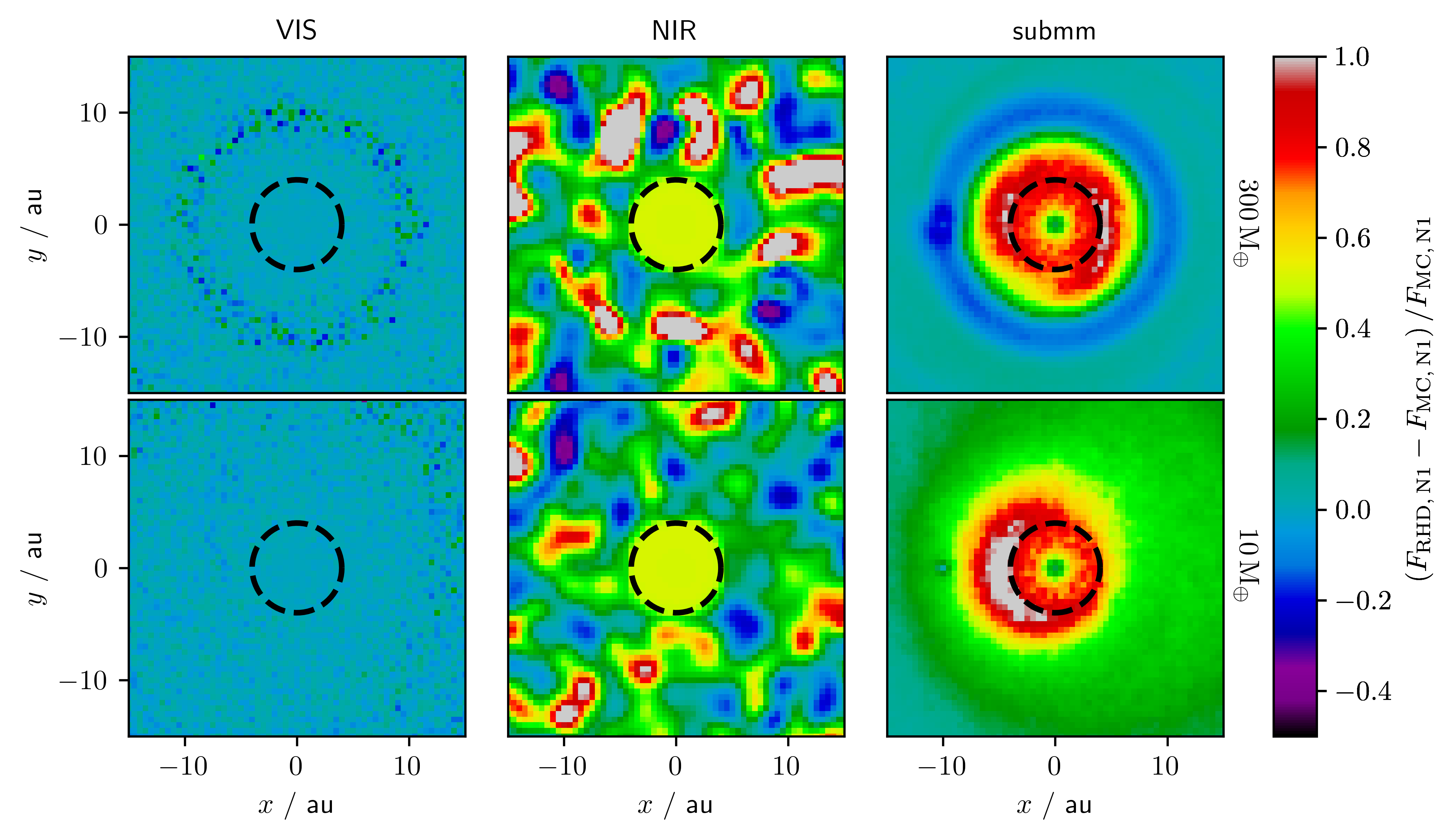}
      \caption{Relative flux difference maps for models based on the resolution \texttt{N1} that simulate a $300\,{\rm M}_\oplus$ (upper row) or $10\,{\rm M}_\oplus$ (lower row) planet when assuming observing wavelengths in the VIS (left column), NIR (central column), or submm (right column) range. For the purpose of better comparability, the color bar has been clipped.}
         \label{fig:flux_reldiff_N1}
   \end{figure*}

    Figure \ref{fig:flux_reldiff_N1} shows the central $30\,{\rm au}\times30\,{\rm au}$ of the relative flux difference maps for models based on the resolution \texttt{N1} with a $300\,{\rm M}_\oplus$ (upper row) or $10\,{\rm M}_\oplus$ (lower row) planet using observing wavelengths in the VIS (left column), NIR (central column), or submm (right column) range. Positive (negative) values of the displayed relative difference $\left( F_{{\rm RHD},\,{\rm \texttt{N1}}}-F_{{\rm MC},\,{\rm \texttt{N1}}} \right)/F_{{\rm MC},\,{\rm \texttt{N1}}}$ correspond increased (decreased) flux values when using $\temphd$ rather than $\temprt$ as underlying temperature distribution. The black dashed circle in the center of each plot additionally highlights the transition from the background model to the high-resolution model at a radius of $4\,$au.  

    In the VIS wavelength range (left column), we find that the flux maps are in great agreement. Most of the differences can be expected to occur as a result of inherent MC noise. The annular feature that can be spotted in the region of the plot where the gap would be located in the flux map is likely merely a region of higher MC noise, that is, with greater variance regarding the determined flux values. The noise arises due to the decreased local density inside the gap and the lower photon interaction probability associated with it. The high level of agreement between both simulation types is expected since scattered stellar flux makes up the majority of the detected flux at this wavelength, which can be seen in the first column of Fig. \ref{fig:flux_maps_reference_model}. As a consequence, the underlying temperature distribution barely affects the observed flux map, and therefore any changes to it are also expected to have only minor effects. 
    In the NIR (central column), we find that the increase in temperature inside the PPD close to the star associated with a change to the RHD temperature distribution, which can for instance be seen in Fig. \ref{fig:temp}, results in a likewise increase in flux from that region. This effect, though, is spatially limited to areas in the central plots well inside the black circles, meaning far from the planet. Inside the high-resolution regions (i.e., outward of the black dashed lines), though, relative differences are completely dominated by MC noise. We note that an increased noise level at similar observing wavelengths is typical for self-scattering simulations of optically very thick PPDs \citep{2021A&A...645A.143K,2023A&A...680A..67K}. However,  the seemingly heightened noise in the relative difference map compared to the actual flux map (see central column in Fig. \ref{fig:flux_maps_reference_model}) is a result of the fact that at greater distances from the star, flux levels tend to decrease. Consequently, even minor variations in the (absolute) flux level appear relatively large. 
    In contrast, in the submm wavelength range (right column), the PPD becomes optically much thinner, and contributions from scattered light to the total flux map can essentially be neglected (compare with Fig. \ref{fig:flux_maps_reference_model}). Considering that the compared models share the same density distribution and only differ in their underlying temperature distributions, we can conclude that flux differences primarily stem from temperature differences in the densest regions of the PPD, namely, the midplane. Comparing the upper-right plot of the relative flux differences in this figure with the upper-right plot of relative temperature differences in Fig. \ref{fig:temp} clearly supports this assertion. As a consequence, regions in the midplane in which the RHD temperature distribution exhibits higher (lower) values than the MCRT temperature distribution are associated with areas in the plot of positive (negative) relative flux differences. Using the RHD temperature distribution of the $300\,{\rm M}_\oplus$ models as a basis to determine flux maps by means of MCRT simulations therefore roughly results in relative flux differences for different regions (from the inside outward): vicinity of the star (${+}10$ to ${+30}\,\%$), region inward of the gap feature  (${+}30$ to ${+}100\,\%$), gap region  (${-}10$ to ${+}0\,\%$), in the circumplanetary region inside the gap (${-}30$ to ${-}10\,\%$), and just outward of the gap feature (${+}0$ to ${+}10\,\%$). For the $10\,{\rm M}_\oplus$ models, the differences are as follows: vicinity of the star (${+}10$ to ${+30}\,\%$), region inward of the planetary orbit  (${+}40$ to ${\gtrsim}100\,\%$), in the circumplanetary region (${+}10$ to ${+}30\,\%$), and the region just outward of the planetary orbit (${+}10$ to ${+}20\,\%$).


    We performed the same analysis for the highest resolution, \texttt{N3}. The maps for the derived relative flux difference $\left( F_{{\rm RHD},\,{\rm \texttt{N3}}}-F_{{\rm MC},\,{\rm \texttt{N3}}} \right)/F_{{\rm MC},\,{\rm \texttt{N3}}}$ are displayed in Fig. \ref{fig:flux_reldiff_N3}. Similar to our results for the lower resolution, we find a general consistency between the RHD and MCRT temperature distributions for the VIS wavelength range. In the NIR, we also find comparable results, which are, first, an area in the plot of increased relative flux differences well within the dashed line and, second, no other clear structure outside the dashed line. However, this region exhibits a generally increased flux difference level compared to the lower resolution case. These elevated values are likely a result of the increased contribution of thermal dust emission associated with the higher resolution. Another expected and confirmed consistency is that the derived relative flux differences within the dashed black line align for all three observing wavelengths across both model resolutions, \texttt{N1} and \texttt{N3}. The greatest differences to the \texttt{N1} models can be found in the submm range, within the high-resolution region beyond a radius of $4\,$au. Similar to the NIR, here, the overall flux difference level is also elevated, which is accompanied by an increased temperature difference in the midplane inside the PPD of the \texttt{N3} models. Additionally, for the $300\,{\rm M}_\oplus$ models, the plot exhibits broad spiral-like features in the region inside the planetary orbit. Quantitatively, we find that using the RHD temperature distribution of the $300\,{\rm M}_\oplus$ models as a basis to determine flux maps by the means of MCRT simulations results roughly in the following relative flux differences: vicinity of the star (${+}10$ to ${+30}\,\%$), region inward of the gap feature  (${+}30$ to ${+}100\,\%$), gap region  (${+}10$ to ${+30}\,\%$), in the circumplanetary region inside the gap (${-}20$ to ${+}0\,\%$), and just outward of the gap feature (${+}20$ to ${+}40\,\%$). For the $10\,{\rm M}_\oplus$ models, the differences are as follows: vicinity of the star (${+}10$ to ${+30}\,\%$), region inward of the planetary orbit  (${+}50$ to ${+}100\,\%$), in the circumplanetary region (${+}10$ to ${+}40\,\%$), and the region just outward of the planetary orbit (${+}30$ to ${+}50\,\%$). Whether these alterations translate to a higher degree of similarity for the \texttt{N3} models between both simulation types is investigated in detail in the following section.

\paragraph*{Impact of the resolution:} 
    
    So far, we have investigated relative flux differences based on results of RHD and MCRT simulations while assuming a fixed resolution. To directly compare both studied resolutions, \texttt{N1} and \texttt{N3}, Fig. \ref{fig:flux_reldiff_resolution} shows exemplarily a map of the relative flux differences between flux maps derived with resolution \texttt{N1} and resolution \texttt{N3}. In  particular, they are given by $\left( F_{{\rm MC},\,{\rm \texttt{N3}}}-F_{{\rm MC},\,{\rm \texttt{N1}}} \right)/F_{{\rm MC},\,{\rm \texttt{N1}}}$, such that positive (negative) values correspond to a case in which the flux values of the \texttt{N3} models exceed (fall below) those of the corresponding \texttt{N1} models. 

    For all the considered observing wavelengths, we find that the region inside the black dashed line in the figure generally yields very consistent results (i.e., differences are close to zero), independent of the particular resolution of the high-resolution region, which reconfirms our previous findings. Outward of $4\,$au, though, the differences can grow significantly, suggesting great differences in the underlying flux maps between models of different resolution. Interestingly, the differences found between RHD and MCRT temperatures for a given resolution seem to be significantly smaller than the differences observed between models of different resolution. Contrary to previous plots, and in order to cope with these large disparities, the color bar therefore allows for the depiction of relative differences of up to ten. For the VIS and NIR for both considered planetary masses, these differences are particularly high in the area of the plot just outside $4\,$au but also at the position of the gap feature. Interestingly, such features did not show up in a comparison of the RHD and MCRT methods in Figs. \ref{fig:flux_reldiff_N1} and \ref{fig:flux_reldiff_N3}, meaning, they are not a feature of the simulation methods but rather of the resolution itself. However, it is important to note that models of different resolutions do not share the same density distribution. Moreover, since observations in the VIS wavelength range primarily originate from scattered stellar light, which is highly dependent on the density distribution in the upper layers of the PPD, these large differences can likely be attributed purely to density variations in these layers. Similar reasoning can be expected to apply for the NIR, and consequently, these differences cannot be solely attributed to a change of resolution. In the submm range, contrary to the results for the VIS and NIR, we find that large relative differences for the $300\,{\rm M}_\oplus$ model, induced by an increase of resolution, are mainly confined to the region of the observed gap feature. Here, the higher-resolution model is accompanied by an increase in flux by a factor of two to ten. In comparison, ideal flux maps of the $10\,{\rm M}_\oplus$ model in the submm range demonstrate that they are only minimally affected by an increase in resolution. This, however, can be expected, as the radial (midplane) temperature profiles of these models, which strongly correlate with their respective flux maps (see Sect. \ref{sec:flux_section}), exhibit their greatest differences in the region of the gap while remaining relatively similar elsewhere. Though this comparison allowed us to assess the precise differences between models with different resolutions, conclusions regarding the impact of the resolution on the similarity of resulting flux maps cannot be drawn from these differences alone, as the impact of the altered density distributions cannot be distinguished from that of the resolution by itself. 


    To reduce the impact of the induced density changes and thereby extract the effect of the resolution itself, Fig. \ref{fig:flux_reldiff_similarity} shows a plot of the quantity $\Delta S{\equiv} \Delta_{\rm N} \left\lVert \left(F_{\rm RHD}-F_{\rm MC}\right)/F_{\rm MC} \right\rVert$, which traces the change in similarity of the results of the RHD and MCRT simulations induced by an increase of resolution. The quantity $\Delta S$ is defined as
    \begin{equation*}
        \Delta S 
        = 
        \left\lvert \frac{F_{\rm RHD,\,\texttt{N1}}-F_{\rm MC,\,\texttt{N1}}}{F_{\rm MC,\,\texttt{N1}}} \right\rvert 
        - 
        \left\lvert \frac{F_{\rm RHD,\,\texttt{N3}}-F_{\rm MC,\,\texttt{N3}}}{F_{\rm MC,\,\texttt{N3}}} \right\rvert, 
    \end{equation*}
    where positive (negative) values correspond to an increase (decrease) in similarity associated with the use of a higher resolution. For instance, a value of $\Delta S=0.1$ ($\Delta S=-0.1$) corresponds to a case in which the flux map, which is based on the temperature distribution $\temphd$, differs from that based on $\temprt$ by being ten percentage points lower (higher) when switching from resolution \texttt{N1} to \texttt{N3}.  

    Using $\Delta S$ as a quantity to measure the change of similarity, we find that within the area in the plot that corresponds to the background model, that is, within the black dashed circle, the similarity of flux maps derived from RHD or MCRT based temperature distributions does not improve with increasing resolution, which is in agreement with our expectations. The VIS and NIR wavelength ranges also do not exhibit clear differences with regard to the similarity. Similar to our previous findings (compare with Fig. \ref{fig:flux_reldiff_N1} and its discussion), first, a slight annular feature can be detected for the VIS range, which again likely arises due to a low simulated interaction rate in the gap region. Second, the map in the NIR can be explained by heightened levels of MC noise likely induced again by small absolute yet large relative flux variations originating from simulated self-scattered thermal dust emission. This indicates a lack of discernible improvements in the similarity of results when using a higher resolution that exceeds any differences induced by the accompanied changes in the density distribution. In the submm range, however, we find large differences between both resolutions. Since observations in this wavelength range trace the midplane temperature of the PPD (compare with discussion of Fig. \ref{fig:flux_reldiff_N1}), the quantity $\Delta S$ can also be understood as a measure of the change of similarity between the temperature distributions $\temphd$ and $\temprt$ induced by a change of resolution. Overall, both plots suggest that the similarity remains unaltered inside the inner region of the background model (within $4\,$au), it increases in the innermost region of the high-resolution model as well as in the circumplanetary region of the $300\,{\rm M}_\oplus$ planet, and it mostly drops elsewhere. For the $300\,{\rm M}_\oplus$ model, in particular, the changes of similarity affect the following regions: the innermost part of the high-resolution region (positive, $0{<}\Delta S{\lesssim}0.3$), the region further out but within the planetary orbit (negative, $0{>}\Delta S{\gtrsim -}0.3$), the gap region (negative, ${-}0.15{<}\Delta S{<-}0.05$), the circumplanetary region (positive, $0.1{<}\Delta S{<}0.2$), and the region just outside the gap (negative, $\Delta S{<-}0.2$). Analogously, for the $10\,{\rm M}_\oplus$ model, the changes affect the following regions: the innermost part of the high-resolution region on the planetary side (positive, $0{<}\Delta S{\lesssim}0.3$), the region further out in the high-resolution region, and the region within the depicted range (negative, $\Delta S{\lesssim-}0.1$). Based on these findings, we conclude that increasing the resolution from \texttt{N1} to \texttt{N3} does not result in significant improvements in similarity for simulations of the $10\,{\rm M}\oplus$ planet. Apart from any change in similarity, though, the qualitative appearance of the systems does change at a higher resolution. Our analyses also suggest that an improvement of the numerical methods, by which the temperature distributions are determined, is crucial for increasing the similarity, as these systematic differences simply cannot be overcome by a higher resolution alone. These results align with our findings for the $300\,{\rm M}_\oplus$ planet. However, in the circumplanetary region, the increase in resolution does lead to greater agreement between the RHD and MCRT temperature distributions in this scenario. 

\subsection{Discussion: Discrepancies between RHD and MCRT}\label{sec:discussion_differences}

In optically thick disk regions up to a few tens of au, the midplane temperature in the RHD simulations is consistently higher than in the MCRT simulations, as shown in Figs. \ref{fig:temp} and \ref{fig:temp_profiles}. This is a consequence of the angle-averaged treatment of radiation in the RHD simulation (Eqs. \eqref{Eq:Frad}-\eqref{Eq:Erad}), which introduces an unphysical interaction between crossing radiation fluxes. In our hydrodynamical disk models, the midplane-directed radiation flux emitted at the $\tau=1$ irradiation surface encounters the flux escaping the disk in the vertical direction. As a result, the cooling produced by the outward diffusion of photons is less efficient in the FLD than in the MCRT simulations, resulting in higher temperatures in the former both in the disk and at the center of the planetary envelope. This also explains the relatively lower hydrodynamical temperatures in the immediate upper layers: the outward radiation flux is lower in the FLD than in MCRT, leading to a lower radiation energy density available to heat those layers. Moreover, this artificial interaction phenomenon explains the good agreement close to the midplane when stellar irradiation is neglected (Fig. \ref{fig:temp_statistics_axisym_models}), since in that case there are no vertically crossing fluxes and frequency-averaged vertical diffusion becomes a good approximation for the radiative flux.

Above the $\tau=1$ irradiation layer, there is only outward transport of radiation, so there is no obstruction of fluxes. Furthermore, the opacity in those regions is low enough that the temperature is only marginally dependent on the diffuse radiation field. The temperature is instead mostly determined by the balance of optically thin cooling and absorption of stellar irradiation, which is similar in both treatments, yielding matching temperatures. We also obtained good agreement in the densest disk regions where viscous heating determines the gradient of the diffuse field, resulting in good accuracy of the radiative diffusion with Rosseland-averaged opacities.

One way to reduce the described discrepancies is by treating radiative transfer in the FLD simulations via multigroup methods, which use different radiation fields to describe photons in different frequency bands \citep[e.g.,][]{Robinson2024}. This way, the long-wavelength outward flux and the short-wavelength inward flux can be treated in different frequency groups, thus preventing them from interacting with each other. We verified that this reduces the temperature discrepancy with MCRT in a multigroup version of the M1 radiative transfer code by \cite{MelonFuksman2021} \citep[article in preparation; see also][]{Robinson2024}. This approach also reduces temperature discrepancies in optically thin outer regions, where the use of frequency-averaged opacities leads to lower temperatures than MCRT, as shown in Fig. \ref{fig:temp_profiles}. However, multigroup methods do not entirely solve the temperature discrepancies, as there is still a significant frequency overlap of the fluxes entering and leaving the disk in the intermediate-wavelength range. Because of this, angular discretization of the radiation field \citep[e.g.,][]{Davis2012,Jiang2021} is also needed to improve the agreement between RHD and MCRT models.

 Further differences between the FLD and MCRT result from the fact that the angle-averaged transport in FLD is unable to describe processes such as anisotropic photon scattering. This becomes particularly noticeable when stellar irradiation is switched off (Fig. \ref{fig:app:temp_axisym_DV}), in which case the scattering of photons emitted at the disk heats up the upper layers. Additionally, the boundary condition for FLD fixing $E_R=a T_\mathrm{min}^4$ with $T_\mathrm{min}=10$ K results in a slight temperature decrease close to the outer radial boundary (Fig. \ref{fig:temp_statistics_axisym_models}). Another relevant approximation is the use of Planck-averaged opacities for stellar irradiation (Eq. \eqref{Eq:Firrad}). This contributes to the sharper temperature transition obtained with FLD compared to MCRT between the midplane and the irradiated layers, as in FLD there is a single $\tau=1$ surface for stellar photons determined by $\kappa_{\rm P}(T_*)$, whereas in MCRT there are as many $\tau=1$ surfaces determined by $\kappa_\lambda$ as there are considered stellar photon frequencies.

Last but not least, some of the obtained discrepancies result from the inability of the MCRT approach to consider velocity-dependent phenomena since this method only takes as input a stationary density distribution. This means that temperature changes due to internal energy advection and PdV work -- that is the terms $\nabla\cdot(e \mathbf{u})$ and $-p\nabla\cdot\mathbf{u}$ in Eq. \eqref{Eq:internal_energy}, respectively -- are ignored. Unlike the viscous heating term, these are not included in MCRT as effective heat sources, as they can be either positive or negative. This limitation becomes particularly significant when these processes are energetically relevant, for instance, in the case of gas compression in spiral wakes. In our simulations, compression heating explains the larger millimeter flux obtained with RHD at the spiral wakes for a planet of $300\,{\rm M}_\oplus$ (Fig. \ref{fig:flux_reldiff_N3}). Conversely, the same figure shows that this flux excess is not significant for the weaker spirals launched by a $10\,{\rm M}_\oplus$ planet. These results show that caution should be practiced before considering MCRT temperatures as "true" values in cases where the mentioned phenomena (or, e.g., time-dependent processes) are significant. To improve the accuracy of MCRT simulations in these cases, a method capable of handling negative energy source terms would be needed.

\section{Conclusions and summary}
\label{sec:summary}
    The goal of this study is to assess the similarity between temperature distributions derived from RHD and MCRT simulations in PPDs that harbor young accreting planets in their late stages of formation. Through comparative analyses at different resolutions (\texttt{N1}, \texttt{N2}, and \texttt{N3}) and accretion parameters (\texttt{A4}, \texttt{A32}, and \texttt{A64}) and for varying planetary masses ($10$ and $300\,{\rm M}_\oplus$), we investigated the impact of these parameters on the resulting temperature distributions in different regions of the PPD and quantified their level of agreement. To that end, we combined results of RHD simulations of an unperturbed protoplanetary disk of $300\,$au in radius (i.e., the background model) with high-resolution simulations that cover a radial domain of $4$ to $25\,$au, including an accreting planet embedded in the disk midplane at a distance of $10\,$au from the central star (i.e., the high-resolution model; see Fig. \ref{fig:model_sketch} for a model sketch as well as Sect. \ref{sec:model_embedement_and_parameters}). Subsequently, we used MCRT simulations to post-process the results of our RHD simulations, using their derived density distributions and self-consistently calculating temperature distributions while also taking into account the full wavelength dependence of the simulated dust in the PPD (see Sect. \ref{sec:rt_simulations}).

    In Sect. \ref{sec:temp_comparison_reference_model}, we describe the impact of the considered numerical and physical parameters on the 3D temperature distribution. Based on that analysis, we assessed the level of agreement between the performed RHD and MCRT simulations, with a particular focus on the high-resolution region and the vicinity of the planet. Our key findings can be summarized as follows:
    \begin{itemize}
        \item The level of agreement varies across different regions of the PPD and is linked to the dominant sources contributing to the local radiation field.
        \item Closer agreement can be found in the optically thickest regions of the PPD, where viscous heating dominates, as well as in the optically thin regions in the upper layers of the PPD, where stellar irradiation dominates. Increased discrepancy in the derived temperatures can be found in a thin layer close to the photosphere at the transition from optically thick to optically thin regions, for example between the circumplanetary disk and the gap, and between the denser PPD layers and higher layers as well as in the far-out regions of the PPD, which are shielded from unscattered stellar irradiation. 
        \item Derived RHD temperatures inside the PPD and particularly the gap region are significantly smoother compared to corresponding MCRT temperatures, while the transition from the dense PPD region to its optically thin upper layers is much sharper.
        \item For all cells in the high-resolution region, the temperature estimates agree across all tested resolutions and accretion parameters by $\lvert \delta_T\rvert \lesssim 0.1-0.13$.
        \item The planetary Hill region in the RHD simulations of the $300\,{\rm M}_\oplus$ ($10\,{\rm M}_\oplus$) planet is  colder than in equivalent MCRT simulations, by $\lvert\langle\delta_T\rangle\rvert
\approx 0.01-0.1$ ($0.02-0.05$)  on average. In this region, the similarity slightly improves with increasing resolution or a decreasing accretion parameter. 
        \item The derived temperature differences are largely systematic, resulting from underlying discrepancies in the capability of RHD and MCRT simulations to properly incorporate all relevant physical processes. 
    \end{itemize}
    To investigate the origin of these discrepancies and improve on the existing methods, we performed a series of HD-FLD simulations of unperturbed PPDs, which included selected physical processes and modifications to the numerical prescription for evaporation and the treatment of low-density regions. The results of these simulations were post-processed using MCRT simulations and the arising temperature differences were analyzed (see Sect. \ref{sec:axisym}) to find the limitations of current simulations (see Sect. \ref{sec:discussion_differences}). The main results of this analysis can be summarized as follows:
    \begin{itemize}
        \item The two simulation
types show excellent agreement overall with regard to the optically thick regions where viscous heating dominates the radiation field. However, in the hydrodynamical simulations, higher layers of the disk exhibit missing contributions of scattered thermal emission that would originate from the hot inner regions of the disk.
        \item The optically thin regions of the disk at $r{\gtrsim}0.1\,$au, where unscattered stellar irradiation dominates the radiation field, show excellent agreement. 
        \item Using an optical depth limiter for the Rosseland opacities leads to overestimation of the gas temperature around the photosphere.
        \item The missing contributions from scattered and re-emitted stellar irradiation lead to insufficient heating of the shadowed regions of the disk.
        \item Differences between MCRT and FLD
        can be largely attributed to both the angle- and frequency-averaging of the radiative intensity in FLD, which introduces an artificial interaction between radiation transported from and into the disk. This effect is subdominant in optically thick regions with significant viscous heating, where diffusion becomes a good approximation, and in upper irradiation-dominated layers where the temperature depends solely on the absorption of stellar photons and optically thin cooling, which are mostly unaffected by the diffuse radiation field. This explains the good agreement in both regions and the increased discrepancy in optically thick regions when viscous heating is switched off.
        \item Further reasons for the discrepancies introduced by the FLD method are the employment of frequency-averaged opacities for both stellar irradiation and reprocessed photons as well as the neglect of photon scattering.
        \item Unlike FLD, our MCRT simulations are not equipped to handle negative energy sources, such as the local PdV work and the transport of internal energy. This results in discrepancies in regions where gas compression is relevant, such as spiral wakes.
    \end{itemize}
    In Sect. \ref{sec:flux_section}, we assess the impact of the previously described temperature differences on ideal flux maps of the planet-harboring models in the VIS, NIR, and submm wavelength ranges and contrast this with the quantitative changes resulting from alteration of the model resolution. The findings of this analysis can be summarized as follows:
    \begin{itemize}
        \item The effect of altering the temperature distributions on observations in the VIS wavelength range is negligible. In the NIR, their alteration leads to differences in the derived flux values from the innermost region in the vicinity of the star. In the submm range, the discrepancies in the temperature distributions affect all regions. 
        \item Overall, the temperature differences translate to flux differences affecting areas near the star, the inner PPD region, the gap region, the circumplanetary region, and even regions radially outward from the gap. Increasing the model resolution changes the qualitative appearance of the PPD notably.
        \item An analysis of the quantity $\Delta S$, which measures the change in similarity of flux estimates, shows no discernible clearly positive changes of similarity associated with a higher resolution,  either in the VIS or in the
NIR. In the submm range, we find positive changes: first for the $300\,{\rm M}_\oplus$ planet simulations in the innermost part of the high-resolution region ($0{<}\Delta S{\lesssim}0.3$) and in the circumplanetary region ($0.1{<}\Delta S{<}0.2$) and, second, for the $10\,{\rm M}_\oplus$ planet simulations in the innermost part of the high-resolution region on the planetary side ($0{<}\Delta S{\lesssim}0.3$). Elsewhere, the change in similarity exhibits negative values, corresponding to a reduction of the similarity between the results of RHD and MCRT simulations induced by an increase in resolution from \texttt{N1} to \texttt{N3}. 
    \end{itemize}

\subsection{Lessons learned}
As long as isothermal or fixed temperature hydro simulations are used as a basis for MCRT post-processing for planet--disk interaction studies, the challenges and problems described in this paper can be avoided. Nevertheless, in this case, important effects that arise from accretion heating around the planet \citep{Garate2021}, spiral wave patterns, and the intricate structure of the circumplanetary disk, or envelope \citep{Szulgyi2016}, will inevitably be missed. Therefore, all the discrepancies laid out in this paper still do not provide sufficient justification to go back to fixed-temperature structures but will hopefully drive technological breakthroughs in radiation hydro simulations \citep{MelonFuksman2021,Muley2024}.
Our overall conclusions are as follows:
    \begin{itemize}
        \item One can never expect the same temperatures from both RHD and MCRT post-processing if adiabatic heating, adiabatic cooling, and advection of thermal energy are not considered in the MCRT simulations and if significant angle- and frequency-averaging is applied in the RHD simulations.
        \item In hydrostatic cases, MCRT is better, as FLD lacks the cooling of the midplane due to the escape of longer-wavelength radiation and does not satisfactorily include the interaction between the flux entering and leaving the disk.
        \item Nevertheless, by choosing the same evaporation prescription in MCRT and RHD, one can get closer to convergence.
        \item Frequency-dependent irradiation leads to only partly better results than Planck irradiation for the heating of the surface layers.
        \item Increasing the model resolution affects the observable planet-induced structures (the gap, the spirals, and the circumplanetary disk)
    up to the highest-tested resolution. More robust conclusions as to the required resolution will be obtained by reconstructing the planet accretion rate and luminosity based on derived flux maps.
        \item Putting temperature distribution discrepancies aside, applying a dedicated MCRT post-processing step for the flux determination remains indispensable.
    \end{itemize}

\subsection{Outlook} 
Our comparison of the temperatures as predicted in a full radiation hydro simulation and MCRT post-processing reveals several discrepancies. Some of the discrepancies are unavoidable, as on one hand a non-iterative MCRT scheme struggles with temperature-dependent opacities and more generally with the dynamical heating and cooling events in the hydro simulations, especially density waves and the fast flows around the planet. On the other hand, FLD is a rather crude approximation, and it leads to correct predictions only for very high and very low optical depths. The transition regions between those regimes are poorly represented and demand more focus in the future.

Nevertheless, we have identified several points we can improve on for the next steps in our ongoing collaboration of radiation hydro simulations and MCRT post-processing in order to derive observables. 
   \begin{itemize}
        \item In the future, we shall use consistent Planck opacities for the dust-evaporation regions for the MCRT versus\ FLD simulations.
        \item We shall develop and implement an MCRT method capable of handling all negative source terms.
        \item Until we can provide realistic values for the gas opacities, we have to make sure that the Planck opacities (for irradiation and radiation balance of dust grains in the FLD simulations) and the frequency-dependent opacities in the MCRT runs use the same evaporation behavior.
        \item The Gaussian transition from dusty opacities to the \citet{Bell1994} opacities is not smooth enough toward extremely low values, as required for consistency with the MCRT simulations. Therefore, we used a sigmoid interpolation with a width of $100$ K around the evaporation temperature, which has also shown much faster convergence for the iterative solver of the FLD system.
        \item We will also use wider radial damping layers for the higher-resolution simulations in order to be able to better compare cases with different resolutions.
        \item Frequency-dependent irradiation does not make a large difference in the temperature structure for the FLD cases, yet this may be different for the disk with a lower optical depth that we have planned for our future scan of the parameter space of disk masses, planet masses, and their locations.
        \item Multi-group methods can potentially improve the temperature structure around the planet \citep{Robinson2024}, allowing us to have incoming and outgoing radiative fluxes at different wavelengths simultaneously. 
        However, whether such a precise yet numerically very expensive scheme will allow high-resolution 3D simulations of planets interacting with their ambient disk is open for exploration.
    \end{itemize}
In a companion paper (Klahr et al., in prep), we explore the effect of numerical parameters, such as resolution and modeling of the gas accretion onto the planet, as well as physical parameters, such as $\alpha$ viscosity and the magnitude of opacity, on the gas-accretion rate and on the amount of gas accumulating in the Hill sphere. In conjunction with the present paper, this will build the foundation for a parameter exploration of a range of planet masses and locations in combination with various disk masses and the prospects of detecting the planet as well as its potential gas accretion.

\begin{acknowledgements}
      We acknowledge the support of the DFG priority program SPP 1992 "Exploring the Diversity of Extrasolar Planets" under contracts WO 857/17-2 and KL 1469/16-1/2. Simulations leading to the presented results were performed on the ISAAC and VERA clusters of the MPIA  hosted at the Max-Planck Computing and Data Facility in Garching (Germany). H.K. also acknowledges support from the DFG via the Heidelberg Cluster of Excellence STRUCTURES in the framework of Germany's Excellence Strategy (grant EXC-2181/1 - 390900948).
\end{acknowledgements}

\bibliography{literature}
\bibliographystyle{aa}

\begin{appendix}
\section{Optical depth map}
\label{sec:app:opt_depth_kappa_rosseland}
\begin{figure}[htp]
   \centering
   \includegraphics[width=\hsize]{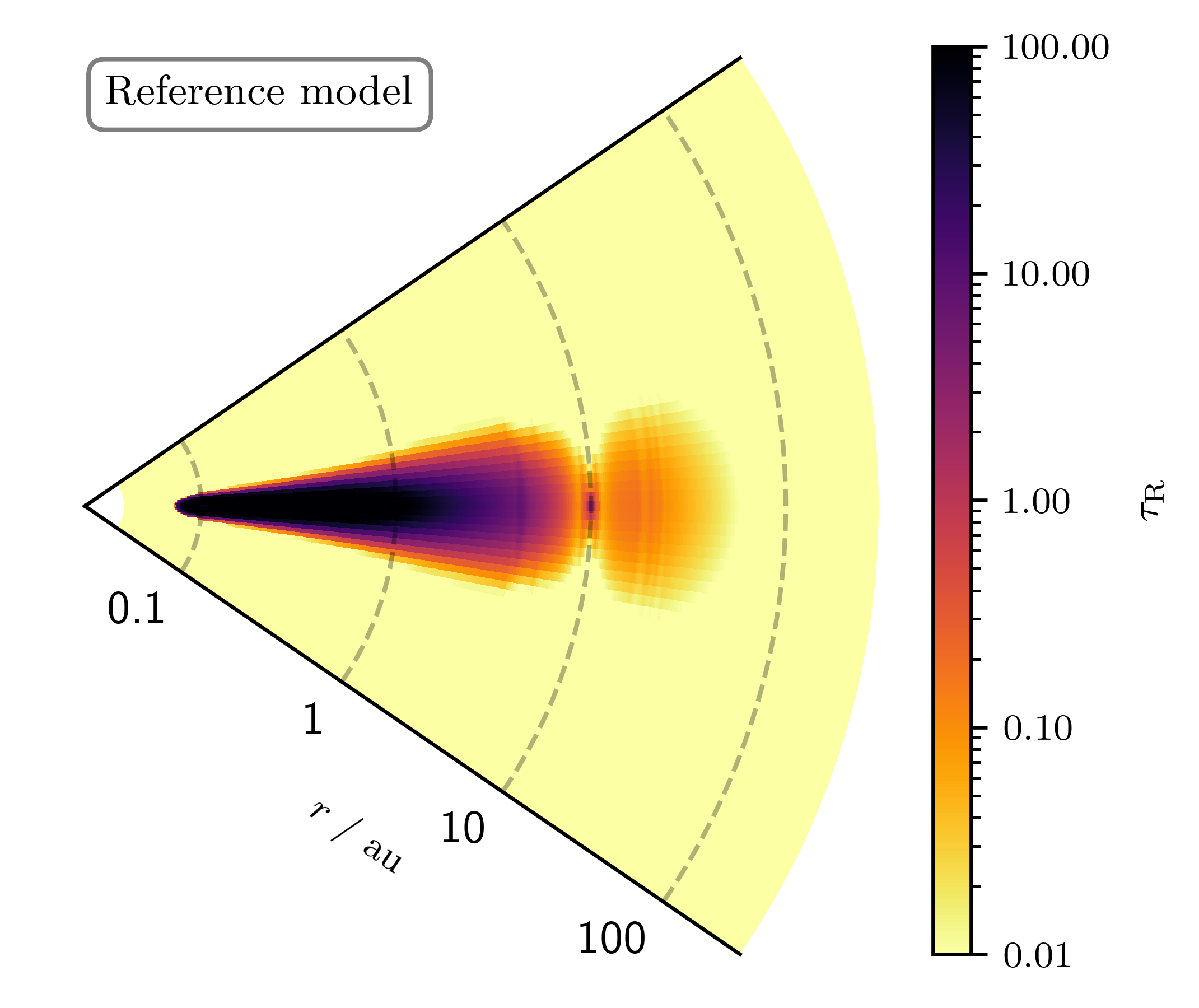}
      \caption{Optical depth map for a vertical cut through the midplane of the reference model. (For details, see Sect. \ref{sec:app:opt_depth_kappa_rosseland}.)}
         \label{fig:app:opt_depth_kappa_rosseland}
   \end{figure}

Figure \ref{fig:app:opt_depth_kappa_rosseland} shows an optical depth map for a vertical cut through the midplane of the reference model. The displayed optical depth $\tau_{\rm R}$ is a calculated cell-wise as the product of the corresponding Rosseland opacity, the gas density, and an approximate size of the cell. For the purpose of improved clarity, the color bar is clipped.

\section{Temperature distribution: \texttt{D+V}}
\label{sec:app:temp_dist_DV}
\begin{figure}[htp]
   \centering
   \includegraphics[width=\hsize]{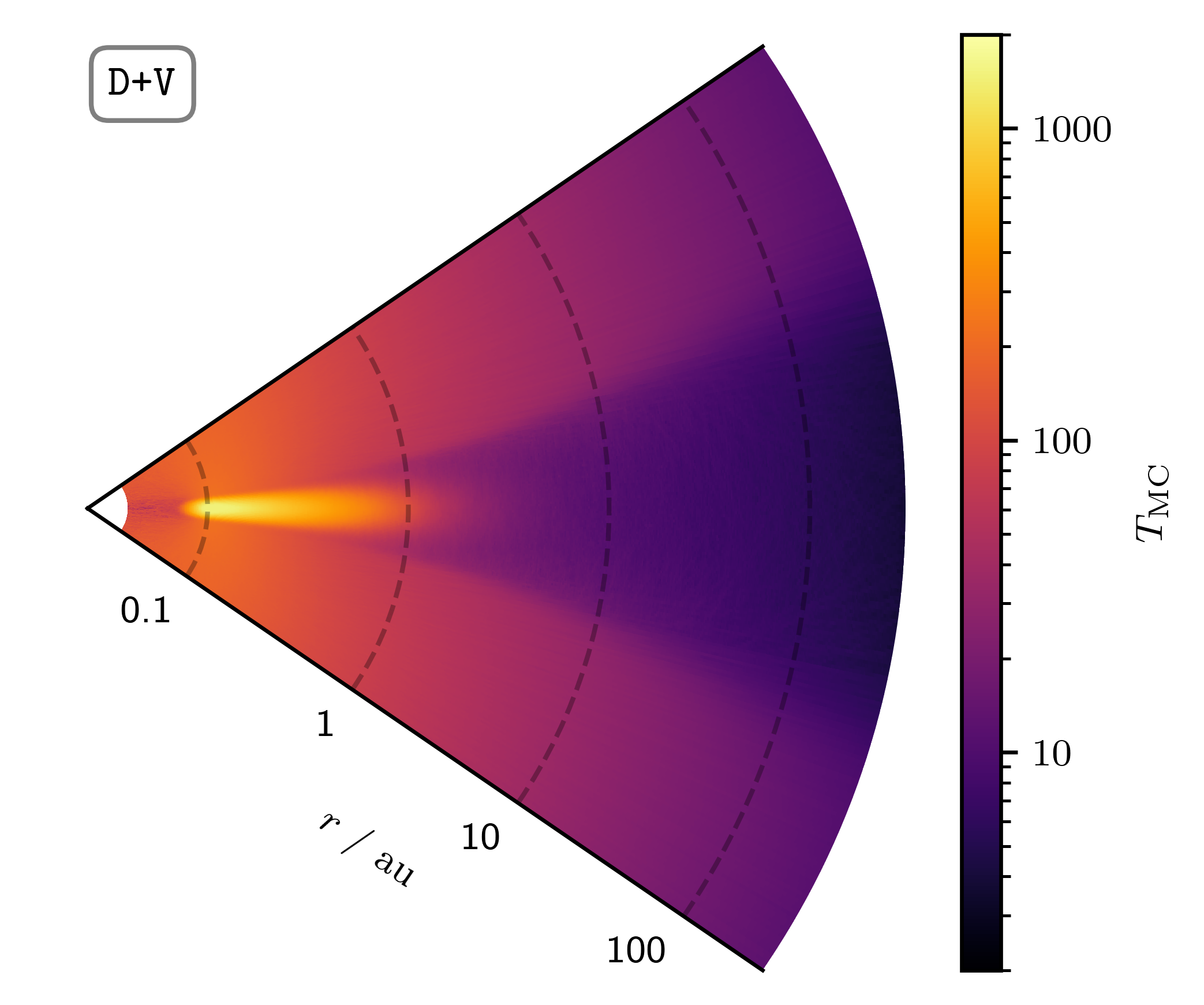}
      \caption{Vertical cut through MCRT temperature distribution of uPPD model \texttt{D+V}. (For details, see Appendix \ref{sec:app:temp_dist_DV}.)}
         \label{fig:app:temp_axisym_DV}
   \end{figure}

Figure \ref{fig:app:temp_axisym_DV} shows the MCRT temperature distribution for a vertical cut through the midplane of a uPPD. The underlying disk model, \texttt{D+V}, is discussed in Sect. \ref{sec:axisym}. For the purpose of improved clarity, the color bar is clipped.

\newpage

\section{Hill sphere - Temperature profiles}
\label{sec:app:profiles}
\begin{figure}[!htb]
   \centering
   \includegraphics[width=\hsize]{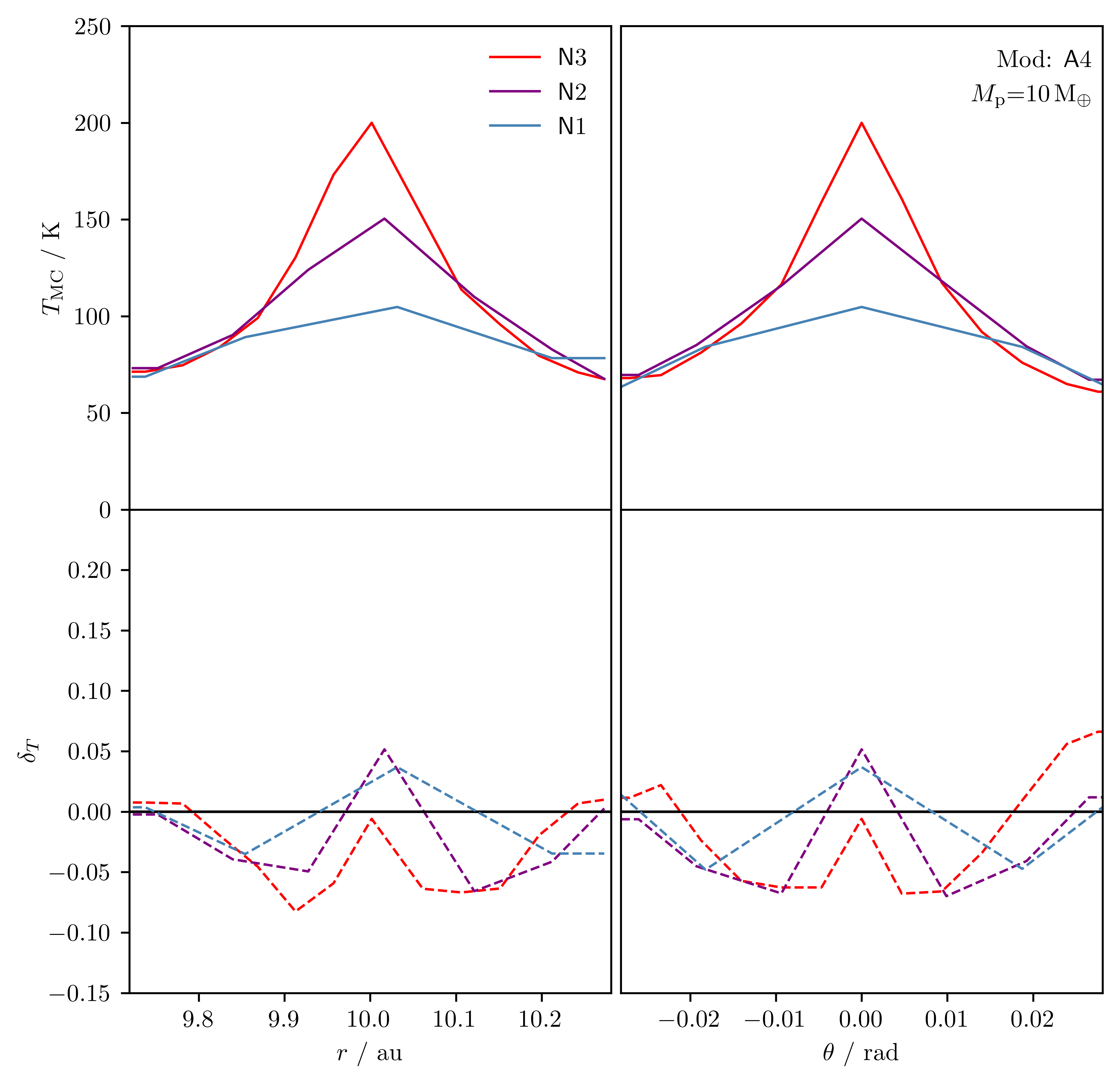}
      \caption{Comparison of temperature profiles. Assumed model parameters are specified in the upper right corner. (For details, see Appendix \ref{sec:app:profiles}.)}
         \label{fig:app:hill_sphere_temp_profiles_1}
   \end{figure}

\begin{figure}
   \centering
   \includegraphics[width=\hsize]{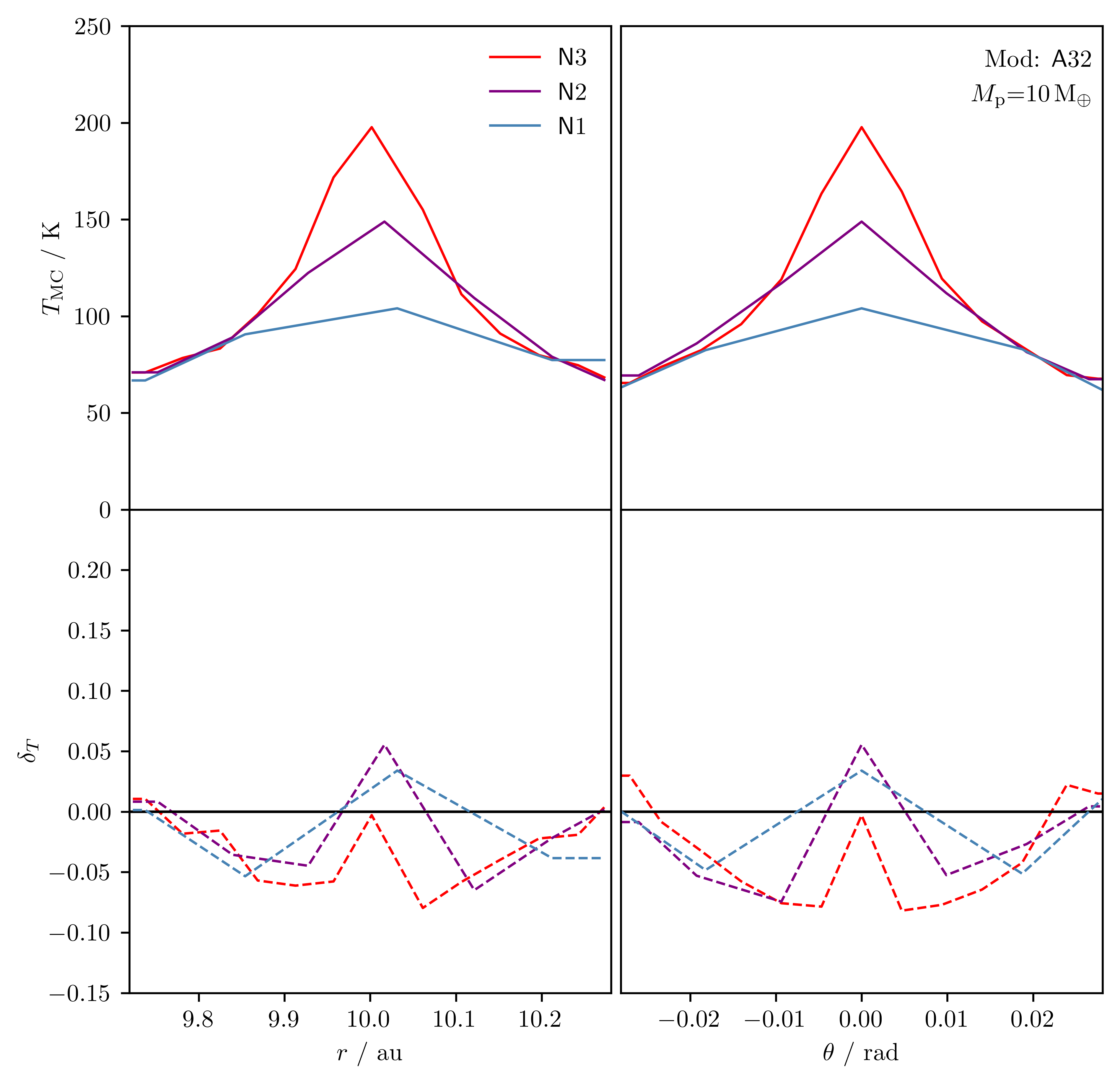}
      \caption{Comparison of temperature profiles. Assumed model parameters are specified in the upper right corner. (For details, see Appendix \ref{sec:app:profiles}.)}
         \label{fig:app:hill_sphere_temp_profiles_2}
   \end{figure}

\begin{figure}
   \centering
   \includegraphics[width=\hsize]{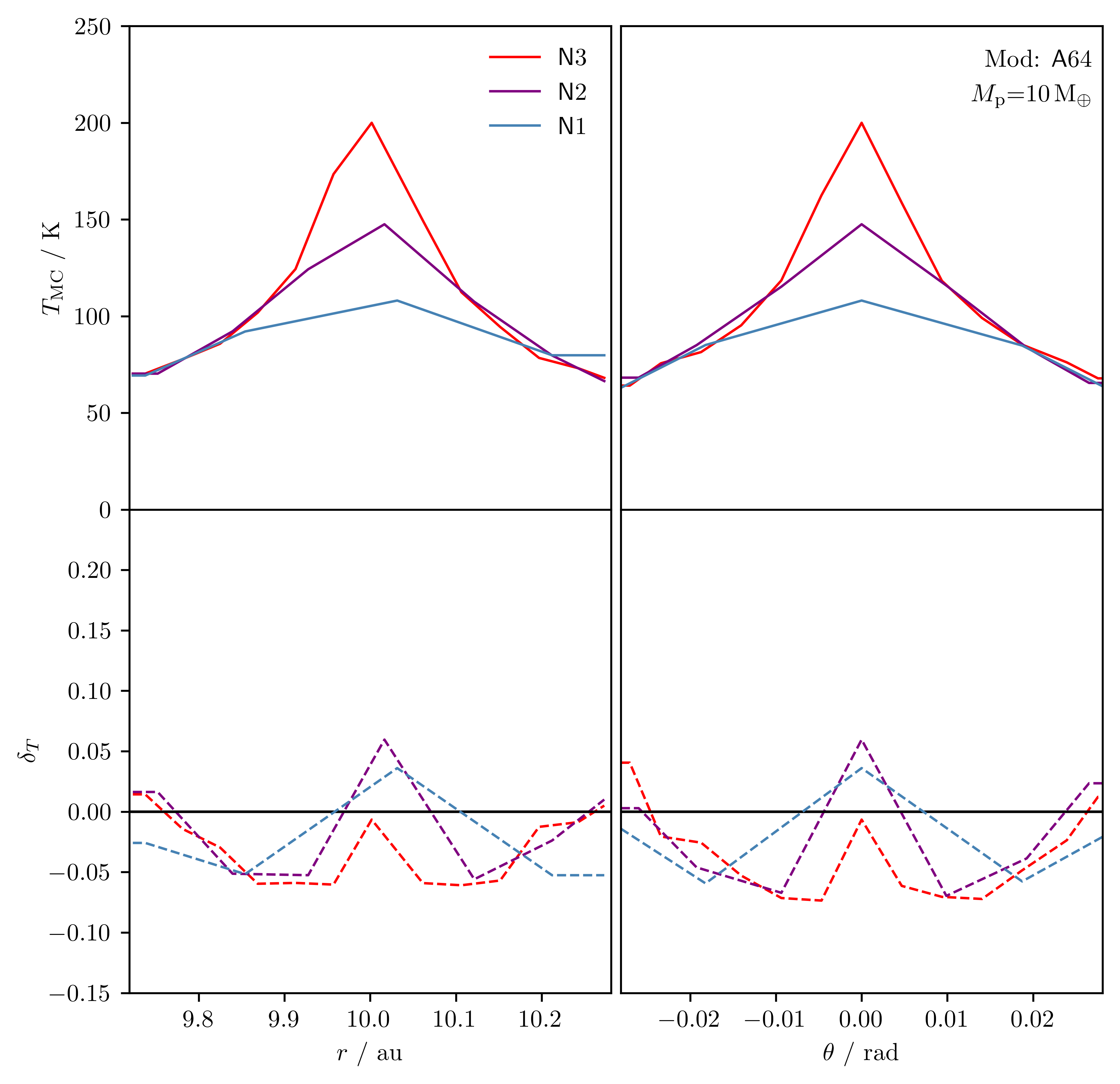}
      \caption{Comparison of temperature profiles. Assumed model parameters are specified in the upper right corner. (For details, see Appendix \ref{sec:app:profiles}.)}
         \label{fig:app:hill_sphere_temp_profiles_3}
   \end{figure}

\begin{figure}
   \centering
   \includegraphics[width=\hsize]{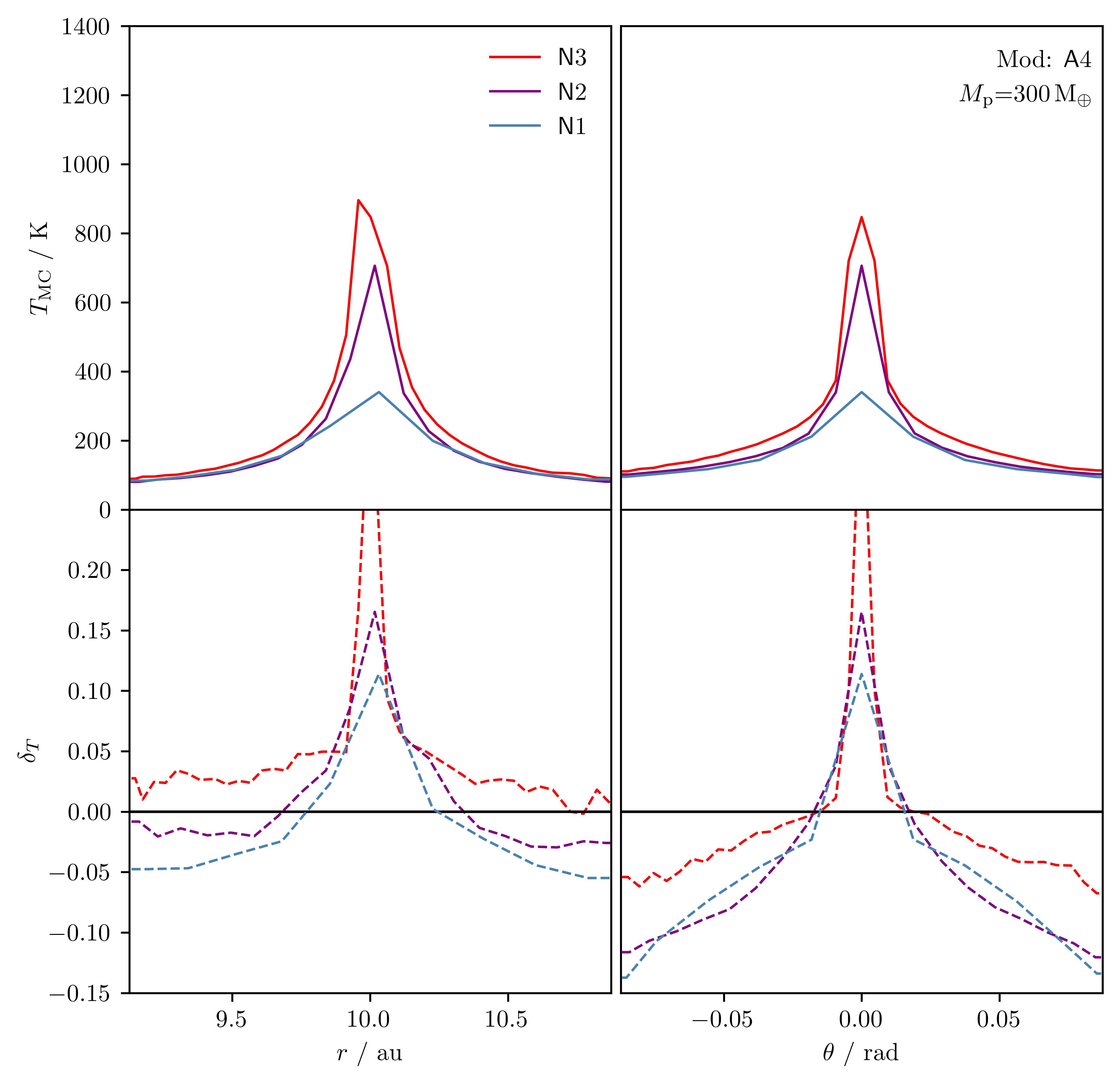}
      \caption{Comparison of temperature profiles. Assumed model parameters are specified in the upper right corner. (For details, see Appendix \ref{sec:app:profiles}.)}
         \label{fig:app:hill_sphere_temp_profiles_4}
   \end{figure}

\begin{figure}
   \centering
   \includegraphics[width=\hsize]{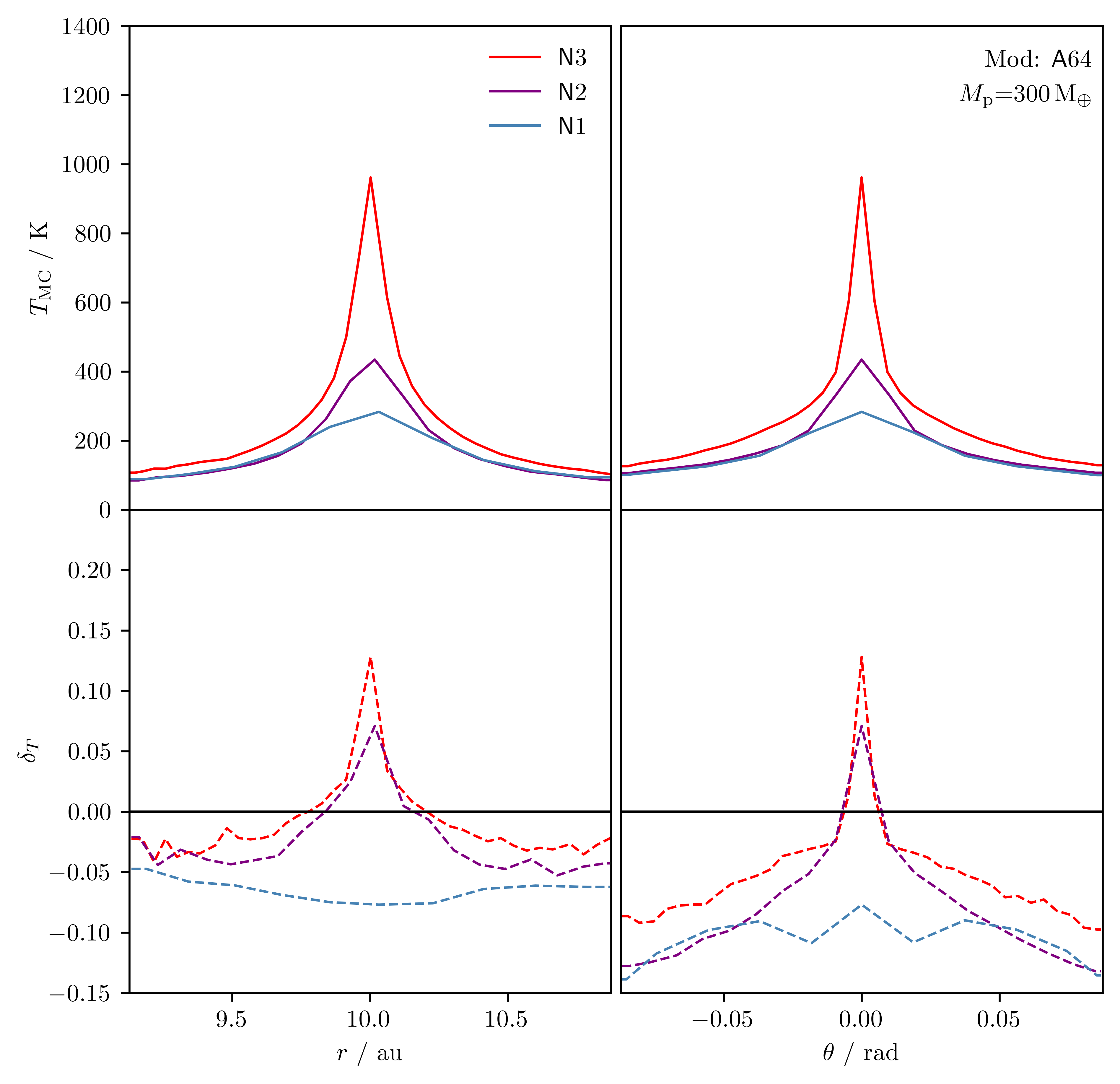}
      \caption{Comparison of temperature profiles. Assumed model parameters are specified in the upper right corner. (For details, see Appendix \ref{sec:app:profiles}).}
         \label{fig:app:hill_sphere_temp_profiles_5}
   \end{figure}

Figures \ref{fig:app:hill_sphere_temp_profiles_1} to \ref{fig:app:hill_sphere_temp_profiles_5} show comparisons of temperature profiles inside the planetary Hill sphere determined by using MCRT simulations (upper row) and their differences $\delta_T$ (lower row) relative to corresponding RHD temperature profiles. Each plot contains three color-coded curves that correspond to three different model resolutions (\texttt{N1}, \texttt{N2}, and \texttt{N3}). Each figure is similar to Fig. \ref{fig:hill_sphere_temp_profiles} and only differs with regard to its assumed model parameters. Left: Radial temperature profiles of the midplane in the azimuthal direction of the planet. Right: Polar temperature profiles of the planetary region. The planetary mass, $M_{\rm p}$, and the accretion parameter, \texttt{A}, are specified in the upper right corner of each figure. The displayed radial and polar range is defined by the size of the corresponding planetary Hill sphere. We note that for the purpose of improved clarity, the relative difference plots of Fig. \ref{fig:app:hill_sphere_temp_profiles_4} were clipped to values below $\delta_T = 0.25$. This, however, only affects the displayed value of a single grid cell, the planetary cell, which has a value of ${\approx}0.37$.

\FloatBarrier
\onecolumn
\section{Flux map comparison results}
\label{sec:app:flux_compare}
\begin{figure*}[!htb]
   \centering
   \includegraphics[width=0.9\hsize]{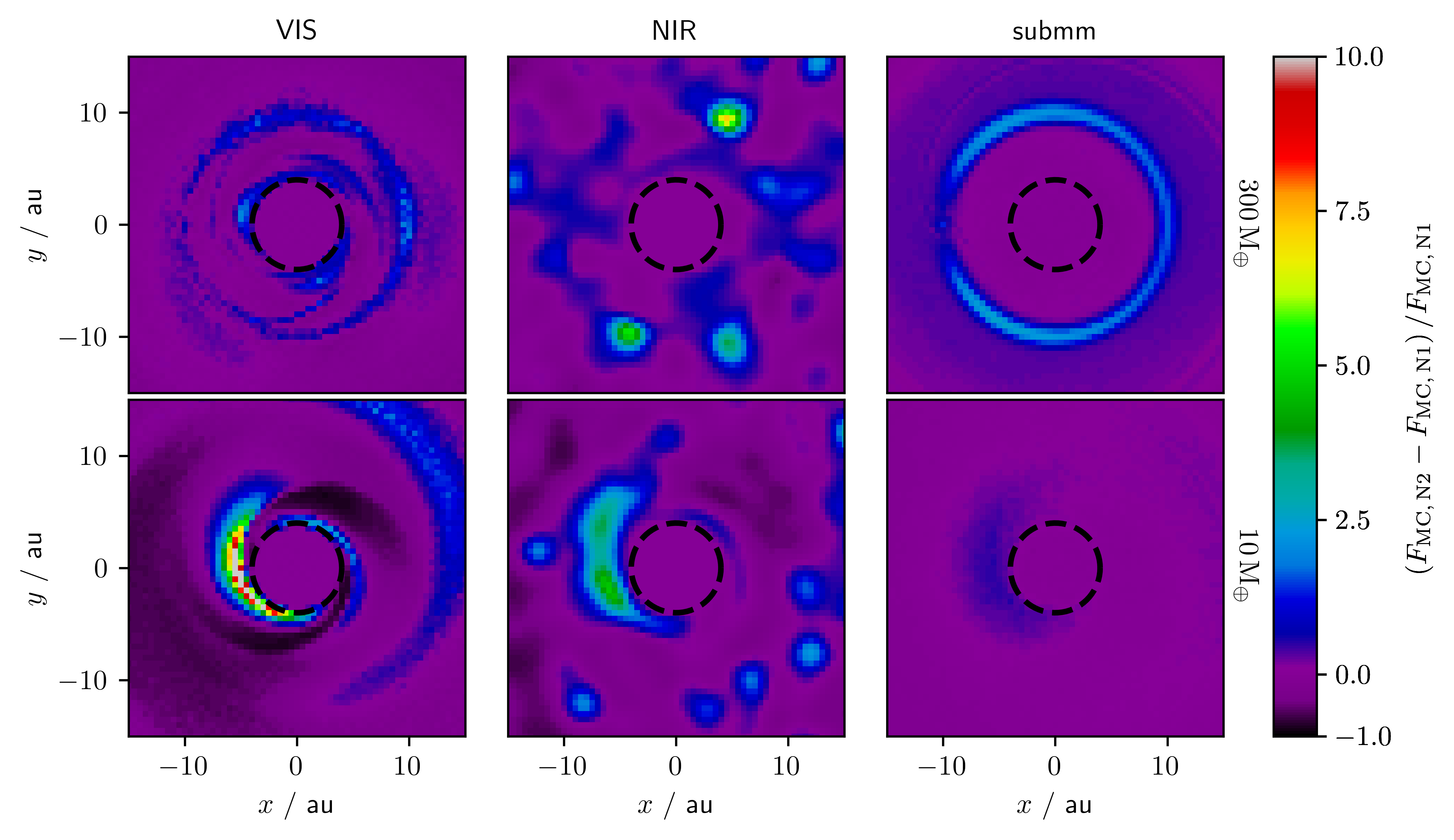}
      \caption{Similar to Fig. \ref{fig:flux_reldiff_resolution} but now for resolutions \texttt{N1} and \texttt{N2}.}
         \label{fig:app:flux_reldiff_resolution12}
   \end{figure*}

\begin{figure*}[!htb]
   \centering
   \includegraphics[width=0.9\hsize]{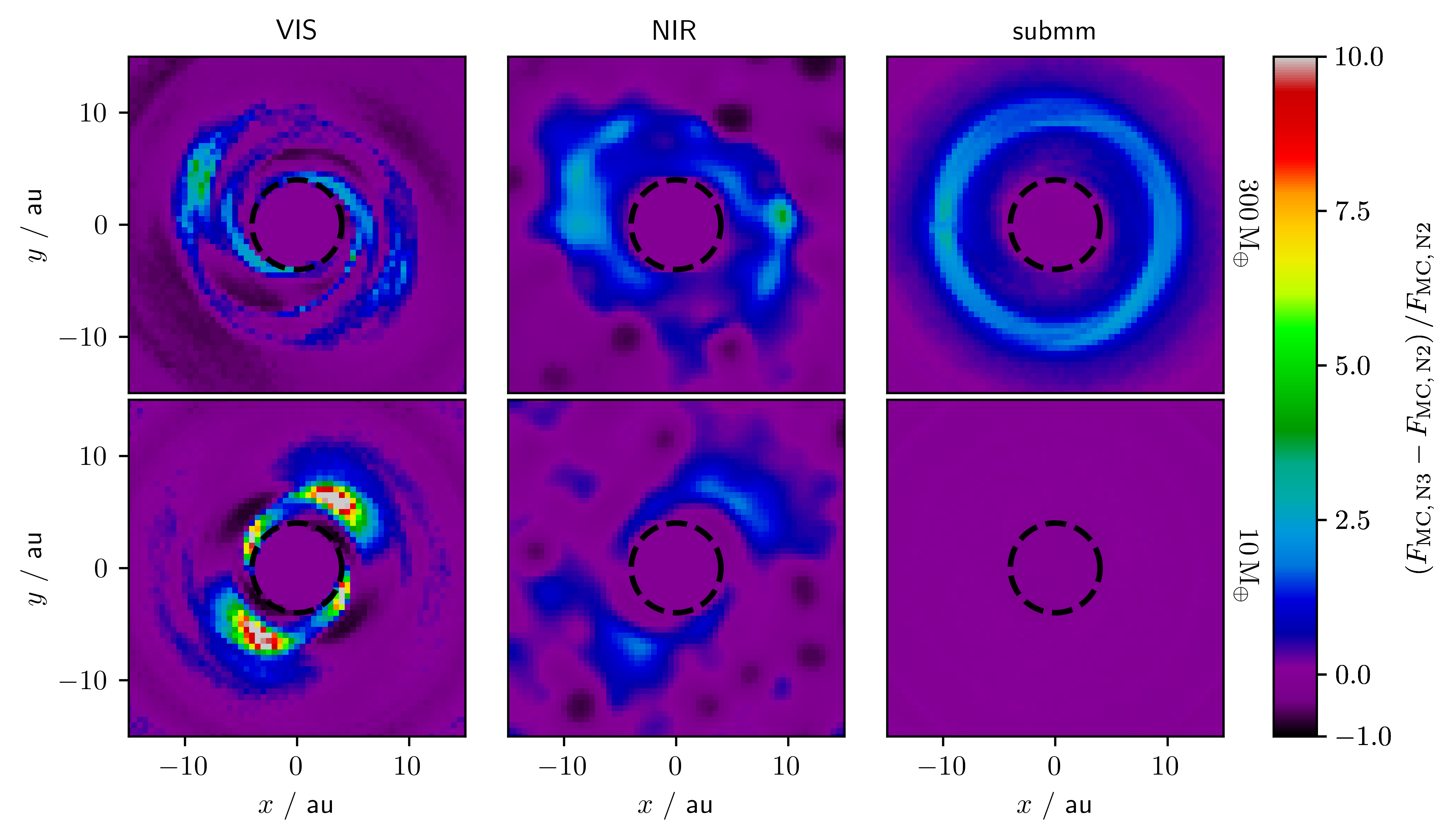}
      \caption{Similar to Fig. \ref{fig:flux_reldiff_resolution} but now for resolutions \texttt{N2} and \texttt{N3}.}
         \label{fig:app:flux_reldiff_resolution23}
   \end{figure*}

\begin{figure*}[b]
   \centering
   \includegraphics[width=0.9\hsize]{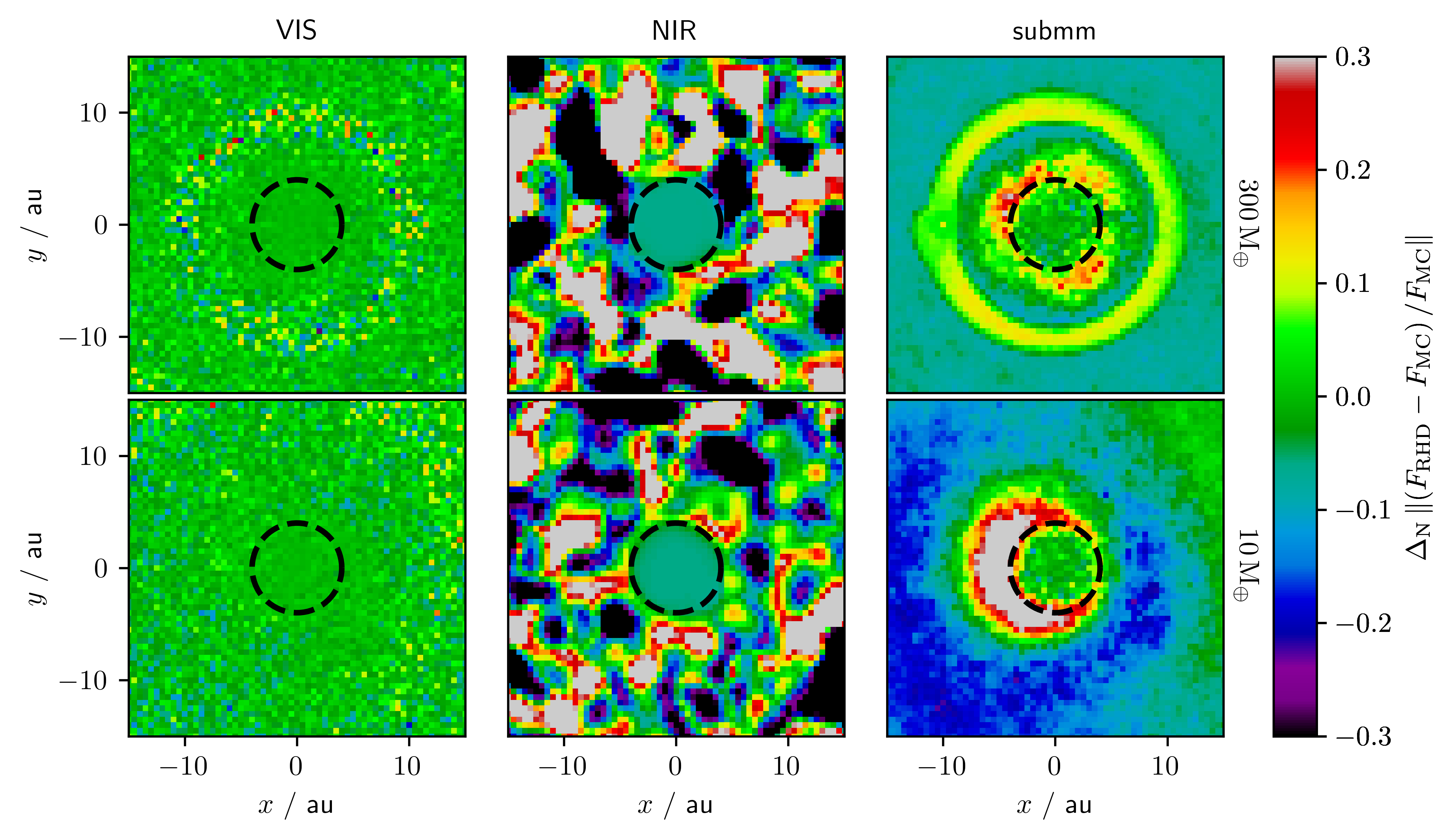}
      \caption{Similar to Fig. \ref{fig:flux_reldiff_similarity} but now for resolutions \texttt{N1} and \texttt{N2}.}
         \label{fig:app:flux_reldiff_similarity12}
   \end{figure*}

\begin{figure*}[b]
   \centering
   \includegraphics[width=0.9\hsize]{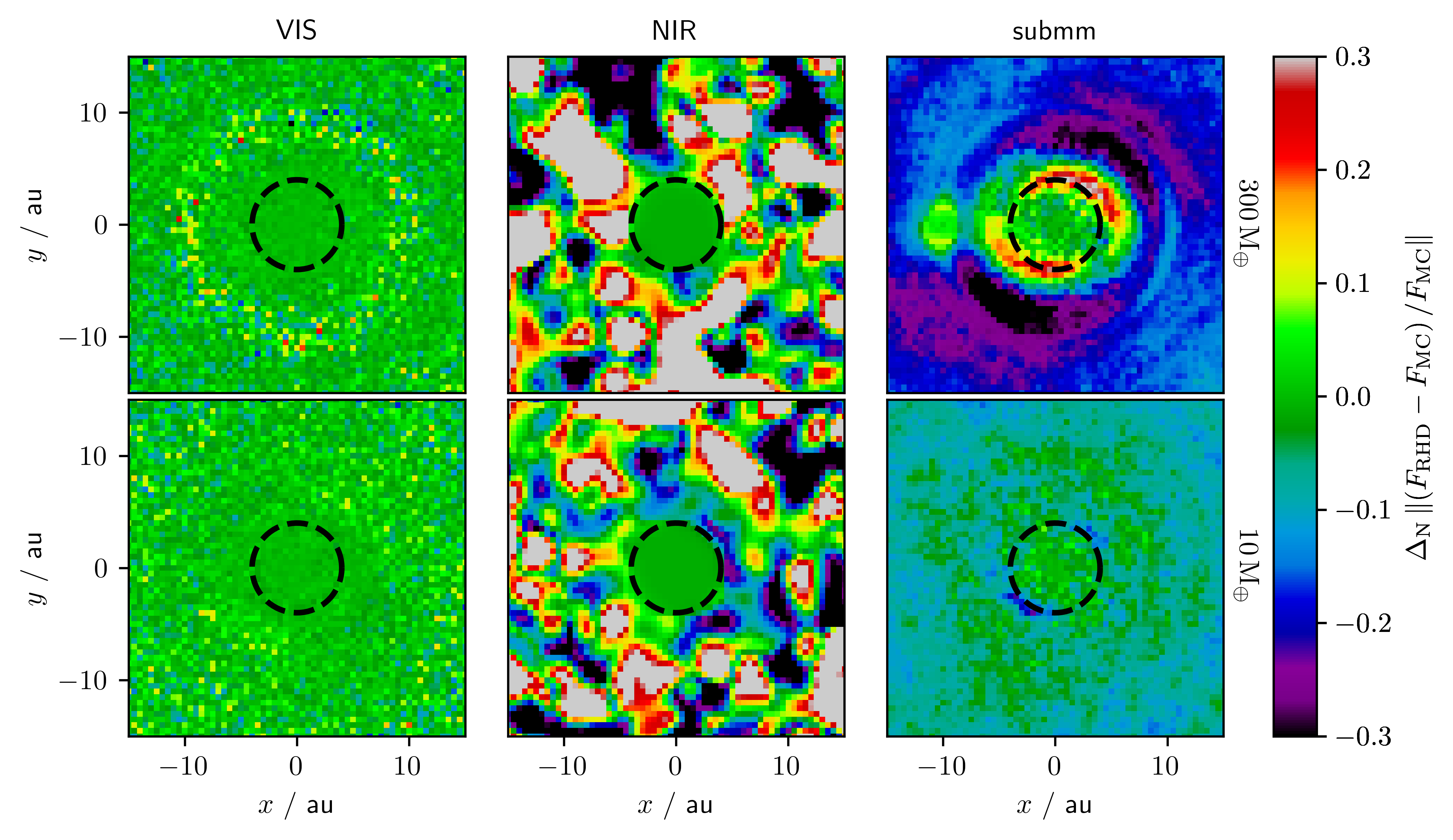}
      \caption{Similar to Fig. \ref{fig:flux_reldiff_similarity} but now for resolutions \texttt{N2} and \texttt{N3}.}
         \label{fig:app:flux_reldiff_similarity23}
   \end{figure*}

Figures \ref{fig:app:flux_reldiff_resolution12} and \ref{fig:app:flux_reldiff_resolution23} show relative flux difference maps based on models of differing resolutions for the case of a $300\,{\rm M}_\oplus$ (upper row) or $10\,{\rm M}_\oplus$ (lower row) planet. The assumed observing wavelengths are in the VIS (left column), NIR (central column), or submm (right column) wavelength range. 
Similar to the previous plots, Figs. \ref{fig:app:flux_reldiff_similarity12} and \ref{fig:app:flux_reldiff_similarity23} now show the similarity change $\Delta S{\equiv} \sum_{\rm N} \left\lVert \left(F_{\rm RHD}-F_{\rm MC}\right)/F_{\rm MC} \right\rVert$ based on flux maps derived with models of differing resolutions and simulation types (RHD and MCRT).
For the purpose of better comparability, the color bar of the plots have been clipped. 
For details, see Sect. \ref{sec:flux_map_comparison}.

\section{Additional figures}
\label{sec:app:additional_figures}
   \begin{figure*}[h!]
   \centering
   \includegraphics[width=0.9\hsize]{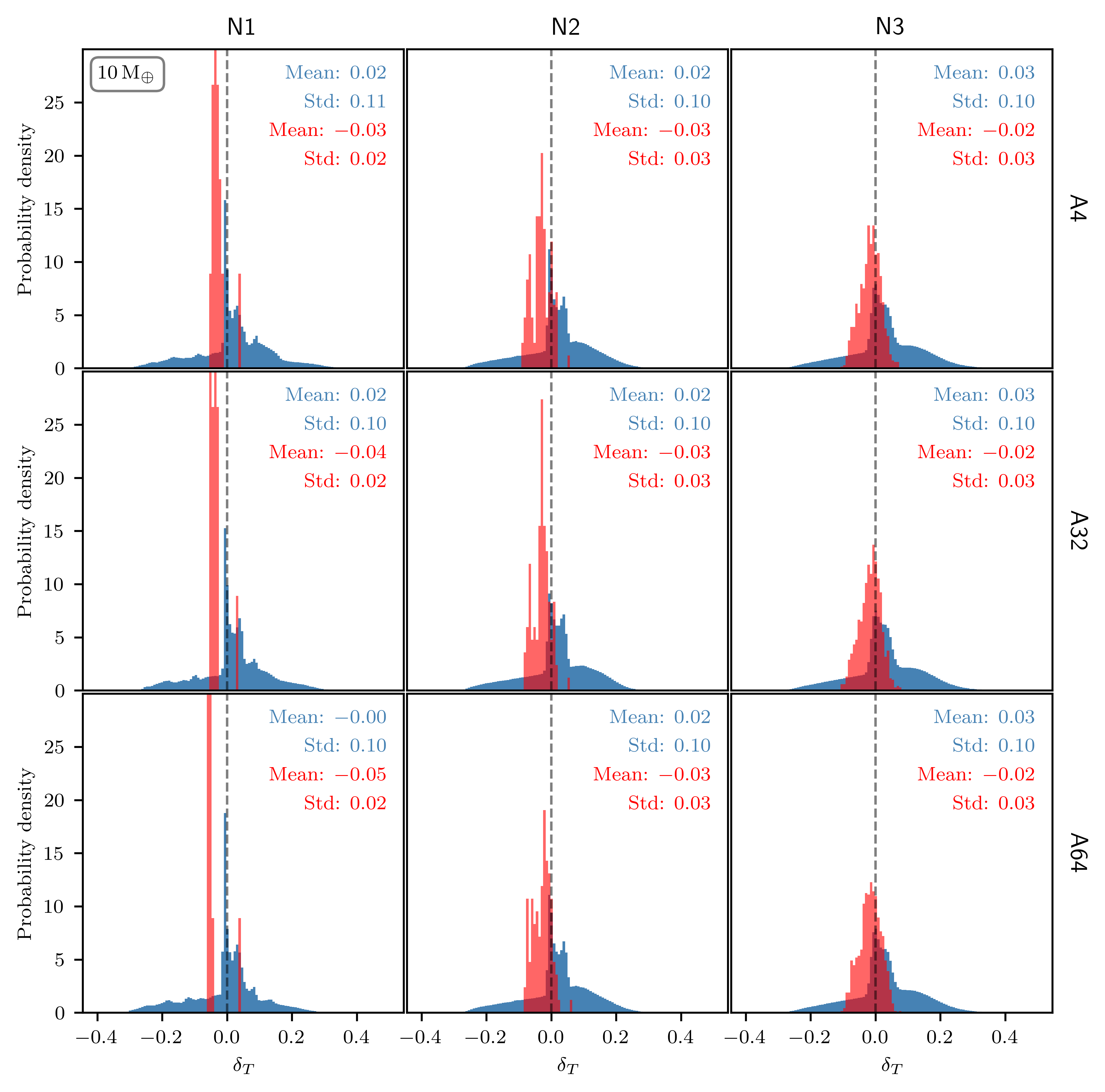}
      \caption{Similar to Fig. \ref{fig:histograms_P300}, but now presenting an overview of normalized histograms illustrating relative temperature differences for the $10\,{\rm M}_\oplus$ models. For the purpose of better comparability, histograms have been clipped at a probability density value of 30.}
         \label{fig:histograms_P10}
   \end{figure*}

\begin{figure*}[h!]
   \centering
   \includegraphics[width=\hsize]{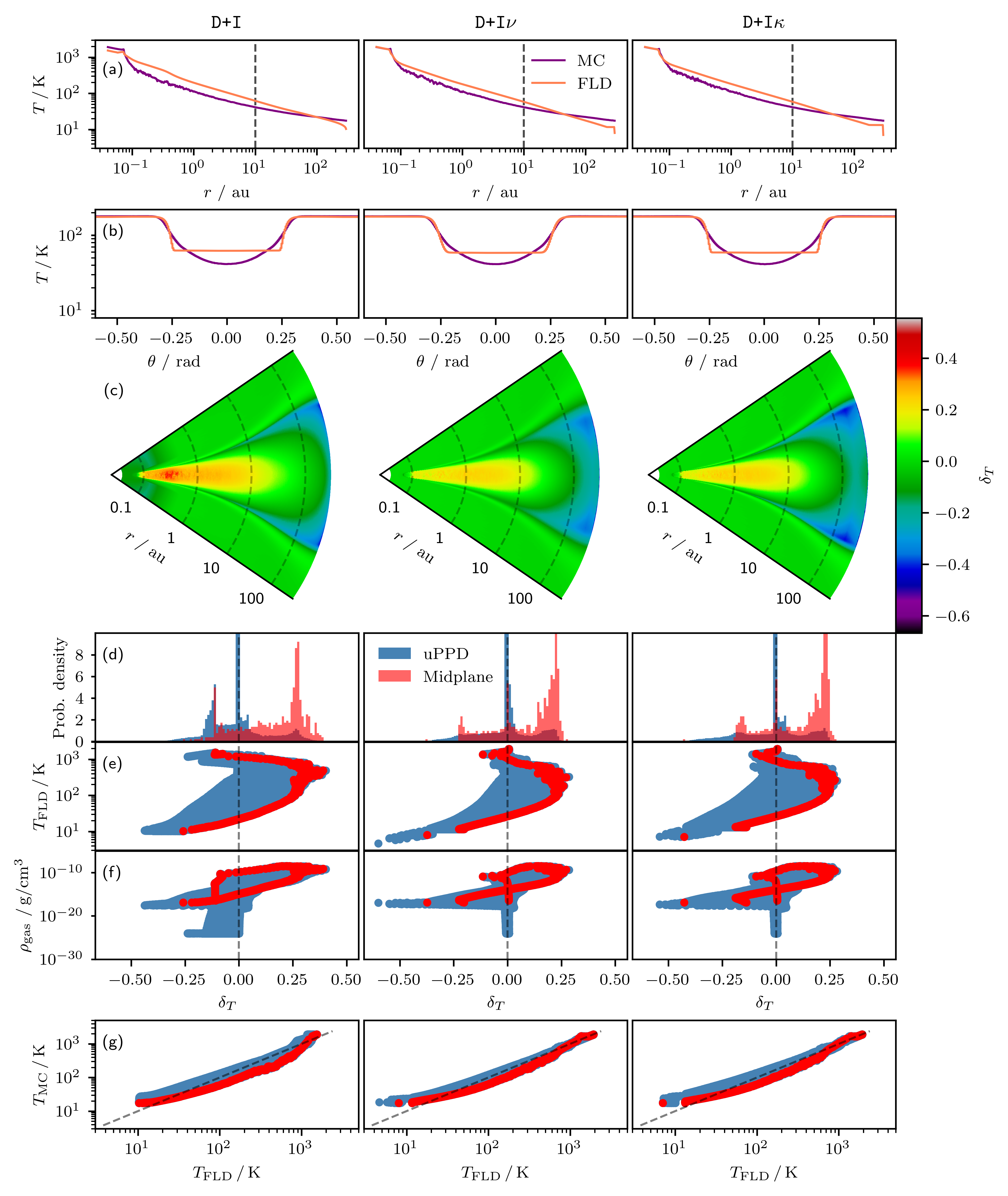}
      \caption{Temperature comparison overview. Similar to Fig. \ref{fig:temp_statistics_axisym_models} but now for models \mdI{}, \mdInu{}, and \mdIkappa{} (left to right column).}
         \label{fig:temp_statistics_axisym_models_IRR}
   \end{figure*}

\begin{figure*}[h!]
   \centering
   \includegraphics[width=\hsize]{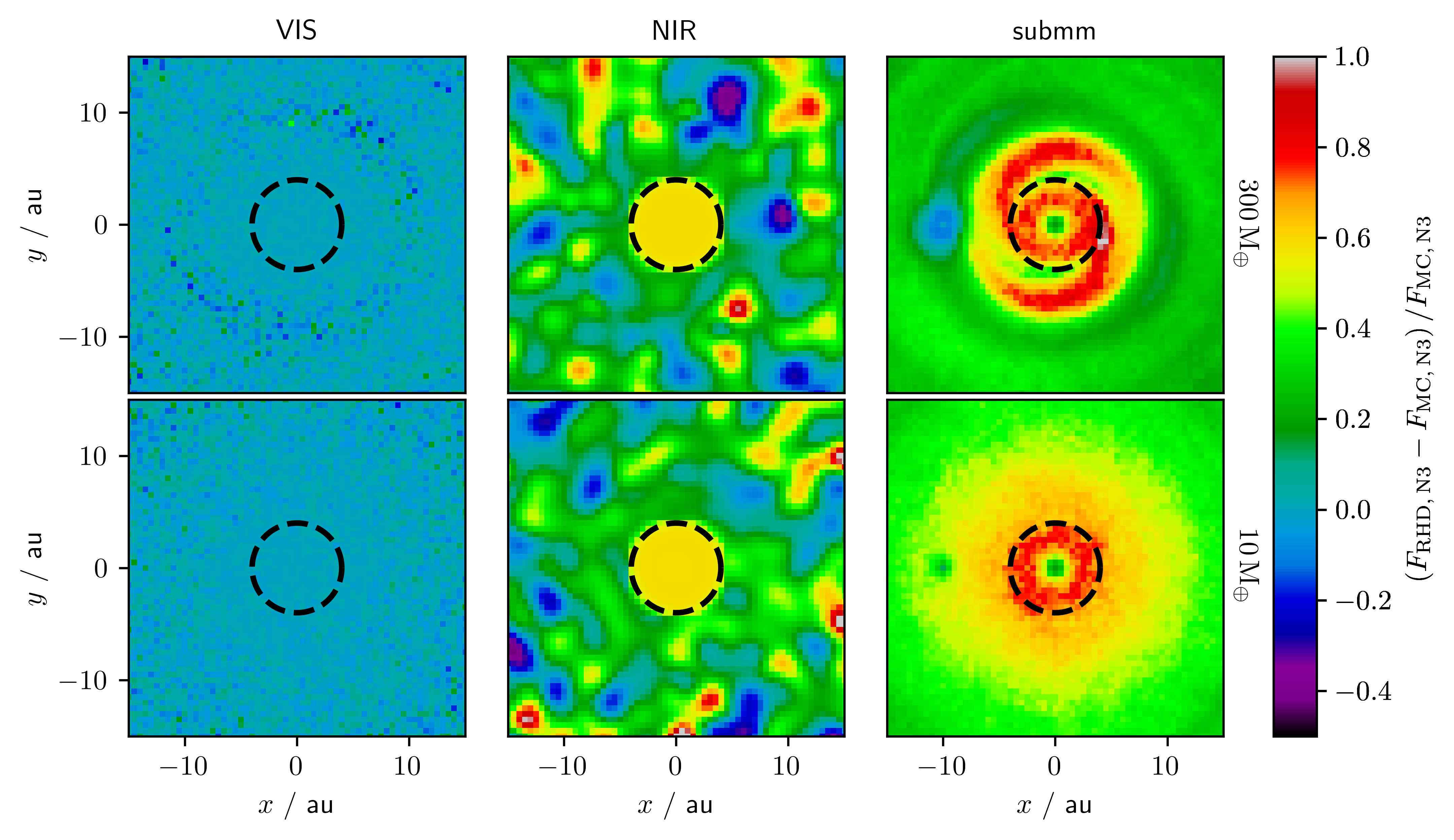}
      \caption{Similar to Fig. \ref{fig:flux_reldiff_N1}, but now assuming a resolution of \texttt{N3}. For the purpose of better comparability, the color bar is clipped.}
         \label{fig:flux_reldiff_N3}
   \end{figure*}

   \begin{figure*}[h!]
   \centering
   \includegraphics[width=\hsize]{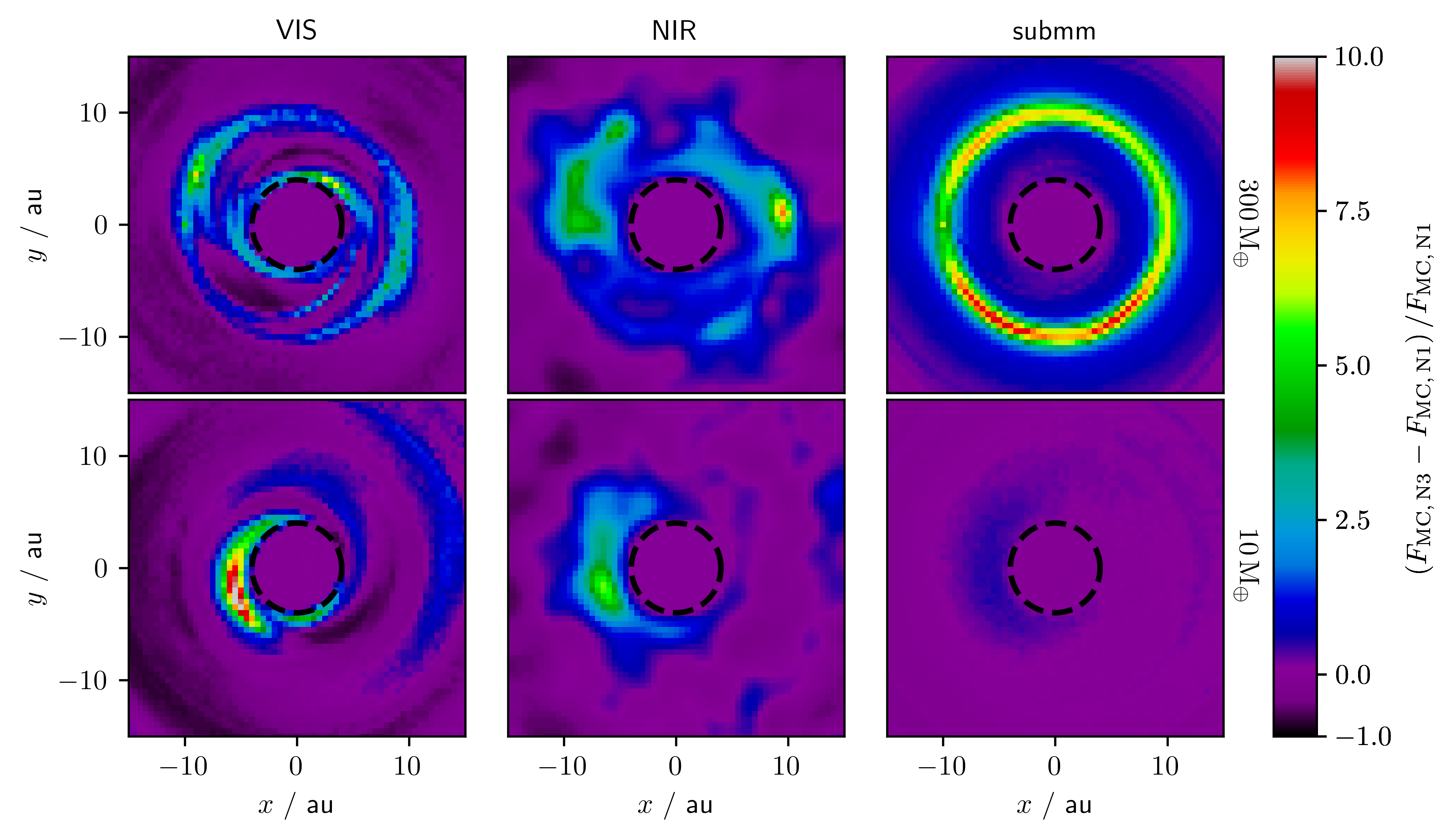}
      \caption{Similar to Fig. \ref{fig:flux_reldiff_N1}, but now the displayed relative flux differences are based on flux maps derived with models of differing resolutions, \texttt{N1} and \texttt{N3}. For the purpose of better comparability, the color bar is clipped.}
         \label{fig:flux_reldiff_resolution}
   \end{figure*}

   \begin{figure*}[h!]
   \centering
   \includegraphics[width=\hsize]{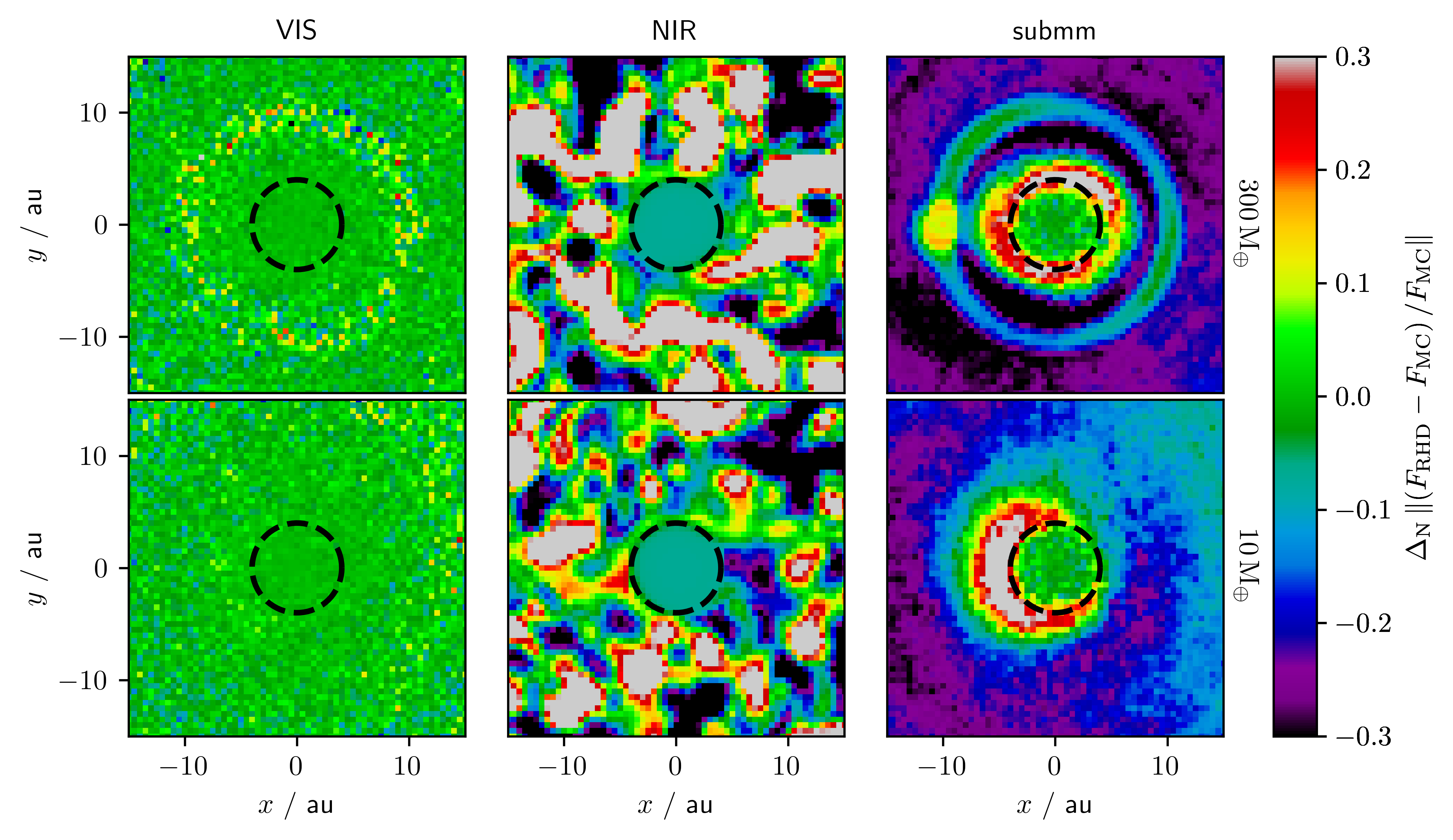}
      \caption{Similar to Fig. \ref{fig:flux_reldiff_N1}, but now showing the similarity change $\Delta S{\equiv} \sum_{\rm N} \left\lVert \left(F_{\rm RHD}-F_{\rm MC}\right)/F_{\rm MC} \right\rVert$ based on flux maps derived with models of differing resolutions, \texttt{N1} and \texttt{N3}, and simulation types (RHD and MCRT). For the purpose of better comparability, the color bar is clipped. (For details, see Sect. \ref{sec:flux_map_comparison}.)}
         \label{fig:flux_reldiff_similarity}
   \end{figure*}
   
\end{appendix}

\end{document}